\documentclass[twocolumn,showpacs,preprintnumbers,amsmath,amssymb,pre]{revtex4-1}
\usepackage{graphicx}
\usepackage{rotating}
\usepackage{bm}

\begin{document} \setcounter{page}{1} \setcounter{page}{1}
\title{Linear aggregation and liquid-crystalline
    order: comparison of  Monte Carlo simulation and analytic theory} 
\author{Tatiana Kuriabova}
\affiliation{Department of Physics and Liquid Crystal Materials
  Research Center, University of Colorado at Boulder} 
\affiliation{Institute for Complex Adaptive Matter, University of
  California, Davis}
\author{M. D. Betterton}
\affiliation{Department of Physics and Liquid Crystal Materials
  Research Center, University of Colorado at Boulder} 
\author{Matthew A. Glaser}
\affiliation{Department of Physics and Liquid Crystal Materials
  Research Center, University of Colorado at Boulder} 

\date{\today}

\begin{abstract}
Many soft-matter and biophysical systems are
  composed of monomers which reversibly assemble into rod-like
  aggregates.  The aggregates can then order into liquid-crystal
  phases if the density is high enough, and liquid-crystal ordering
  promotes increased growth of aggregates.  Systems that display
  coupled aggregation and liquid-crystal ordering include wormlike
  micelles, chromonic liquid crystals, DNA and RNA, and protein
  polymers and fibrils.  Coarse-grained molecular models that capture
  key features of coupled aggregation and liquid-crystal ordering
  common to many different systems are lacking; in particular, the
  role of monomer aspect ratio and aggregate flexibility in
  controlling the phase behavior are not well understood.  Here we
  study a minimal system of sticky cylinders using Monte Carlo
  simulations and analytic theory.  Cylindrical monomers interact
  primarily by hard-core interactions but can stack and bind end to
  end. We present results for several different cylinder aspect ratios
  and a range of end-to-end binding energies.  The phase diagrams are
  qualitatively similar to those of chromonic liquid crystals, with an
  isotropic-nematic-columnar triple point.  The location of the triple
  point is sensitive to the monomer aspect ratio.We find that the
  aggregate persistence length varies with temperature in a way that
  is controlled by the interaction potential; this suggests that the
  form of the interaction potential affects the phase behavior of the
  system.  Our analytic theory shows improvement compared to previous
  theory in quantitatively predicting the I-N transition for
  relatively stiff aggregates, but requires a better treatment of
  aggregate flexibility.
\end{abstract}
\maketitle 
\vspace{0.6cm}

\section{Introduction}

Self-assembly of aggregates is ubiquitous in soft matter and
biophysics \cite{hamley07}.  Aggregation requires only a pool of
monomers with some type of attractive interaction. Reversible
aggregation occurs when monomers interact via relatively weak,
non-covalent attractions that are typical of soft matter; in this case
the aggregates are in equilibrium with the pool of monomers.
Reversible aggregation is also called equilibrium polymerization or
supramolecular polymerization \cite{ciferri02}. While different
geometries of aggregates are possible (depending on the nature of the
attraction), in many cases anisotropic interactions favor linear or
filamentous aggregates.  These rod-like aggregates can form
liquid-crystal (LC) phases; the liquid-crystal order then couples to
the aggregation, often promoting the formation of longer aggregates in
the LC phase \cite{taylor93,ciferri99}. The liquid-crystal ordering of
the aggregates can occur even when the monomers alone do not form
liquid-crystal phases.

Since the ingredients required for coupled aggregation and LC ordering
are so simple, this basic physics occurs for a variety of systems,
including worm-like micelles and microemulsions, chromonic liquid
crystals, nucleic acids, proteins, and protein assemblies. Some of the
first work on coupled aggregation and LC order was inspired by
experiments on sickle-cell hemoglobin protein, a fiber-forming protein
important in sickle-cell anemia \cite{herzfeld81,hentschke89}; later
work has considered other peptides and proteins that form fibers or
fibrils \cite{aggeli03,ciferri07,lee09a,lee09b,park09,jung10}.  Other
early work considered worm-like micelles, formed either of amphiphilic
molecules in water or microemulsions of water and oil which are
stabilized by amphiphilic molecules. If the micelles are relatively
stiff they can display liquid-crystal phases \cite{khan96}; these
systems have been the subject of extensive experimental study
\cite{charvolin79,hendrikx81,hendrikx83,boden85,boden90,quist92,leaver93,furo95,johannesson96,vonberlepsch96,angelico00,pereira00,fischer01,goodchild07,kuntz08}.
Liquid-crystal phases have also been observed in folic acid salts
\cite{ciuchi94}, guanosine derivatives \cite{franz94,spindler02}, and
nucleosome core particles \cite{mangenot03}.  Recent work on chromonic
liquid crystals and nucleic acids has renewed interest in coupled
aggregation and LC order.  Chromonic LCs are formed from relatively
flat, disk-like dye molecules.  When the dye molecules are suspended
in water, hydrophobic interactions drive the stacking and formation of
rod-like aggregates that can form LC phases
\cite{tiddy95,harrison96,lydon98,vonberlepsch00,lydon04,horowitz05,nastishin04,nastishin05,prasad07,edwards08,park08,joshi09,wu09}.
Hydrophobically-driven end stacking also causes short pieces of DNA
and RNA to assemble into aggregates and form LC phases
\cite{nakata07,zanchetta08a,zanchetta08b}.

Extensive work on the analytic theory of aggregation and LC ordering
has been done since the early work of Herzfeld and Briehl
\cite{herzfeld81}. Much of the initial work focused on perfectly rigid
aggregates
\cite{herzfeld81,gelbart85,mcmullen85,herzfeld88a,herzfeld88b,hentschke89,taylor90,taylor91}.
However, when aggregates are assumed perfectly rigid and excluded
volume interactions are treated in the second-virial approximation, a
``nematic catastrophe'' occurs in which the aggregates in the nematic
phase grow infinitely long \cite{odijk87,vanderschoot94a}. Later work
therefore emphasized adding aggregate semiflexibility to the models
\cite{hentschke91a,hentschke91b,vanderschoot94b,vanderschoot94c,vanderschoot95,vanderschoot96,odijk96,kindt01,lu04,lu06}.
Analytic theory qualitatively reproduces results of experiments and
simulations, including a first-order I-N transition and longer
aggregates in the nematic phase. Some work has also predicted an
isotropic-nematic-columnar triple point \cite{taylor91}. However,
quantitative agreement between simulation and theory remains lacking,
particularly for aggregates with higher flexibility \cite{lu04,lu06}.

Surprisingly little simulation work has addressed coupled aggregation
and LC order. Some papers have focused on the molecular details that
promote aggregate formation in specific systems and considered single
aggregates\cite{bast96a,bast96b,tieleman00,mohanty06,yakovlev07,turner10}.
Simulating larger systems of interacting aggregates is too
computationally costly for atomistic simulation and coarse-grained
models are required.  Edwards, Henderson, and Pinning did pioneering
Monte Carlo simulations of disks formed from hard spheres, with
interactions that promoted disk stacking. They observed isotropic,
nematic, columnar, and crystalline phases \cite{edwards95}.  A similar
approach was used by Maiti et al.~to simulate formation of columnar
aggregates in chromonic liquid crystals. This work focused on
conditions required to find columnar aggregates and didn't study the
phase behavior \cite{maiti02}.  Rouault used a coarse-grained 2D
lattice model of wormlike micelles to study how varying the stiffness
affects the ordering transition; stiffer molecules display higher
orientational ordering \cite{rouault98b}.

More recently, a number of simulation papers have considered hard
spheres with an added anisotropic interaction that promotes linear
aggregation.  Hentschke and coworkers performed molecular dynamics and
Monte Carlo simulations of spheres with anisotropic Lennard-Jones
potential \cite{fodi00,ouyang07}.  They observed isotropic, nematic,
and columnar phases, and observed the narrowing and eventual
disappearance of the nematic phase in a I-N-C triple point as the
temperature is increased.  Similarly, Chatterji and Pandit added an
anisotropic interaction to hard spheres via a 3-particle potential
that favors linear aggregation \cite{chatterji03}. They observe a
first-order I-N transition in Monte Carlo simulations.  Some of the
most detailed simulation work of a simplified model was done by L\"u
and Kindt, who performed Monte Carlo simulation of spheres with sticky
``patches'' that allow assembly into linear aggregates
\cite{lu04,lu06}.  Their simulations did not show the columnar LC
phase but did see the I-N transition.

The key physical ingredients required for aggregation coupled to LC
order are (1) monomers with an anisotropic attractive interaction that
promotes the formation of long, thin filaments and (2) excluded volume
interactions between aggregates that promote liquid-crystalline order
at high aggregate density. While other physical effects such as
electrostatic interactions are clearly important in some systems
\cite{aggeli03,mangenot03,kroeger07,kuntz08,jung10}, a minimal
physical picture which incorporates filament formation and
excluded-volume interactions is a valuable starting point to
understand common features of the diverse experimental systems.

In this work we develop a coarse-grained model of aggregation-induced
LC order, and study the model using Monte Carlo simulations and
analytic theory.  
Our goal is to ultimately compare the simulation and
theoretical results to experimental data on chromonic and DNA liquid
crystals, so we have chosen the form of the model and the
approximations used to facilitate this comparison. 
Motivated by recent work on nucleic acids and chromonic liquid
crystals, we consider molecules with anisotropy both in shape and
interactions.  Therefore, our monomer is a cylinder of length $L$ and
diameter $D$, so the aspect ratio $r=L/D$. The cylinders are sticky:
if two cylinders stack so that their circular ends are sufficiently close
together, they experience an attractive interaction. Other than this
binding energy, the cylinders experience only a hard-core repulsion.
This coarse-grained model captures the main physical effects governing
linear aggregation and mesophase formation in these systems, namely
excluded volume interactions and end-to-end association.  For the
appropriate range of binding energy, density of cylindrical monomers,
and temperature, the Monte Carlo simulations of sticky cylinders find
linear aggregates and a range of phases, including isotropic, nematic,
columnar liquid crystal, and columnar crystal phases.

The hard cylinder system (corresponding to zero binding energy between
cylinders) has been studied previously in simulations by Blaak et al.
\cite{blaak99} and by Ibarra-Avalos et al.  \cite{ibarra-avalos07},
but to our knowledge this is the first simulation study of hard
cylinders with attractive interactions.

In simple (non-aggregating) LC systems, the rod aspect ratio is a
primary determinant of possible phases and the location of the phase
transitions. The role of monomer aspect ratio in controlling the phase
behavior of aggregating LC systems is not well understood.  Our
sticky-cylinder model enables systematic investigation of the
dependence of aggregation and phase behavior on monomer aspect ratio
$L/D$ for systems ranging from disk-like chromonic liquid crystals
($L/D \ll 1$) to duplex DNA oligomers ($L/D \gtrsim 1$).

In section \ref{sim_methods} we discuss the simulation model and the
Monte Carlo simulation methods, including cluster moves we used to
improve configurational sampling. Section \ref{theory_methods} outlines the
analytic theory, the free energy minimization, and the I-N phase
coexistence equations.  In section \ref{sim_results} we present the
simulation results, including the phase behavior, order parameter and
pair correlation functions, aggregate length distributions, and
aggregate persistence length. The comparison of analytic theory and
simulation for the I-N transition is presented in section
\ref{theory_results}, with a comparison both to our simulations and to
the work of L\"u and Kindt \cite{lu04}. Section \ref{conclusion} is
the conclusion. Appendix \ref{calc} describes some calculation details
of the analytic theory.

\section{Simulation methods}
\label{sim_methods}


The monomers are hard cylinders of length $L$ and diameter $D$ with
short-range attractive interactions between cylinder ends.  The
short-range attraction between cylinder ends is a generalized ramp
potential,
\begin{equation}
U(r) = \left\{
\begin{array}{ll}
- E_{\rm bond} \left[1 - \left( r / r_c \right)^{\gamma} \right], & r < r_c \\
0, & r \geq r_c
\end{array}
\right. ,
\label{sticky_potential}
\end{equation}
where $r$ is the distance between the centers of the circular end
faces of two neighboring cylinders. An exponent of $\gamma = 1$
corresponds to a linear ramp potential, while the limit $\gamma
\rightarrow \infty$ corresponds to a square well potential. We chose
$\gamma = 2$ and $r_c = D / 2$, parameter values that we empirically
found promote the formation of linear (as opposed to branched)
aggregates; larger values of $r_c$ and $\gamma$ give rise to branched
networks.

This model depends on three dimensionless parameters: cylinder aspect
ratio $L/D$, packing fraction $\phi = v_0 / v$ (or, alternatively,
dimensionless pressure $\beta P D^3$), and dimensionless binding
energy $\beta E_{\rm bond}$. Here $P$ is the pressure, $\beta = 1 /
(k_B T)$ is the inverse temperature, $v_0 = \pi D^2 L / 4$ is the
cylinder volume, and $v = V / N$ is the volume per particle (inverse
number density).  Unless otherwise noted, the cylinder diameter $D$ is
the unit of length.

We investigated systems of sticky cylinders with three aspect ratios,
$L/D = 0.5$, 1, and 2. We carried out $NPT$ Monte
Carlo simulations over a range of pressures for binding energies
ranging from $\beta E_{\rm bond} = 0$ to $\beta E_{\rm bond} = 12$.

The persistence length $l_p$ of sticky cylinder aggregates is
implicitly determined by the cylinder-cylinder pair interaction
potential; no explicit bond angle bending potential is included. An
implicit dependence of $l_p$ on the nature and strength of effective
intermolecular interactions is a key feature of nucleic acid and
chromonic aggregates, which may influence the temperature dependence
of the I-N coexistence curve. In this respect our model is more
specific than that of L\"{u} and Kindt \cite{lu04}, which included a
bond angle bending potential, enabling the binding energy and
persistence length to be varied independently.

\subsection{Monte Carlo Simulations}

We carried out Monte Carlo (MC) simulations of sticky cylinder systems
with periodic boundary conditions, with fixed monomer number $N$,
pressure $P$, and temperature $T$ \cite{frenkel01}.  Our simulations
employed several types of MC move, including small displacements and
rotations of single particles and changes in simulation cell volume
and shape under an applied pressure.  To improve configurational
sampling, we also used cluster moves (see below), and flip moves
($\pi/2$ reorientation of individual cylinders). An MC cycle includes
$N$ trial single-particle displacement/rotation moves, one trial
volume-changing move, and (in cases where these are used) $N/10$ trial
cluster moves and/or $N/10$ trial flip moves.

We simulated systems with between $N = 720$ and $N = 1920$ cylinders.
For each aspect ratio and binding energy we performed both an
expansion run (a series of simulations for decreasing pressure
starting from a high-density columnar crystal) and a compression run
(a series of simulations for increasing pressure starting from a
low-density isotropic fluid). For each value of the pressure, we
simulated $10^6$ MC cycles, and measured thermodynamic and structural
properties during the final $5 \times 10^5$ MC cycles.  To accommodate
crystalline and LC phases of arbitrary symmetry without defects or
strain, we utilized a fully flexible simulation cell, except in cases
where this led to highly elongated or deformed simulation cells (e.g.,
at low densities).

\subsubsection{Cluster moves.~~}

Long aggregates undergo slow effective translational and rotational
diffusion in simulations that utilize only single-particle MC moves.
To improve configurational sampling of aggregates, we used cluster
moves: simultaneous displacements and rotations of groups of
neighboring cylinders. Similar cluster moves have been used in
simulations of ionic fluids \cite{orkoulas94}. Clusters are defined by
proximity: any two cylinders with $r < r_{\rm clust}$ belong to the
same cluster.  The cluster cutoff $r_{\rm clust}$ can be varied to
control the average cluster size $\langle n_{\rm clust} \rangle$.  For
the simulations described here, we adjusted $r_{\rm clust}$ to
maintain $\langle n_{\rm clust} \rangle \approx 3$ for all pressures,
binding energies, and aspect ratios.  

A trial cluster MC move consists of the following five steps. 1.
Select a root cylinder at random, and find all cylinders in the
cluster containing the root cylinder. 2. Generate a trial displacement
and rotation of the cylinders in the cluster. The cluster is rotated
as a rigid body about an axis passing through the center of the root
cylinder. We performed two types of rotation move with equal
probability: spin moves, for which the axis of rotation is the axis of
the root cylinder, and tilt moves, for which the axis of rotation is a
randomly selected axis perpendicular to the root cylinder axis.  3.
Reject the trial move if cylinder-cylinder overlaps are found in the
new configuration.  4. Reject the trial move if the number of
cylinders in the cluster has changed (i.e., if cylinders not in the
original cluster are members of the cluster in the new configuration),
as such moves violate detailed balance.  5. If no overlaps are found,
and if the cluster size is unchanged, accept the move with probability
${\cal P} = \min[1, \exp(-\beta \Delta U)]$, where $\Delta U = U_{\rm
  final} - U_{\rm initial}$ is the change in potential energy.

Cluster moves were used in all simulations of cylinders having a
nonzero binding energy ($\beta E_{\rm bond} \geq 2$).  Flip moves were
previously introduced by Blaak et al. \cite{blaak99} to improve
configurational sampling for hard cylinders at high density, in the
vicinity of a partially ordered (cubatic-like) intermediate phase. We
used flip moves for small binding energy ($E_{\rm bond} \leq 2$) for
the same reason.

\begin{figure}
\includegraphics[width=0.45 \textwidth]{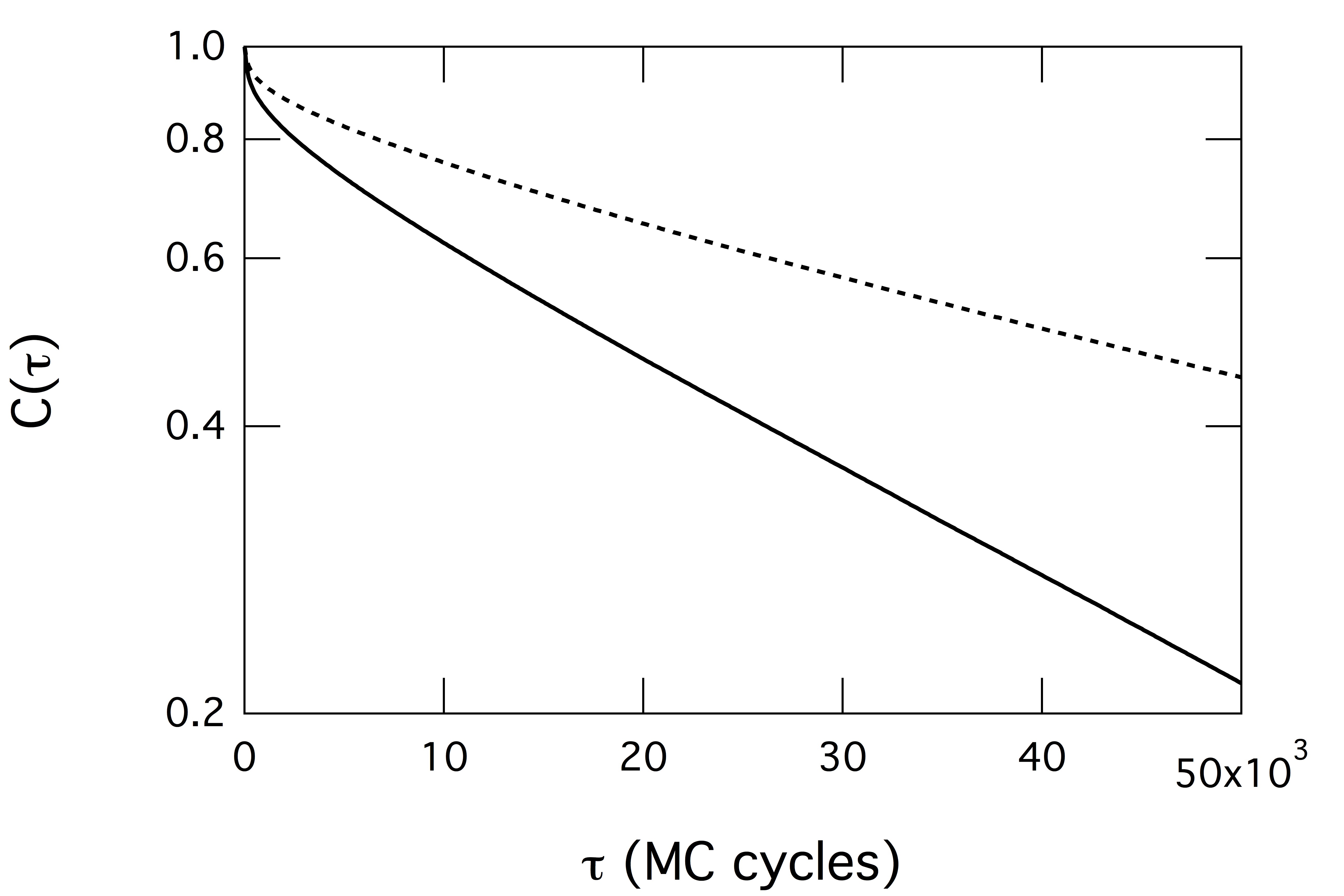}
\caption{\label{P1_acf} Orientational autocorrelation function
  $C(\tau)$ for an isotropic fluid of sticky cylinders with aspect
  ratio $L/D = 0.5$ and binding energy $\beta E_{\rm bond} = 12$ at a
  pressure of $\beta P D^3 = 0.8$. Results are shown for simulations
  with (solid curve) and without (dashed curve) cluster MC moves.
  Cluster moves decrease the orientational decorrelation time by
  approximately a factor of two.}
\end{figure}

Cluster moves moderately improve sampling efficiency, as measured by
the cylinder orientational decorrelation rate. The orientational
autocorrelation function is
\begin{equation}
C(\tau) = {{1} \over {N}} \sum_{i = 1}^N \left[ \left\langle \hat{\bf u}_i(t) \cdot \hat{\bf u}_i(t + \tau) \right\rangle
  -  \left\langle \hat{\bf u}_i(t) \right\rangle \cdot \left\langle \hat{\bf u}_i(t + \tau) \right\rangle \right],
\label{autocorrelation_function}
\end{equation}
where $\hat{\bf u}_i$ is a unit vector along the axis of cylinder $i$,
and the angle brackets denote an average over all time origins $t$
(with ``time'' measured in units of MC cycles).  In
Figure~\ref{P1_acf} we show the orientational autocorrelation function
$C(\tau)$ for an example simulation in the isotropic phase. In this
simulation, cylinders assemble into aggregates containing about 10
cylinders on average, which leads to slow effective rotational
diffusion.  Note that $C(\tau)$ decays exponentially for large $\tau$.
Cluster moves reduce the orientational correlation time by
approximately a factor of two, from $7.6 \times 10^4$ MC cycles in a
simulation with only single-particle moves (dashed curve) to $3.8
\times 10^4$ MC cycles in a simulation incorporating cluster moves
(solid curve). This decrease occurs despite the fact that the
acceptance rate for cluster moves is rather low (around $3.5\%$ in
this case).  Performing $N/10$ cluster moves per MC cycle adds
approximately $50\%$ to the computational cost of the simulation, so
the gain in efficiency is around $35\%$.  Similar performance was
found for other parameter values. More substantial improvements in
sampling efficiency will likely require more sophisticated schemes
such as configurational bias Monte Carlo \cite{frenkel01} or the
cluster cleaving method of Whitelam and Geissler \cite{whitelam07,
  whitelam08}.

\subsubsection{Overlap test.~~}

The test for cylinder-cylinder overlaps is more time-consuming than
the corresponding test for hard spheres, spherocylinders or
ellipsoids. We use a procedure similar to that used in previous
simulation studies of hard cylinders \cite{blaak99, ibarra-avalos07},
and that consists of three tests: 1. Spherocylinder overlap: test for
overlaps between the enclosing spherocylinders (obtained by adding
hemispherical endcaps to hard cylinders). If the spherocylinders don't
overlap, then neither do the hard cylinders.  2. Disk-disk overlap: if
an overlap is found in the first step, test for overlaps between the
flat end faces of the cylinders.  3. Disk-cylinder overlap: if no
overlaps are found in the second step, test for overlaps between the
flat end faces of one cylinder and the cylindrical surface of the
other.

If the enclosing spherocylinders overlap, but no cylinder-cylinder
overlaps are found, then the two cylinders may interact via the
short-range attractive potential (Eq.~\eqref{sticky_potential}).
Computing the attractive interaction between sticky cylinders requires
minimal additional effort, because the distance between the centers of
the circular end faces of the cylinders is available from the
spherocylinder overlap test.

\section{Theory}
\label{theory_methods}

To compare the I-N transition in simulation and theory, we extended
the work of van der Schoot and Cates \cite{vanderschoot94a}. This
analytic theory considers cylindrical monomers (of length $L$ and
diameter $D$) that reversibly assemble into linear aggregates. The
form of the free energy is determined from three key assumptions.
First, the aggregates are treated as nearly rigid cylinders. The ratio
of aggregate length $\ell$ to the persistence length $l_p$ is assumed
small, $\ell/l_p \ll 1$.  Second, the aggregates interact via steric
repulsion. Only pairwise interactions are taken into account within
the second virial approximation, but the Parsons-Lee approximation is
used to extend the theory to higher density.  Third, the assembly of
monomers into aggregates is driven by an energetic penalty $E_{\rm
  bond}$ for monomers to have free ends.  We assume that $E_{\rm
  bond}$ is independent of the aggregate length $\ell$.

The assumption of nearly rigid aggregates does not hold in our
simulations: the results shown in this paper have a ratio of aggregate
lengths to persistence length $\ell/l_p\sim 1$. We discuss below the
limitations of the theory in this flexibility regime.

The work of van der Schoot and Cates
considered two-particle interactions in the second virial
approximation. Since this approximation is accurate only in the dilute
limit, we used the Parsons-Lee approximation to extend the theory to
the regime of higher packing fractions \cite{parsons79}. The
Parsons-Lee method effectively inlcudes contributions from higher
virial coefficients in the interaction free energy.

The total free energy of the system depends on the density of monomers
$\rho_\ell({\bf u})$ that belong to a subset of aggregates of contour
lengths between $\ell$ and $\ell + d\ell$ and have orientations within
the solid angle $\Omega_{\bf u}$ and $\Omega_{\bf u} + d\Omega_{\bf
  u}$ with respect to some specified direction $\hat{\bf n}$.  The
monomers have length $L$, diameter $D$, aspect ratio $r=L/D$, and
volume $v_0=\pi D^2 L/4$.  The free energy per unit volume (in units
of $k_B T$) is
\begin{eqnarray}
\frac{F}{V} &=& \int d\ell\, \frac{d\Omega_{\bf u}}{4\pi}\, \frac{L}{\ell}\,\rho_\ell({\bf u})\,
\left[\log\left(\frac{v_0 L^2}{\ell}\rho_\ell({\bf u})\right)-1 \right]\nonumber\\
&&{}-\frac{2L}{3l_p}\,\int d\ell\, \frac{d\Omega_{\bf
    u}}{4\pi}\,[\rho_\ell({\bf u})]^{1/2} 
\nabla^2 [\rho_\ell({\bf u})]^{1/2}\nonumber \\
&&{}+ L^2\,D\, \eta(\phi)\, \int d\ell_1\,d\ell_2\,\frac{d\Omega_{{\bf u}_1}}{4\pi}\,\frac{d\Omega_{{\bf u}_2}}{4\pi}\rho_{\ell_1}({\bf u}_1)\times \nonumber\\
&&\rho_{\ell_2}({\bf u}_2)|\sin\gamma|
+ E_{\rm bond} \int d\ell\,\frac{d\Omega_{\bf u}}{4\pi}\,\frac{L}{\ell}\,\rho_\ell({\bf u}).
\label{eq-free-en}
\end{eqnarray}
All length integrals (over $\ell$) are from zero to infinity and all
angular integrals (over $\Omega$) are over $4 \pi$ of solid angle. The
first term is the free energy of a mixture of ideal gases of
aggregates.  Aggregates of different lengths and orientations are
treated as different chemical species. The second term describes
semiflexibility of aggregates; it is a perturbative correction to the
orientational degrees of freedom (implicitly included in the first
term) due to the flexibility of aggregates \cite{khokhlov82}. The
third term describes the excluded-volume interaction between pairs of
monomers with relative orientations described by $\cos \gamma\equiv
{\bf u}_1 \cdot {\bf u}_2$. The prefactor $\eta(\phi)$ reflects the
Parsons-Lee correction for higher virial coefficients
\cite{parsons79,cinacchi04,cuetos07} based on the Carnahan-Starling
expression for the correlation function of hard spheres \cite{carnahan69},
\begin{equation}
\label{eq-Parsons-factor}
\eta(\phi)=\frac{1}{4}\,\left(\frac{4-3\phi}{(1-\phi)^2}\right),
\end{equation}
where $\phi=\rho v_0$ is the packing fraction of monomers. In the
usual second-virial approximation $\eta(\phi)=1$.  The last term in
Eq.~(\ref{eq-free-en}) is the free energy of polymerization, which is
proportional to the total number of aggregates in the system.  In
writing Eq.~(\ref{eq-free-en}) we assumed that in equilibrium the
distribution of monomers is spatially homogenous and that the
conformational degrees of freedom of aggregates decouple from the
interaction degrees of freedom.

For a closed system the total number of monomers $N$ is fixed,
leading to the normalization condition
\begin{equation}
\label{eq-rho-norm}
\int d\ell \,\frac{d\Omega_{\bf u}}{4\pi}\,\rho_\ell({\bf
  u})=\frac{N}{V}. 
\end{equation}

To determine the I-N phase diagram we take the standard approach of
requiring mechanical and diffusive equilibrium of the two phases at
coexistence. We calculate the osmotic pressure $p^{(i,n)}$ and
chemical potential per monomer $\mu^{(i,n)}$ in each phase.
Coexistence requires
\begin{equation}
\label{eq-equil-conditions}
p^{(i)}=p^{(n)}, \quad \mu^{(i)}=\mu^{(n)}.
\end{equation}
The solution of Eqs.~(\ref{eq-equil-conditions}) provides information
about the packing fractions $\phi^{(i)}$ and $\phi^{(n)}$ of monomers
at the coexistence boundaries, the mean aggregation number $\langle n
\rangle$ (and distribution of aggregation numbers), as well as the
order parameter $S=\frac{1}{2}(\langle 3 \cos^2\theta\rangle - 1)$,
where $\theta$ is the angle between the monomer's axis and the
director $\hat{\bf n}$, and averaging is performed over all monomers
in the system.

In the sections below we outline the calculations and resulting
equations for the isotropic and nematic phases as well as the
numerical methods.
\subsection{Isotropic phase}
In the isotropic phase the monomers are distributed uniformly both in
space and in orientation, so $\rho^{(i)}_\ell({\bf
  u})=\rho_\ell^{(i)}$ independent of angle.  The free energy
functional Eq.~(\ref{eq-free-en}) for the total free energy of the
system in the isotropic phase reduces to
\begin{eqnarray}
\label{eq-free-enI}
 f^{(i)}&\equiv&F^{(i)}\cdot\frac{v_0}{V}=
\int d\ell\, \frac{L}{\ell}\rho^{(i)}_\ell\left[\log\Big( \frac{v_0
    L^2}{\ell} \rho^{(i)}_\ell\Big) -1\right] 
\nonumber\\
&&{}+ r \eta(\phi) \phi^2 + E_{\rm bond}\int d\ell
\frac{L}{\ell}\rho^{(i)}_\ell,  
\end{eqnarray}
where the monomer aspect ratio $r=L/D$ and the monomer packing
fraction $\phi=\rho v_0$.

Functional minimization of the free energy is subject to the
normalization condition Eq.~(\ref{eq-rho-norm}), which can be handled
with a Lagrange multiplier $\lambda$:
\begin{equation}
\label{eq-min-fi}
\frac{\delta (f^{(i)} + \lambda \int d\ell \rho_\ell^{(i)})}{\delta
  \rho^{(i)}_\ell} =0.
\end{equation}
This results in a length distribution of aggregates that is, as
assumed, independent of orientation
\begin{equation}
\rho^{(i)}_\ell =\frac{\ell\,\phi}{v_0 L^2 M_0^2}\, e^{-\ell/(L M_0)},
\end{equation}
with
\begin{equation}
\label{eq-M0}
M_0=\sqrt{\phi}e^{E_{\rm bond}/2}.
\end{equation}
This solution corresponds to an exponential distribution for the
density of aggregates $N^{(i)}_{\rm agg}(\ell)/V$ as a function of
aggregate length $\ell$:
\begin{equation}
\frac{N^{(i)}_{\rm agg}(\ell)}{V}=\frac{L}{\ell}\rho_\ell^{(i)}= \frac{\phi}{v_0 L M_0^2}\, e^{-\ell/(L M_0)}.
\end{equation}
The chemical potential per monomer is given by
\begin{equation}
\label{mu-is}
\mu^{(i)}=\frac{\partial f^{(i)}}{\partial \phi} =
r\,\phi\,(2\eta(\phi) + \phi\,\eta^\prime(\phi)) -\frac{e^{-E_{\rm
      bond}/2}}{\sqrt{\phi}}, 
\end{equation}
where $\eta^\prime(\phi)$ is the derivative of $\eta(\phi)$ with respect to $\phi$.
The osmotic pressure is
\begin{eqnarray}
\label{p-is}
p^{(i)}&=& -\frac{1}{v_0}(f^{(i)}- \phi\mu^{(i)}) =\frac{1}{v_0}\Big[ r\,\phi^2\,(\eta(\phi) +\phi\, \eta^\prime(\phi))
\nonumber\\
&&{} + e^{-E_{\rm bond}/2}\,\sqrt{\phi}\, \Big].
\end{eqnarray}

\begin{figure*}
\includegraphics[width=0.95 \textwidth]{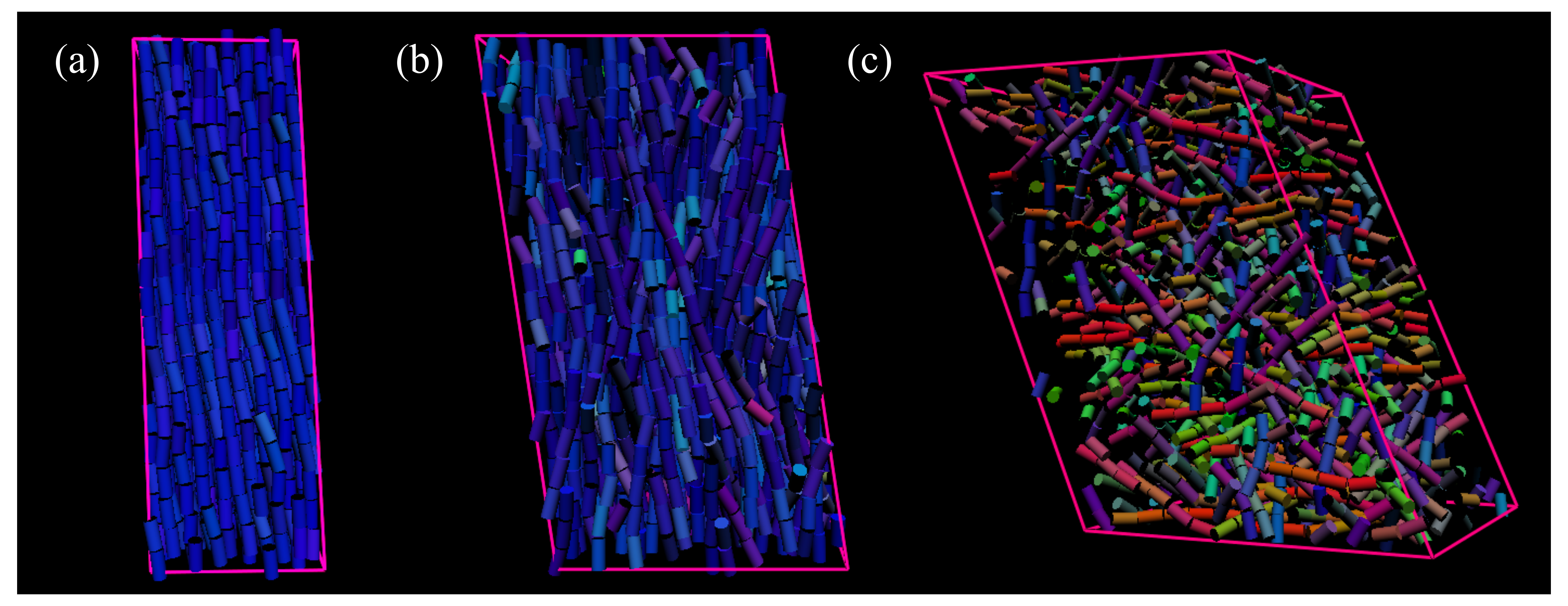}
\caption{\label{snapshots} Instantaneous configurations from $NPT$ MC
  simulations of $N = 1440$ $L/D = 2$ sticky cylinders with binding
  energy $\beta E_{\rm bond} = 12$ in various phases: (a) columnar
  LC phase, $\beta PD^3 = 0.7$; (b) nematic phase, $\beta
  PD^3 = 0.28$; (c) isotropic fluid phase, $\beta PD^3 = 0.18$.
  Cylinder color encodes orientation: the RGB color index of cylinder
  $i$ is $(|\hat{u}_{ix}|, |\hat{u}_{iy}|,|\hat{u}_{iz}|$), where
  $\hat{\bf u}_i$ is a unit vector along the cylinder axis.}
\end{figure*}

\begin{figure*}
\includegraphics[width=1.0 \textwidth]{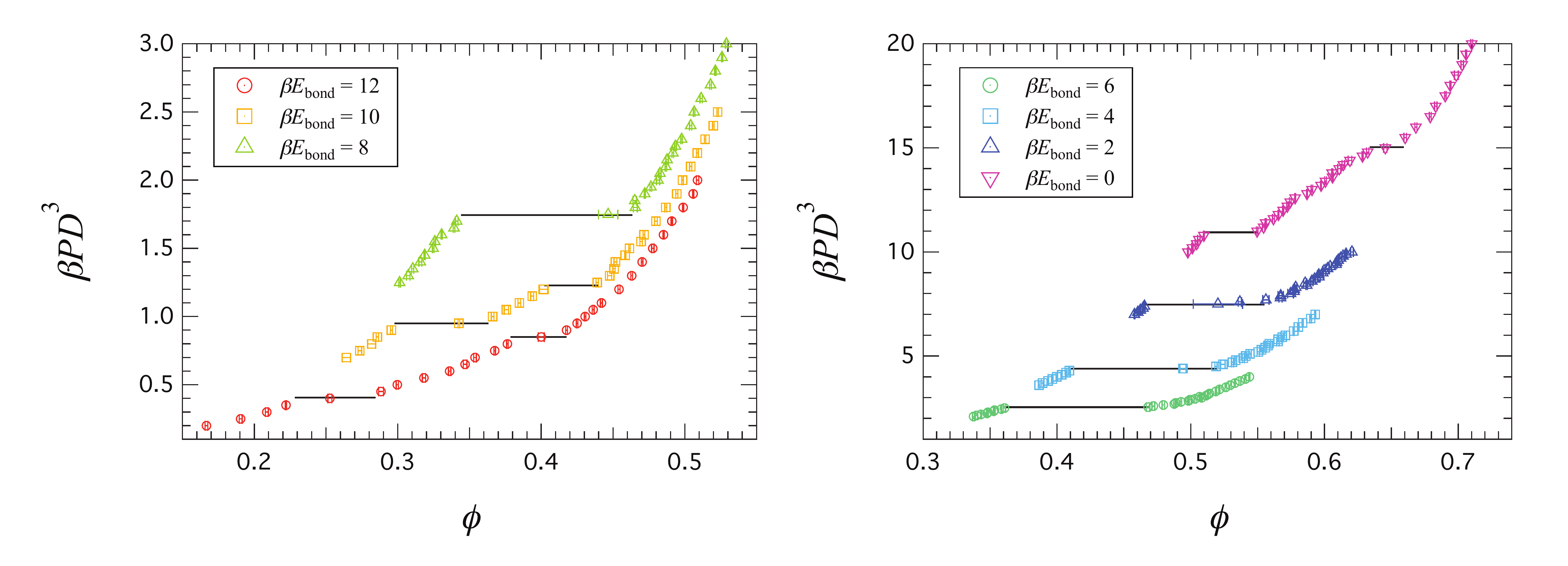}
\caption{\label{eos_L_1} Equation of state for $L/D = 1$ measured on
  expansion from a high-density columnar crystal state (symbols). The
  horizontal lines indicate the approximate location of first-order
  phase transitions.}
\end{figure*}
\subsection{Nematic phase}
In the nematic phase the system is spatially homogenous but
orientationally ordered due to alignment of aggregates along a
spontaneously chosen direction $\hat{\bf n}$. The system preserves
azimuthal symmetry with respect to rotations about $\hat{\bf n}$.
Therefore, in spherical coordinates the density $\rho_\ell({\bf u})$
is independent of the polar angle $\varphi$ and
\begin{equation}
\int \frac{d\Omega_{\bf u}}{4\pi}\, \rho_\ell({\bf u}) =
\frac{1}{2}\int_{-1}^1 dx\,\rho_\ell (x), 
\end{equation}
with $x=\cos\theta$, where $\theta$ is the angle to the director
$\hat{\bf n}$.  As in the previous section, we minimize the system's
free energy with respect to the orientation-dependent density function
$\rho_\ell(x)$,
\begin{equation}
\frac{\delta (F^{(n)} + \lambda \int d\ell \int_{-1}^1 dx \rho_\ell(x) )}{\delta \rho_\ell(x)}=0.
\end{equation}
This results in the integro-differential equation
\begin{eqnarray}
0&=&\lambda  + \frac{E_{\rm bond}}{2}\,\frac{L}{\ell}+ 
\frac{1}{2}\,\frac{L}{\ell}\log\left(\frac{v_0 L^2}{\ell}\rho_\ell(x)
\right)\nonumber \\
&-& 
\frac{L}{3l_p}\,\frac{\partial^2_x
  [\rho_\ell(x)]^{1/2}}{[\rho_\ell(x)]^{1/2}} + \frac{L^2
  D\,\eta(\phi)}{2} \times \nonumber\\ 
&{}& \int d\ell'\int_{-1}^1 dx'
\rho_{\ell'}(x')\,K(x',x). \label{eq:f-deriv} 
\end{eqnarray}
The interaction kernel is
\begin{equation}
\label{eq:sin-expand2}
K(x_1,x_2)=\int_0^{2\pi}\frac{d\phi_1}{2\pi}\frac{d\phi_2}{2\pi}|\sin\gamma|.
\end{equation}

Solving Eq.(\ref{eq:f-deriv}) for the function $\rho_\ell (x)$ that
minimizes the free energy is a difficult task that requires either
numerical solution or introduction of a trial function.  We chose a
mixed approach that starts with a trial function and uses numerical
solution of the resulting algebraic equations. We adopted the trial
function proposed in reference \cite{vanderschoot94a}:
\begin{equation}
\label{eq:trial}
\rho^{(n)}_\ell (x) = \frac{\ell\phi}{v_0 L^2 M_0^2}e^{-\ell/(L M(x))}, 
\end{equation}
where the parameter $M_0$ is given by Eq.~(\ref{eq-M0}) and $M(x)$ is
an orientation-dependent aggregation number.

The free energy of the nematic phase as a functional of $M(x)$ becomes
\begin{eqnarray}
f^{(n)}&\equiv &F^{(n)}\cdot\frac{v_0}{V}=
-\frac{\phi}{M_0}\int_{-1}^1 dx\, \bar{M}(x) \nonumber\\
&&{}{}+ \frac{r \phi^2}{\pi}\int_{-1}^1\!\!\int_{-1}^1 dx_1 dx_2  \bar{M}^2(x_1)\,\bar{M}^2(x_2)\,K(x_1,x_2)
\nonumber\\
&&-\frac{\phi}{6\bar{l}_p}\int_{-1}^1 dx \bar{M}(x)\partial_x^2 \,\bar{M}(x)
\label{free-en-M}.
\end{eqnarray}
The bars denote dimensionless variables: $\bar{l}_p=l_p/L$ and
$\bar{M}(x)\equiv M(x)/M_0$.  The next step is functional minimization
of the free energy to determine an integro-differential equation for
$\bar{M}(x)$. We find a series solution for $\bar{M}(x)$ by expanding
in Legendre polynomials, which turns the equation for $\bar{M}(x)$
into a system of algebraic equations that we solve numerically. For
details of the calculations, see Appendix A. Here we summarize our
results for the chemical potential per monomer $\mu^{(n)}$ and osmotic
pressure $p^{(n)}$ in the nematic phase:
\begin{eqnarray}
\mu^{(n)}&=&\frac{\partial f^{(n)}}{\partial\phi} = -\frac{e^{-E_{\rm bond}/2}}{2\sqrt{\phi}}\, I_1 -
\frac{1}{6\bar{l}_p}\,I_2\nonumber \\
&&{} +  \frac{\phi r}{\pi}(2\eta(\phi)+\phi\eta^\prime(\phi))I_3 \label{eq-muN}\\
p^{(n)}&=&-\frac{1}{v_0}(f^{(n)}-\phi\mu^{(n)})=\frac{1}{v_0}
\Big[\frac{\sqrt{\phi}}{2}\, 
e^{-E_{\rm bond}/2}I_1
\nonumber\\
&&{}+\frac{\phi^2\,r}{\pi}(\eta(\phi) + \phi\eta^\prime(\phi))I_3  \,\Big],
\end{eqnarray}
where $I_1$--$I_3$ are integrals involving $\bar{M}(x)$ that are defined in
Eqs.~(\ref{eq-M-I1})-(\ref{eq-M-I3}).

\begin{figure*}
\includegraphics[width=1.0 \textwidth]{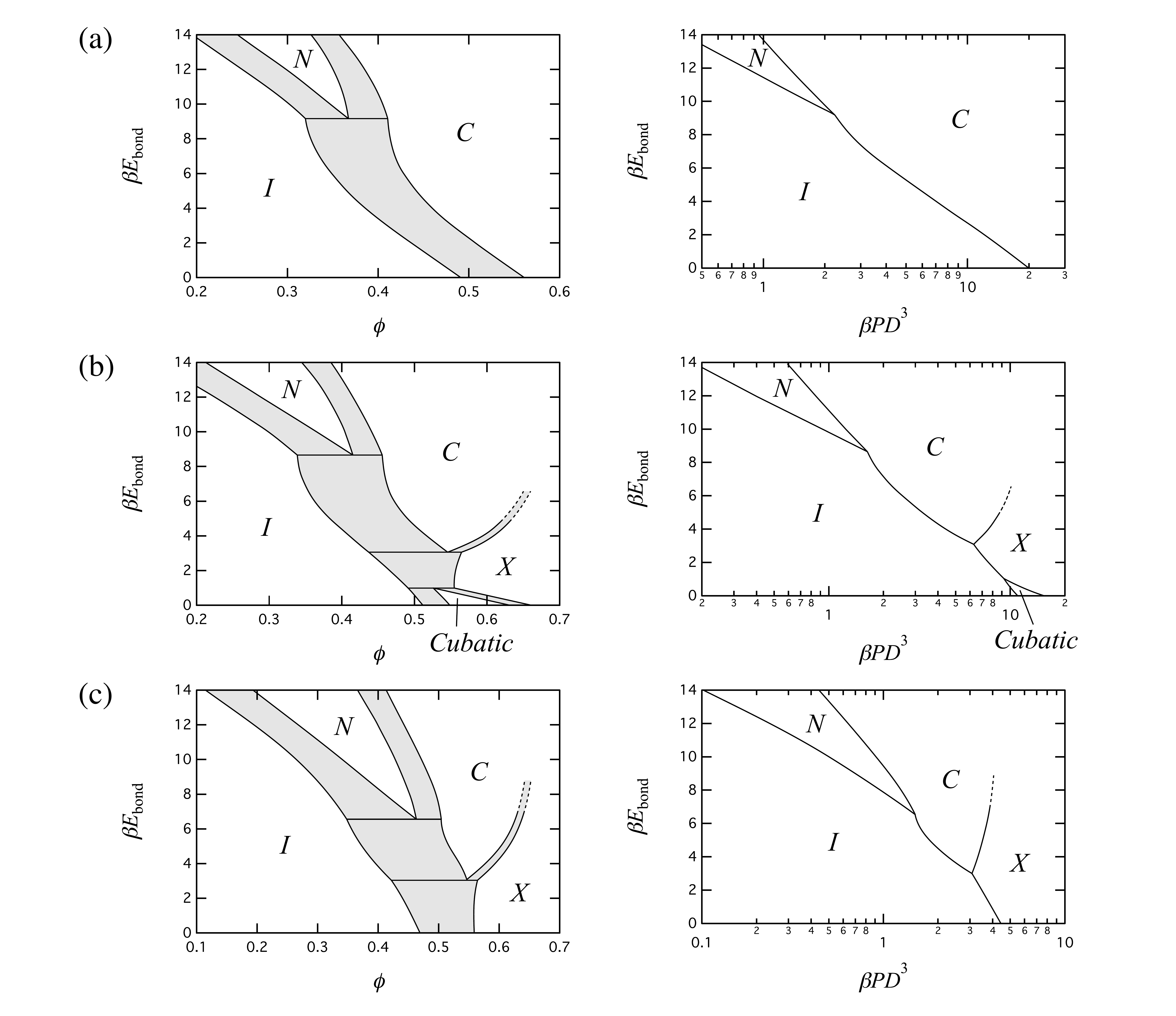}
\caption{\label{phase_diagrams} Phase diagrams of the sticky cylinder
  system as a function of binding energy and packing fraction (left)
  and binding energy and pressure (right), for aspect ratios $L/D =
  0.5$ (a), $L/D = 1$ (b), and $L/D = 2$ (c). A variety of phases are
  observed, including isotropic ($I$), nematic ($N$), columnar liquid
  crystal ($C$), columnar crystal ($X$), and cubatic-like ($Cubatic$)
  phases. Shaded areas correspond to regions of two-phase
  coexistence.}
\end{figure*}

The mean aggregation number in the nematic phase is 
\begin{equation}
\langle n \rangle=\frac{1}{L}
\frac{ \int d\ell \,\ell\, N^{(n)}_{\rm agg}(\ell)}
{\int d\ell N^{(n)}_{\rm agg}(\ell) },
\end{equation}
where 
\begin{equation}
N^{(n)}_{\rm agg}(\ell) = (L/\ell)\int_{-1}^1 \rho^{(n)}_\ell(x)dx.
\end{equation}
is the distribution function of aggregates of length $\ell$ in the
nematic phase.  The order parameter of aggregates of length $\ell$ is
\begin{equation}
S_\ell = \frac{\int_{-1}^1 dx P_2(x)\rho_\ell(x)}{\int_{-1}^1 dx
  \rho_\ell(x)}, 
\end{equation}
where $P_2(x)$ is the second-order Legendre polynomial.  The order
parameter (averaged over all monomers in the system) is
\begin{equation}
S=\frac{\int d\ell \int_{-1}^1 dx P_2(x)\rho_\ell(x)}{\int d\ell
  \int_{-1}^1 dx \rho_\ell(x)}. 
\end{equation}

\begin{figure*}
\includegraphics[width=1.0 \textwidth]{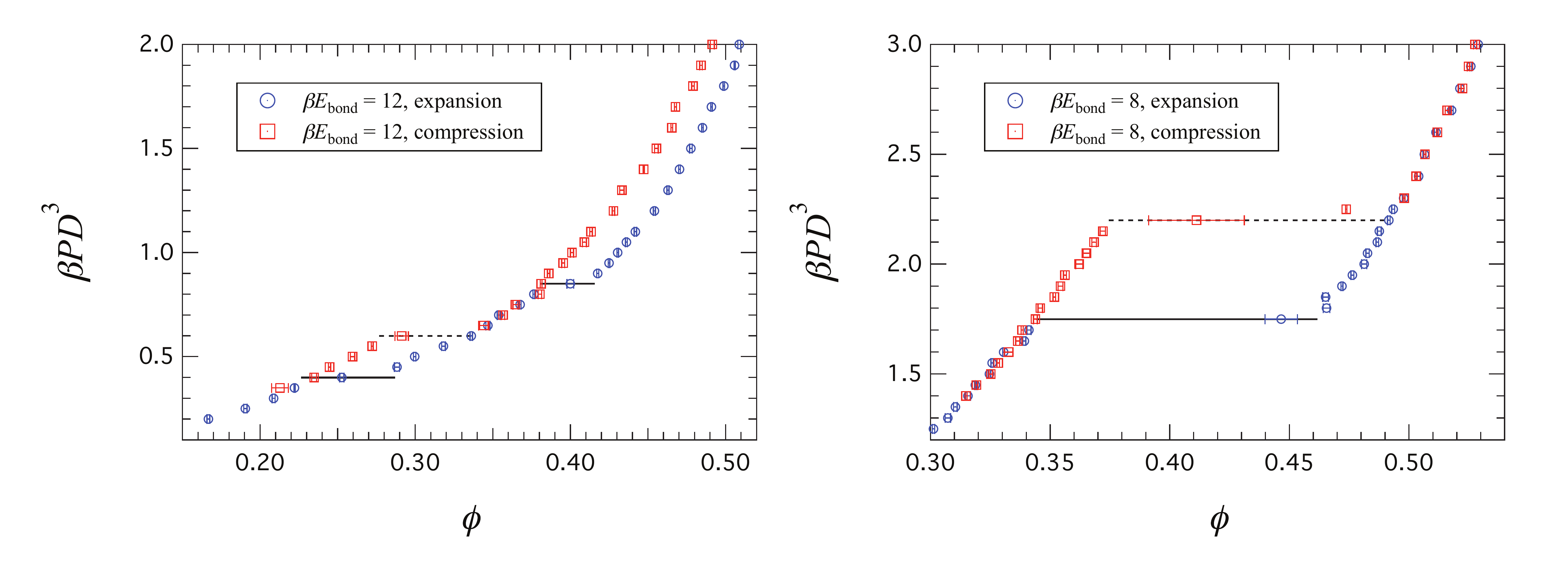}
\caption{\label{hysteresis_L_1} Hysteresis in the equation of state
  for $L/D = 1$ with $\beta E_{\rm bond} = 12$ (left) and $\beta
  E_{\rm bond} = 8$ (right). The horizontal lines indicate the
  approximate location of first-order phase transitions observed on
  expansion from a high-density columnar crystal state (solid lines)
  and on compression from a low-density isotropic fluid state (dashed
  lines).}
\end{figure*}
\begin{figure}
\includegraphics[width=0.5 \textwidth]{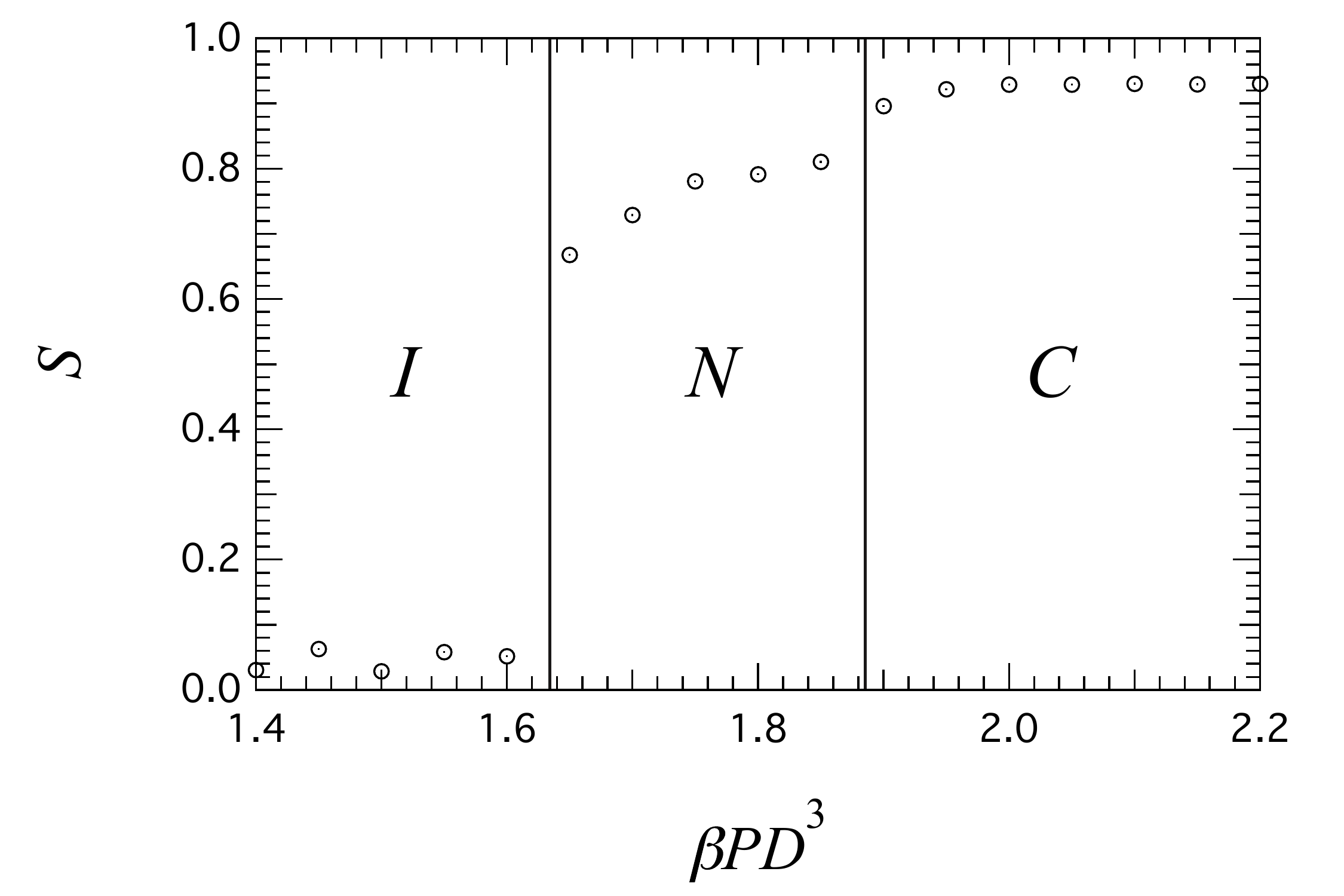}
\caption{\label{nematic_order_L_0p5} Nematic order parameter $S$ as a
  function of pressure for $L/D = 0.5$ and $\beta E_{\rm bond} = 10$.}
\end{figure}

\section{Simulation results}
\label{sim_results}

In Figure~\ref{snapshots} we illustrate the phase behavior with sample
snapshots of the simulations. Instantaneous configurations from
simulations of the $L/D = 2$ sticky-cylinder system with binding
energy $\beta E_{\rm bond} = 12$ are shown. This figure displays the
typical phase sequence observed upon expansion from a high-density
columnar crystal for large binding energy, which includes the columnar
LC phase, the nematic phase, and the isotropic fluid phase.

To determine phase boundaries, we determined the locations of plateaus
in the equation of state obtained on expansion from a high density
columnar crystal state. Such plateaus indicate first-order phase
transitions.  Phase identification was based on structural properties
such as the pair-correlation function and nematic order parameter, as
discussed below. For example, the equation of state for $L/D = 1$ and
various values of $\beta E_{\rm bond}$ is shown in
Figure~\ref{eos_L_1}, with the approximate location of first-order
phase transitions indicated by horizontal solid lines.

Phase diagrams of the sticky-cylinder system for aspect ratios $L/D =
0.5$, 1 and 2 are shown in Figure~\ref{phase_diagrams}.  All three
phase diagrams share the same key features. First, both nematic ($N$)
and columnar LC ($C$) phases are observed for large $\beta E_{\rm
  bond}$. The range of nematic stability (in terms of pressure or
packing fraction) increases with increasing binding energy. Second, an
isotropic-nematic-columnar triple point is observed at intermediate
binding energies. For binding energies below the triple point, there
is a direct transition from the isotropic phase to the columnar LC
phase.

We also find a columnar crystal ($X$) phase at high density, and the
$L/D = 1$ system exhibits a cubatic-like phase for small $\beta E_{\rm
  bond}$. The cubatic-like phase was previously described by Blaak et
al. \cite{blaak99} in the $L/D = 0.9$ hard cylinder system. As the
focus of this paper is on LC phases, we have not investigated the
crystalline or cubatic phases in detail, and the phase boundaries
involving these phases should be regarded as schematic.

Comparison of the equation of state obtained on compression from an
isotropic fluid state with that obtained on expansion from a
high-density columnar crystal state reveals considerable hysteresis
(Figure~\ref{hysteresis_L_1}). This hysteresis becomes increasingly
pronounced with decreasing $\beta E_{\rm bond}$, and compression from
the low-density isotropic state for small $\beta E_{\rm bond}$
typically leads to a high-density disordered (jammed) state, likely
due to the dominance of hard-core interactions in this limit. For
large $\beta E_{\rm bond}$, the I-N transition occurs at relatively
low packing fractions, which facilitates annealing into a well-ordered
nematic state starting from an isotropic state. Because we generally
observed slow kinetics of formation of an ordered state starting from
disordered (isotropic) initial configuration, we used isotherms
obtained on expansion (e.g., Figure~\ref{eos_L_1}) to construct the
phase diagrams in Figure~\ref{phase_diagrams}. This implies that the
phase boundaries shown are lower limits (both in density and
pressure), and should be considered provisional. In future work we
will refine the phase diagram using free energy calculations.

\begin{figure*}
\includegraphics[width=1.0 \textwidth]{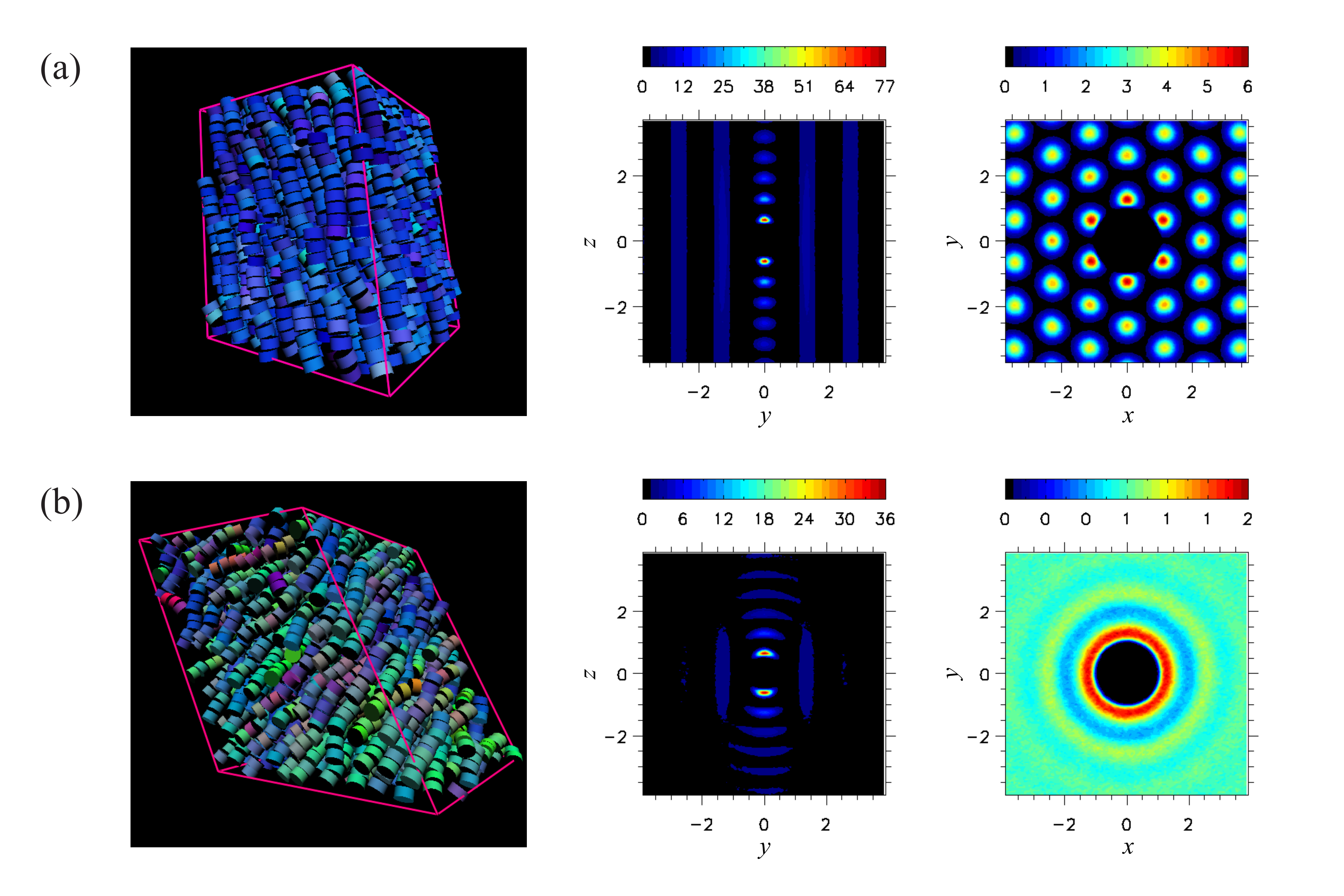}
\caption{\label{pair_dist_L_0p5_be_12} Instantaneous configurations
  (left) and pair distribution functions in planes containing the
  nematic director $\hat{\bf z} \equiv \hat{\bf n}$ ($g(0, y, z)$,
  center) and perpendicular to the nematic director ($g(x, y, 0)$,
  right) for $L/D = 0.5$ and $\beta E_{\rm bond} = 12$. (a) Columnar
  LC phase at $\beta PD^3 = 1.8$, obtained by expansion
  from a high-density columnar crystal.  There are no long-range
  correlations in the $z$ positions of cylinders, as shown by $g(0, y,
  z)$, and a well-defined hexagonal lattice of columns is apparent in
  $g(x, y, 0)$. (b) Nematic LC phase at $\beta PD^3 =
  1.1$, obtained by compression from a low-density isotropic fluid.
  Again, there are no long-range correlations in the $z$ positions of
  cylinders, as shown by $g(0, y, z)$. The plot of $g(x, y, 0)$
  reveals isotropic and rapidly decaying positional correlations in
  the plane perpendicular to $\hat{\bf z}$, indicative of nematic
  order.}
\end{figure*}

\subsection{Order parameter and correlation functions}

We measured nematic order by calculating the tensor
\begin{equation}
  Q_{\alpha \beta} = {1 \over N} \sum_{i = 1}^N \left( {3 \over 2}
    \hat{u}_{i \alpha} \hat{u}_{i \beta}  - {1 \over 2} \delta_{\alpha
      \beta} \right), 
\end{equation}
where $\alpha$ and $\beta$ refer to cartesian components, $i$ indexes
the sum over monomers, and $\delta_{\alpha \beta}$ is the Kronecker
delta function. The nematic order parameter $S$ is the largest
eigenvalue of the average order tensor $\langle Q_{\alpha \beta}
\rangle$, and the nematic director $\hat{\bf n}$ is the corresponding
eigenvector. Typical results for $S$ as a function of pressure are
shown in Figure~\ref{nematic_order_L_0p5}, for $L/D = 0.5$ and $\beta
E_{\rm bond} = 10$. The order parameter jumps abruptly from a small
value ($S \approx 0.05$) in the isotropic phase to $S \approx 0.7 -
0.8$ in the nematic phase, and exhibits a further abrupt jump to $S
\approx 0.9$ in the columnar LC phase. We list values of the nematic
order parameter in the nematic or columnar liquid crystal phases at
coexistence ($S_N$ or $S_C$) with the isotropic phase in
Table~\ref{averages}. The values of $S$ are generally large ($> 0.7$),
indicating a high degree of orientational order in all LC phases
investigated.

To distinguish between the various orientationally ordered phases we
computed the three-dimensional pair distribution function $g({\bf r})$
and examined positional correlations as a function of pair separations
parallel and perpendicular to the nematic director.  The pair
distribution function $g({\bf r})$ is defined as the normalized
probability of finding a pair of particles with separation ${\bf r}$,
\begin{equation}
  g({\bf r}) = {1 \over {\rho N}} \left\langle \sum_{i = 1}^N \sum_{j
        \neq i} \delta({\bf r} + {\bf  r}_j - {\bf r}_i)
  \right\rangle, 
\end{equation}
where $\rho = N/V$ is the average number density of particles and
$\delta({\bf r})$ is the Dirac delta function.

Figure~\ref{pair_dist_L_0p5_be_12} shows example plots of pair
distribution functions.  The nematic director defines the $z$
direction ($\hat{\bf z} \equiv \hat{\bf n}$), so $g(0, y, z)$ is in a
plane containing the nematic director and $g(x, y, 0)$ is in a plane
perpendicular to the nematic director. For this set of simulations,
$L/D = 0.5$ and $\beta E_{\rm bond} = 12$. The columnar LC
phase shown in Figure~\ref{pair_dist_L_0p5_be_12}(a) is characterized
by a well-defined hexagonal lattice of columns and by the absence of
long-range correlations in the $z$ positions of cylinders. The nematic
LC phase shown in Figure~\ref{pair_dist_L_0p5_be_12}(b)
exhibits isotropic and rapidly decaying positional correlations in the
plane perpendicular to $\hat{\bf n}$.

\begin{figure}
\includegraphics[width=0.5 \textwidth]{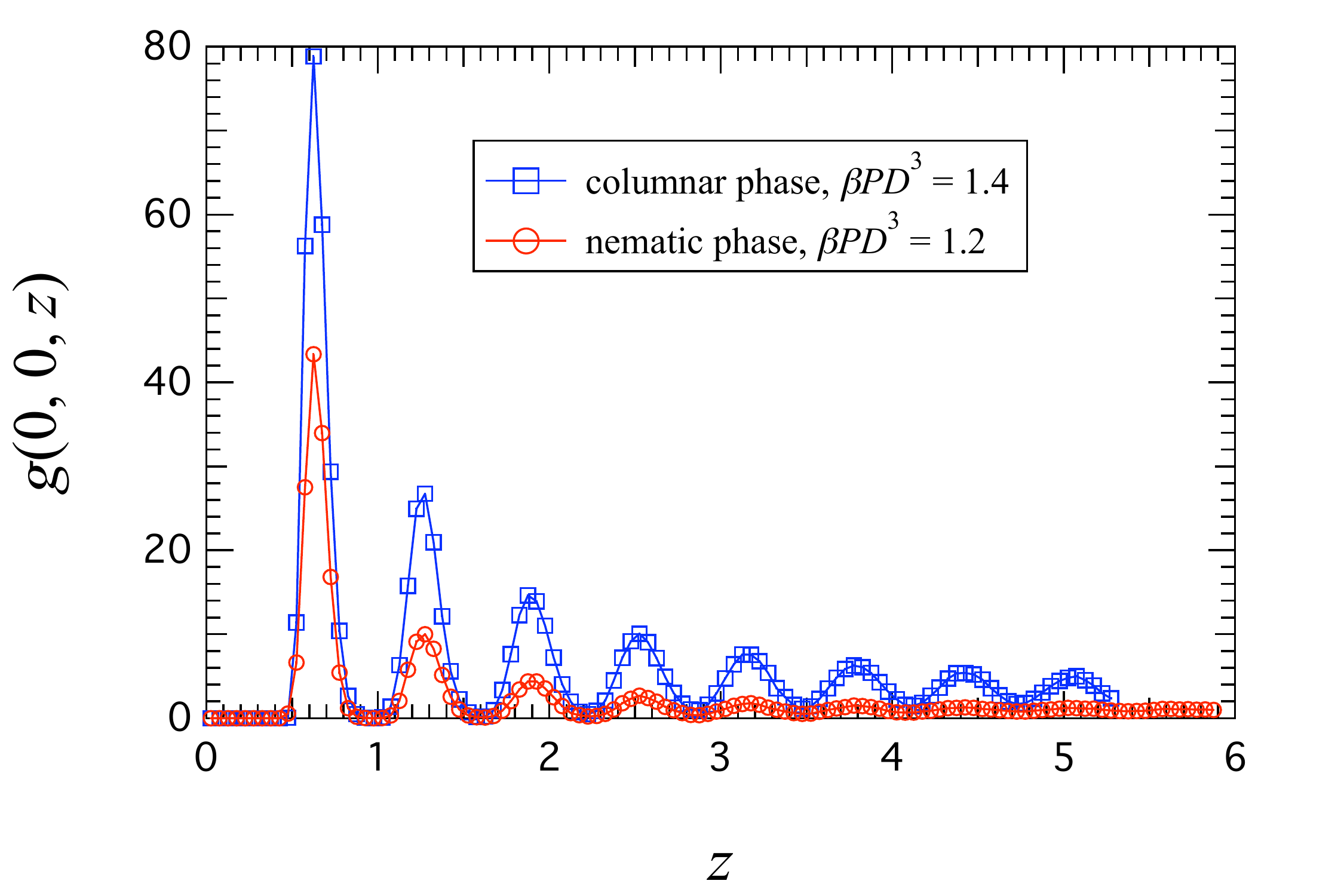}
\caption{\label{g_r_par_L_0p5} One-dimensional pair distribution
  function $g(0, 0, z)$ along a line parallel to the nematic director
  in the columnar LC phase (blue squares) and the nematic
  phase (red circles) for $L/D = 0.5$ and $\beta E_{\rm bond} = 12$.}
\end{figure}

We computed the positional correlation length $\xi_\parallel$ along
the nematic director in both the nematic and columnar LC phases from
the one-dimensional pair distribution function along a line parallel
to the nematic director, $g(0, 0, z)$.  Figure~\ref{g_r_par_L_0p5}
shows typical results for the columnar LC and nematic phases for $L/D
= 0.5$ and $\beta E_{\rm bond} = 12$.  We find exponential decay of
correlations to a constant in both phases (in the nematic phase $g(0,
0, z)$ approaches 1 for large $z$, while in the columnar LC phase we
estimated that $g(0, 0, z)$ approaches 3.5 for large $z$). The
correlation length determined by fitting the exponential decay is
$\xi_\parallel = 1.3$ in the nematic phase at $\beta PD^3 = 1.2$ and
$\xi_\parallel = 1.8$ in the columnar LC phase at $\beta PD^3 = 1.4$.
In the same simulations, the mean aggregate lengths are 44 in the
nematic phase and 352 in the columnar LC phase. This gives a ratio of
aggregate length to correlation length that is large:
$\ell/\xi_\parallel \approx 33$ in the nematic phase and
$\ell/\xi_\parallel \approx 200$ in the columnar LC phase.  (For more
discussion of aggregate lengths, see Section~\ref{aggregates} below.)
In other words, we find that the correlation length can differ from
the mean aggregate length by 1-2 orders of magnitude.  This in turn
suggests that it may not be feasible to determine the mean aggregation
number from X-ray scattering experiments that measure correlation
lengths, such as the experiments of references \cite{park08,joshi09}.

\begin{figure*}
\includegraphics[width=1.0 \textwidth]{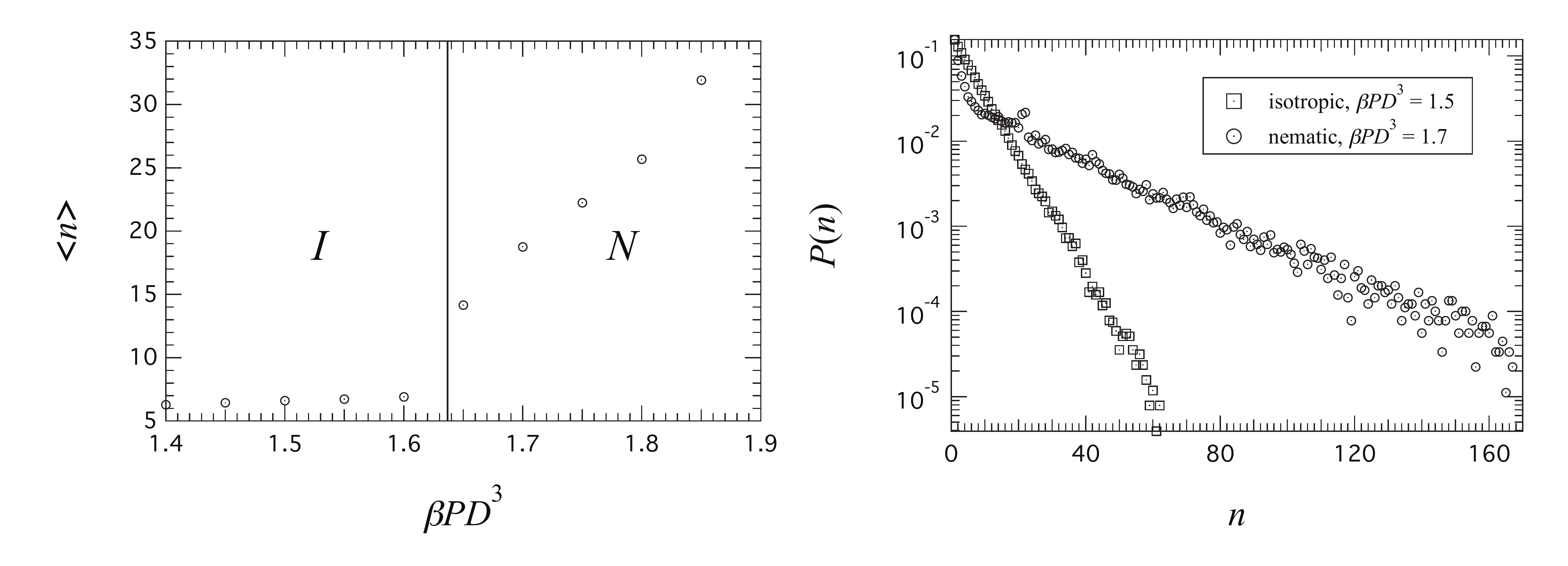}
\caption{\label{agg_num} Mean aggregation number $\langle n \rangle$
  as a function of pressure (left) and aggregation number
  distributions $P(n)$ (right) for $L/D = 0.5$ and $\beta E_{\rm bond}
  = 10$, measured on expansion from a high-density columnar crystal
  state. The nematic phase $P(n)$ displays fast exponential decay for
  small $n$ crossing over to slow exponential decay for large $n$
  (bi-exponential distribution). The small peak near $n = 20$ in the
  nematic aggregation number distribution is a finite-size artifact,
  associated with aggregates that connect to themselves across the
  periodic boundaries.}
\end{figure*}

\subsection{Aggregate Statistics}
\label{aggregates}

We measured the aggregation number distribution function $P(n)$ and
its first moment, the mean aggregation number $\langle n \rangle$, in
the isotropic, nematic, and columnar LC phases. Here $n$
is the aggregation number (the number of monomers in an aggregate).
Aggregates are defined by proximity: any pair of cylinders with
endpoint separation smaller than $r_c$ belongs to the same aggregate.

The use of a finite simulation volume imposes a cutoff on the
aggregate-length distribution. This finite-size effect is particularly
evident in the limit of large binding energies, where the mean
aggregate length can exceed the maximum linear dimension of the
periodic simulation cell. To ameliorate this effect, the initial
columnar crystal configurations used for expansion runs were
constructed with an oblique orientation of columns with respect to the
edge vectors of the periodic simulation cell, so that a given column
traverses the periodic box multiple times before intersecting itself.
With this expedient, we are able to measure aggregation number
distributions over a wide range of conditions, although small
finite-size artifacts associated with aggregates that connect to
themselves across the periodic boundaries are observed in many of the
expansion runs (see Figure \ref{agg_num}).

We illustrate the typical aggregation behavior in
Figure~\ref{agg_num}, which shows the mean aggregation number $\langle
n \rangle$ as a function of pressure (left) and aggregation number
distributions $P(n)$ in the isotropic and nematic states (right) for
$L/D = 0.5$ and $\beta E_{\rm bond} = 10$, measured upon expansion
from a high-density columnar crystal state.  The mean aggregation
number jumps discontinuously at the I-N transition, and increases
strongly with increasing pressure in the nematic phase. This behavior
is qualitatively consistent with that observed in previous simulation
and theoretical studies of semiflexible linear aggregates.  The
distribution $P(n)$ has an exponential dependence on $n$ in the
isotropic phase, but displays biexponential behavior in the nematic
phase, i.e., fast exponential decay for small $n$ crossing over to
slow exponential decay for large $n$. Such biexponential behavior has
been noted previously by L\"u and Kindt \cite{lu04}, who argued that
the short length scale behavior of $P(n)$ comes from a distinct
population of short aggregates with relatively low orientational
order. Biexponential behavior in the nematic phase is also predicted
by the analytic theory; see Figure \ref{lu_distrib}.  In the
simulations we also observed a biexponential aggregation number
distribution in the columnar LC phase.

Mean aggregation numbers in the isotropic phase at coexistence
($\langle n \rangle_I$) and in the nematic or columnar LC phase at
coexistence with the isotropic phase ($\langle n \rangle_N$ or
$\langle n \rangle_C$) are listed in Table~\ref{averages}. The mean
aggregation number at coexistence tends to increase with increasing
$\beta E_{\rm bond}$, although the trend in the nematic or columnar LC
phase at coexistence isn't strictly monotonic.  We note that the mean
aggregation numbers at coexistence decrease with increasing cylinder
aspect ratio for a given $\beta E_{\rm bond}$, possibly because the
corresponding packing fractions at coexistence decrease with
increasing aspect ratio.

\begin{table*}
  \caption{\label{averages} Properties of the sticky-cylinder system
    measured in $NPT$ MC simulations. Columns contain the
    cylinder aspect ratio $L/D$, the dimensionless binding energy $\beta
    E_{\rm bond}$, the number of particles $N$, the measured persistence
    length $l_p$, the mean intra-aggregate bond length $\langle R
    \rangle$, the mean aggregation number in the isotropic phase at
    coexistence $\langle n \rangle_I$, the mean aggregation number in
    the nematic or columnar LC phase at coexistence with the isotropic
    phase ($\langle n \rangle_N$ or $\langle n \rangle_C$), and the
    nematic order parameter in the nematic or columnar phase at
    coexistence with the isotropic phase ($S_N$ or $S_C$).} 
\begin{ruledtabular}
\begin{tabular}{cccccccccc}
$L/D$ & $\beta E_{\rm bond}$ & $N$ & $l_p$ & $\langle R \rangle$ & $\langle n \rangle_I$ & $\langle n
\rangle_N$ & $\langle n \rangle_C$ & $S_N$ & $S_C$ \\
\hline \hline
0.5 & 12 & 1920 & 14.9 & 0.66 & 9.96 & 24.08 & - & 0.704 & - \\
       & 10 & 1680 & 13.5 & 0.67 & 6.91 & 18.75 & - & 0.729 & - \\
       & 8 & 1680 & 11.4 & 0.68 & 4.06 & - & 45.90 & - & 0.908 \\
       & 6 & 1440 & 9.2 & 0.69 & 2.74 & - & 27.44 & - & 0.895 \\
       & 4 & 1440 & 7.2 & 0.70 & 2.10 & - & 15.85 & - & 0.872 \\
       & 2 & 1200 & 5.4 & 0.70 & 1.78 & - & 14.96 & - & 0.869 \\
       & 0 & 1200 & - & - & - & - & - & - & - \\
1 & 12 & 1440 & 31.4 & 1.15 & 5.65 & 14.98 & - & 0.761 & - \\
    & 10 & 1200 & 27.3 & 1.16 & 4.15 & 16.38 & - & 0.841 & - \\
    & 8 & 1200 & 22.3 & 1.17 & 2.83 & - & 58.42 & - & 0.947 \\
    & 6 & 1200 & 17.7 & 1.17 & 1.89 & - & 12.95 & - & 0.903 \\
    & 4 & 960 & 13.3 & 1.18 & 1.55 & - & 11.04 & - & 0.902 \\
    & 2 & 960 & - & - & - & - & - & - & - \\
    & 0 & 720 & - & - & - & - & - & - & - \\
2 & 12 & 1440 & 62.6 & 2.14 & 3.80 & 11.88 & - & 0.840 & - \\
    & 10 & 1200 & 51.0 & 2.15 & 2.54 & 11.15 & - & 0.873 & - \\
    & 8 & 1200 & 44.4 & 2.16 & 1.82 & 7.99 & - & 0.870 & - \\
    & 6 & 960 & 32.2 & 2.16 & 1.43 & - & 19.58 & - & 0.965 \\
    & 4 & 960 & 24.3 & 2.17 & 1.22 & - & 9.31 & - & 0.956 \\
    & 2 & 720 & - & - & - & - & - & - & - \\
    & 0 & 720 & - & - & - & - & - & - & - \\
\end{tabular}
\end{ruledtabular}
\end{table*}

\subsection{Persistence Length}
\label{persistence}

As discussed above, the persistence length of sticky-cylinder
aggregates is determined by cylinder-cylinder interactions. We
measured the persistence length from the bond orientational
correlation function,
\begin{equation}
C_O(s) = \left\langle \hat{\bf R}_i \cdot \hat{\bf R}_{i+m} \right\rangle,
\end{equation}
where $\hat{\bf R}_i$ is a unit vector along the $i$th bond (the
vector joining the centers of two neighboring cylinders) in an
aggregate, the contour length $s = m \langle R \rangle$, and $\langle
R \rangle$ is the average bond length. In the isotropic phase,
$C_O(s)$ decays exponentially with increasing $m$, with a decay
constant that can be identified with the persistence length $l_p$,
$C_O(s) = \exp(- s / l_p)$.  By fitting this exponential decay to the
simulation data, we determined the persistence length.
For a given $L/D$ and $\beta E_{\rm bond}$, we find that the
calculated persistence length is nearly independent of pressure within
the isotropic phase, which gives us confidence that this procedure
measures the intrinsic bending rigidity of aggregates.
Table~\ref{averages} lists the persistence length and average bond
length as a function of $L/D$ and $\beta E_{\rm bond}$, where both
quantities are averaged over isotropic state points. 

\begin{figure}
\includegraphics[width=0.5 \textwidth]{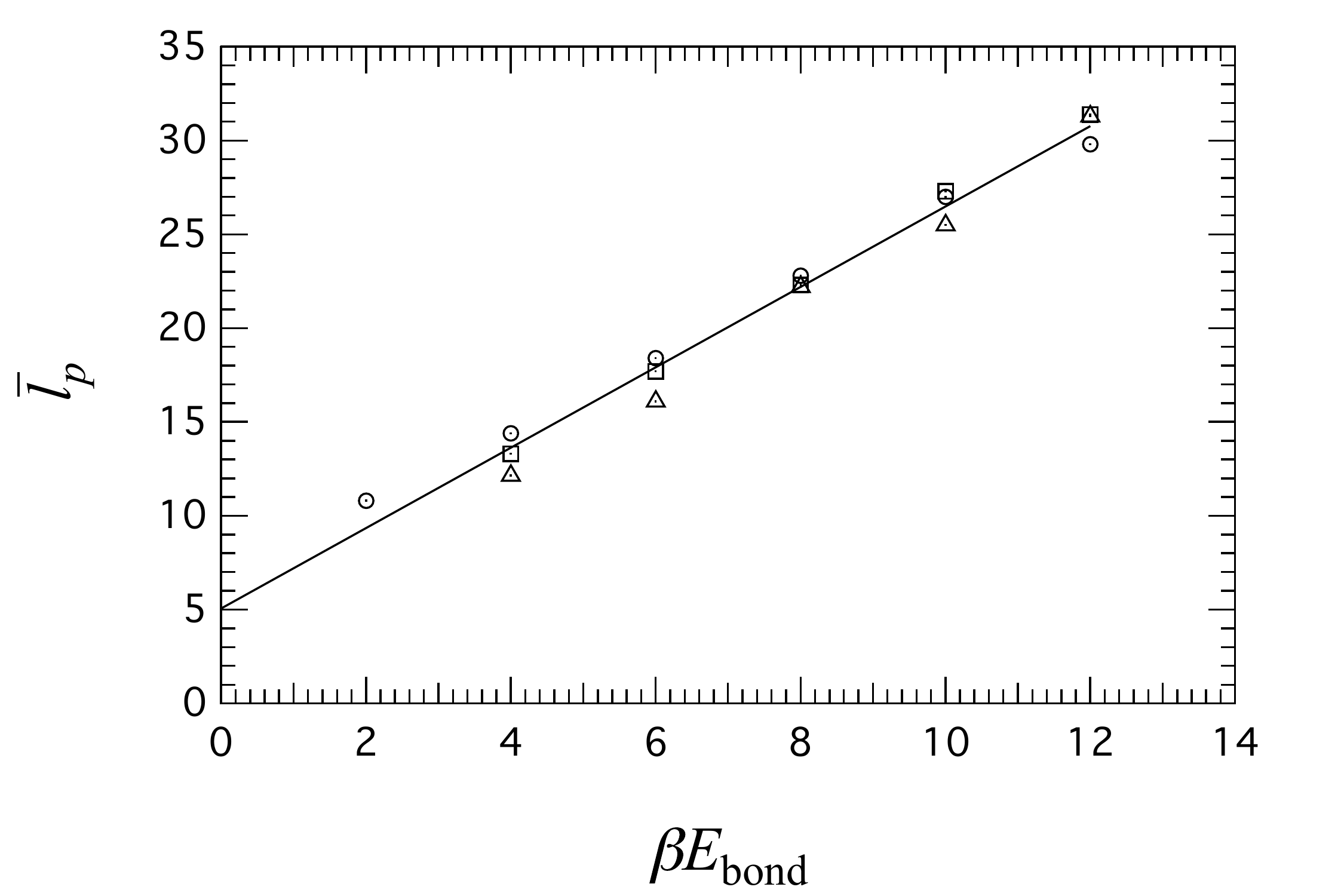}
\caption{\label{red_persist_length} Ratio of persistence length to
  monomer length $\bar{l}_p = l_p / L$ as a function of $\beta E_{\rm
    bond}$ for aspect ratios $L/D = 0.5$ (circles), $L/D = 1$
  (squares), and $L/D = 2$ (triangles). The data for all three aspect
  ratios is well approximated by a universal linear function,
  $\bar{l}_p = 5.07 + 2.14 \beta E_{\rm bond}$, which represents the
  best linear fit to the data (solid line).}
\end{figure}

The persistence length increases approximately linearly with
increasing binding energy, and is also approximately proportional to
cylinder aspect ratio $L/D$ for a given binding energy. Because of
this, the reduced persistence length $\bar{l}_p = l_p/L$ is reasonably
well approximated as a universal linear function of $\beta E_{\rm
  bond}$, independent of cylinder aspect ratio, as shown in
Figure~\ref{red_persist_length}.  The solid line in
Figure~\ref{red_persist_length} is the best linear fit to the data,
$\bar{l}_p = 5.07 + 2.14 \beta E_{\rm bond}$. The ratio of persistence
length to monomer length approaches a nonzero constant in the limit of
zero binding energy. This indicates that there is a significant
entropic contribution to the persistence length due to hard-core
interactions between neighboring cylinders in an aggregate.

We expect that the precise way that $\bar{l}_p$ varies with $\beta
E_{\rm bond}$ to depend strongly on the form of the intermolecular
potential. This dependence can be estimated by simple dimensional
analysis arguments. When two bound cylinders interact, thermal
fluctuations will typically cause their separation to vary to about 1
$k_B T$ above their minimum interaction energy $E_{\rm min} = -E_{\rm
  bond}$.  Therefore a typical separation $r_t = R/\sqrt{\beta E_{\rm
    bond} }$. This suggests a characteristic angle defined by the
distance $r_t$ and the cylinder radius $R$: $\tan \theta = r_t/R =
1/{\sqrt{\beta E} }$. The typical angle is therefore $ \cos \theta =
1/\sqrt{ 1 + 1/(\beta E_{\rm bond})}$.  This gives an estimate
of the persistence length, based on the relation $\langle \cos \theta
\rangle = e^{-L/l_p}$ and assuming $\beta E_{\rm bond} \gg 1$, of
$\bar{l}_p = 2 \beta E_{\rm bond}$.  In other words, the persistence
length should vary linearly with $\beta E_{\rm bond}$, with a slope of 2, as
observed in simulations.

For other forms of the potential, this dependence would change. Using
the same dimensional analysis argument, we would predict that for an
exponent of $\gamma$ in the interaction potential (Eqn.
\eqref{sticky_potential}), the scaling is $\bar{l}_p = 2 (\beta E_{\rm
bond})^{2/\gamma}$. In the limit of a square well ($\gamma \to
\infty$) the persistence length would become independent of $\beta
E_{\rm bond}$.  Because the phase behavior of aggregates is strongly
dependent on the persistence length, this implies a strong dependence
of the phase behavior on the \textit{form} of the monomer-monomer
interaction potential, in addition to the obvious dependence on the
binding energy. In the future it would be interesting to investigate
the interplay of these two effects, focusing how sticky-cylinder phase
behavior varies with the form of the interaction potential.

The aggregates in this study are relatively flexible: the ratio of
persistence length to mean aggregate length for the nematic phase at
coexistence ranges from $\approx 0.8$ for $L/D = 0.5$ and $\beta
E_{\rm bond} = 12$ to $\approx 2.9$ for $L/D = 2$ and $\beta E_{\rm
  bond} = 12$.  

\section{Theory-simulation comparison}
\label{theory_results}

We compared our analytic theory results both to the simulations of
L\"u and Kindt \cite{lu04} and to our simulations. These comparison
illustrate the key role of aggregate flexibility in determining the
theory-simulation agreement.

\begin{figure}
  \includegraphics[width=0.5
  \textwidth]{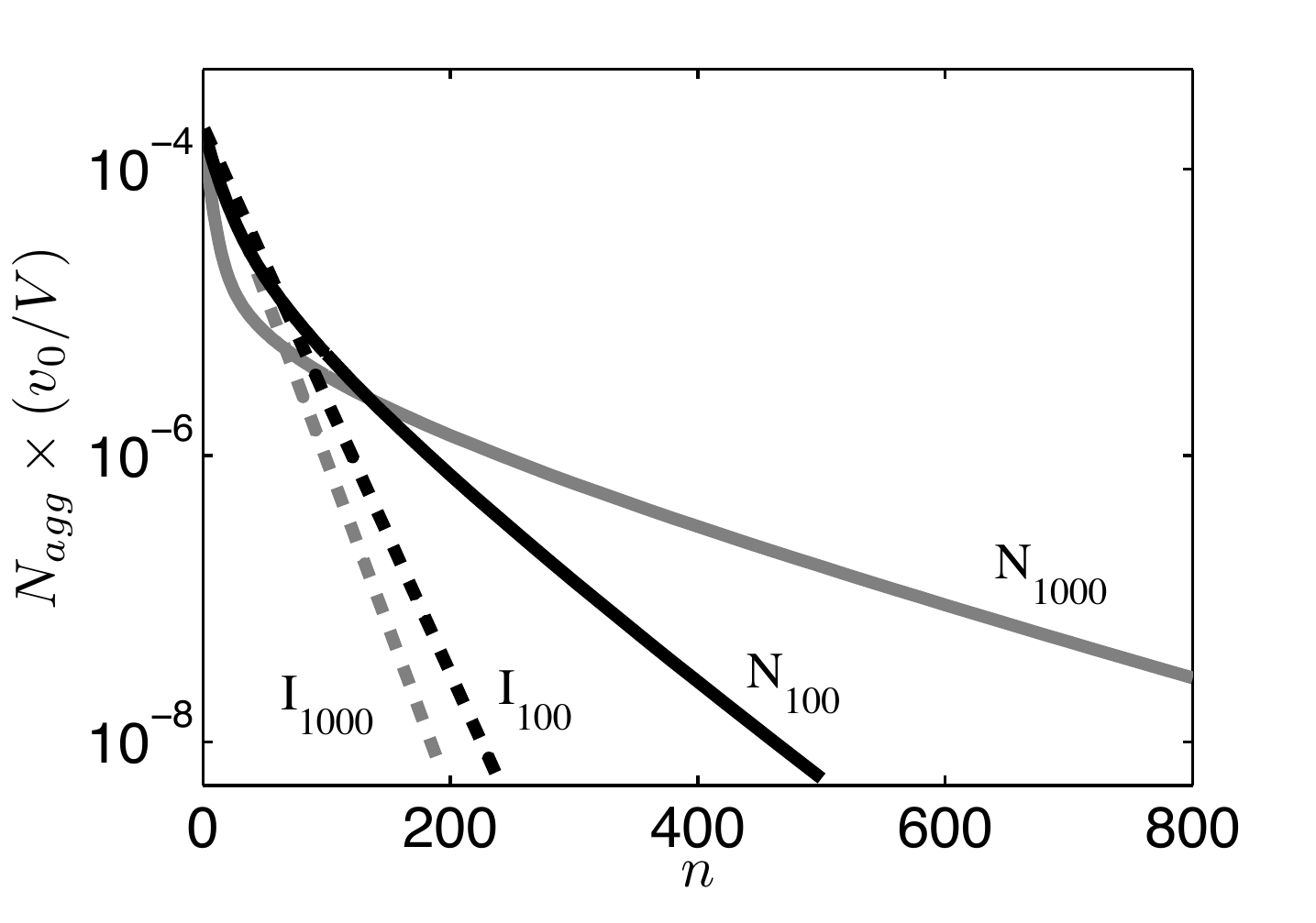}
  \includegraphics[width=0.5
  \textwidth]{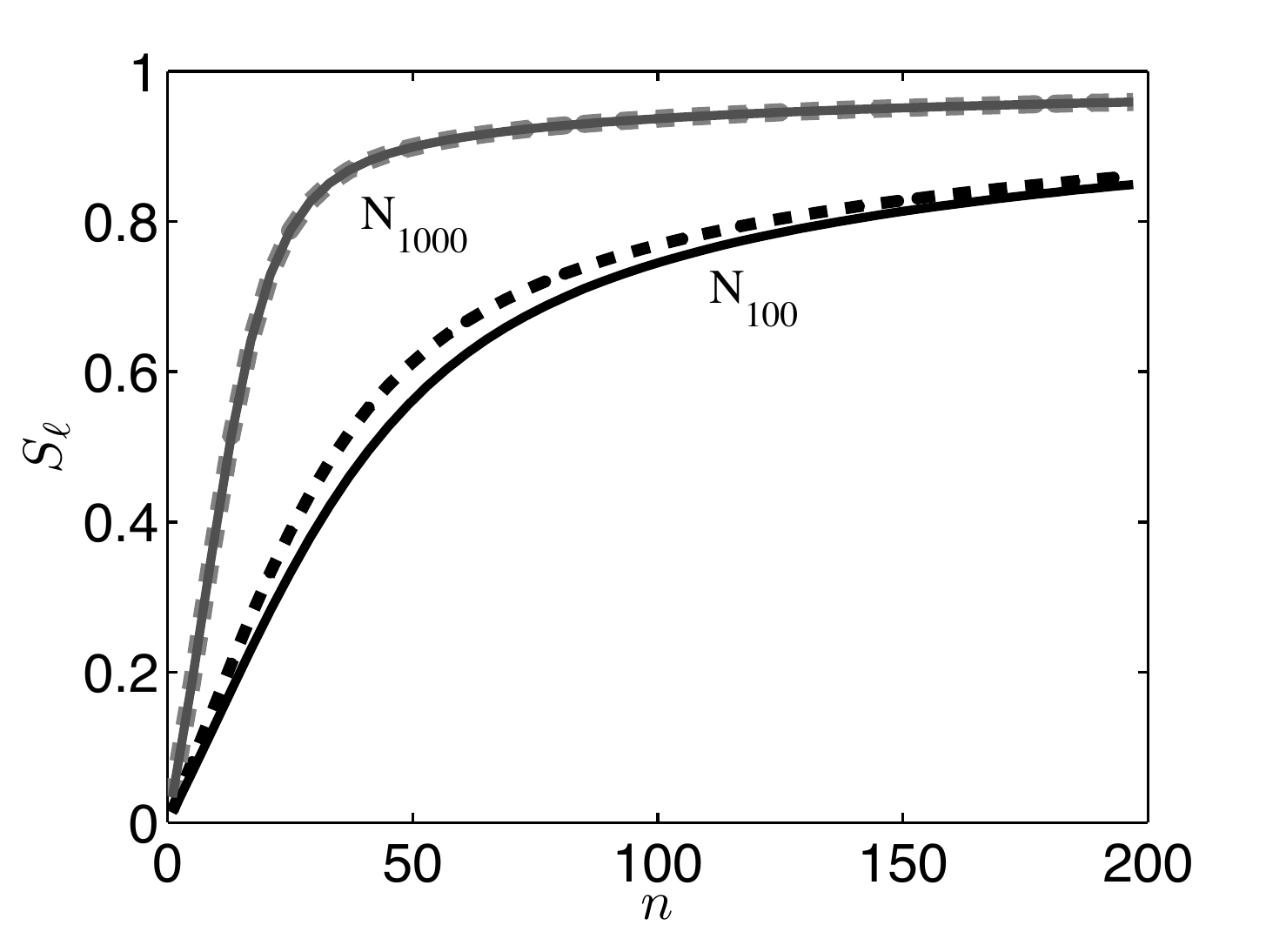}
  \caption{\label{lu_distrib} Theoretical predictions of aggregate
    length and order parameter distributions. Top, aggregate length
    distributions at I-N coexistence of the isotropic (I, dashed
    curves) and nematic (N, solid curves) phases with $\bar{l}_p=1000$
    and $\bar{l}_p=100$ (subscripts).  Note that in this figure the
    predictions of the simple second-virial approximation and the
    Parsons-Lee approximation overlap. Bottom, length-dependent
    aggregate order parameter at I-N coexistence in the nematic phase
    with $\bar{l}_p=1000$ and $\bar{l}_p=100$ (subscripts). Solid
    curves: simple second-virial approximation.  Dashed curves:
    Parsons-Lee approximation. In both cases, the association constant
    $K = e^{\beta E_{\rm bond}} = 5000$, for comparison with the
    simulations of L\"u and Kindt \cite{lu04}.}
\end{figure}

  \begin{figure}
  \includegraphics[width=0.5 \textwidth]{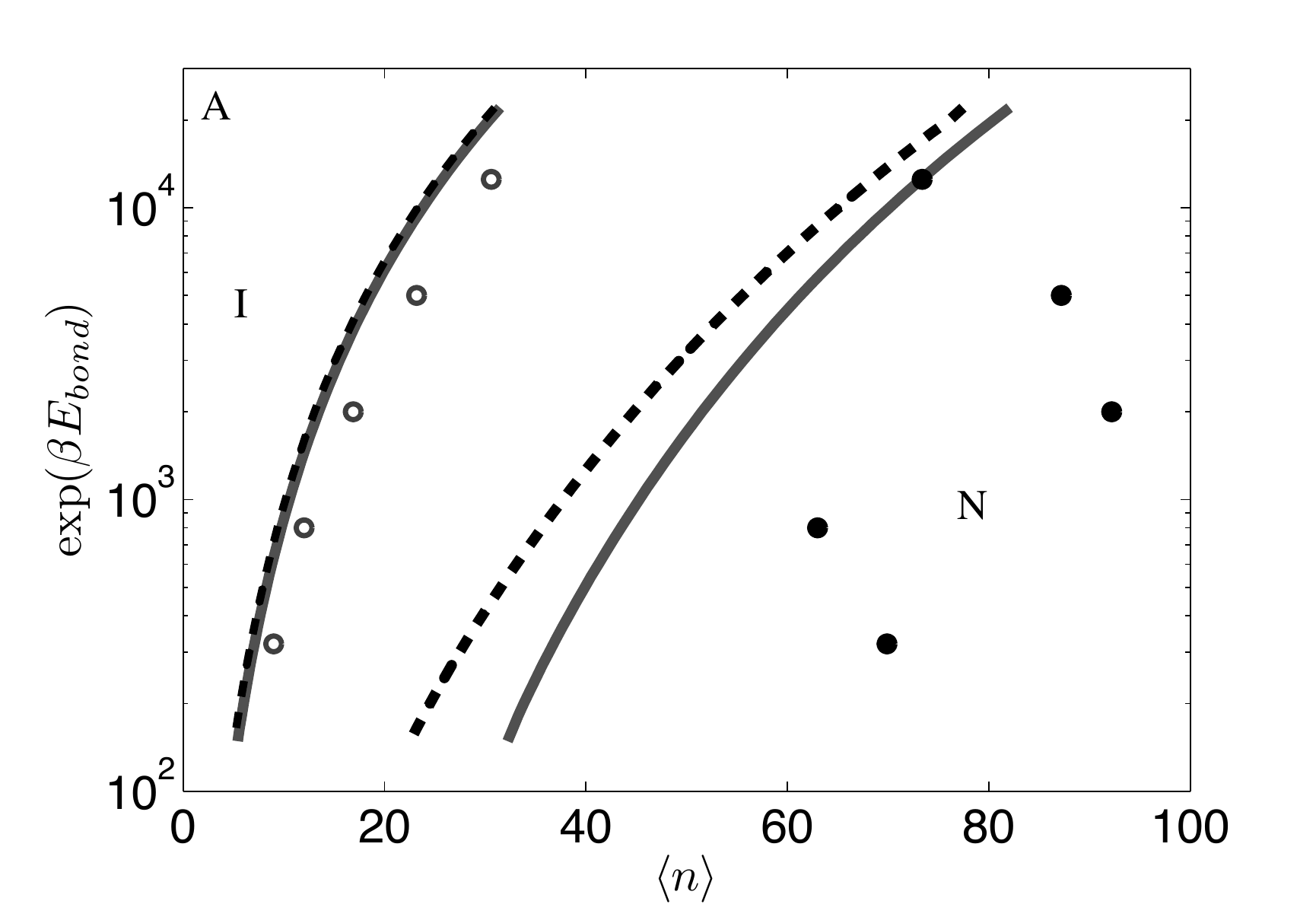}
  \includegraphics[width=0.5 \textwidth]{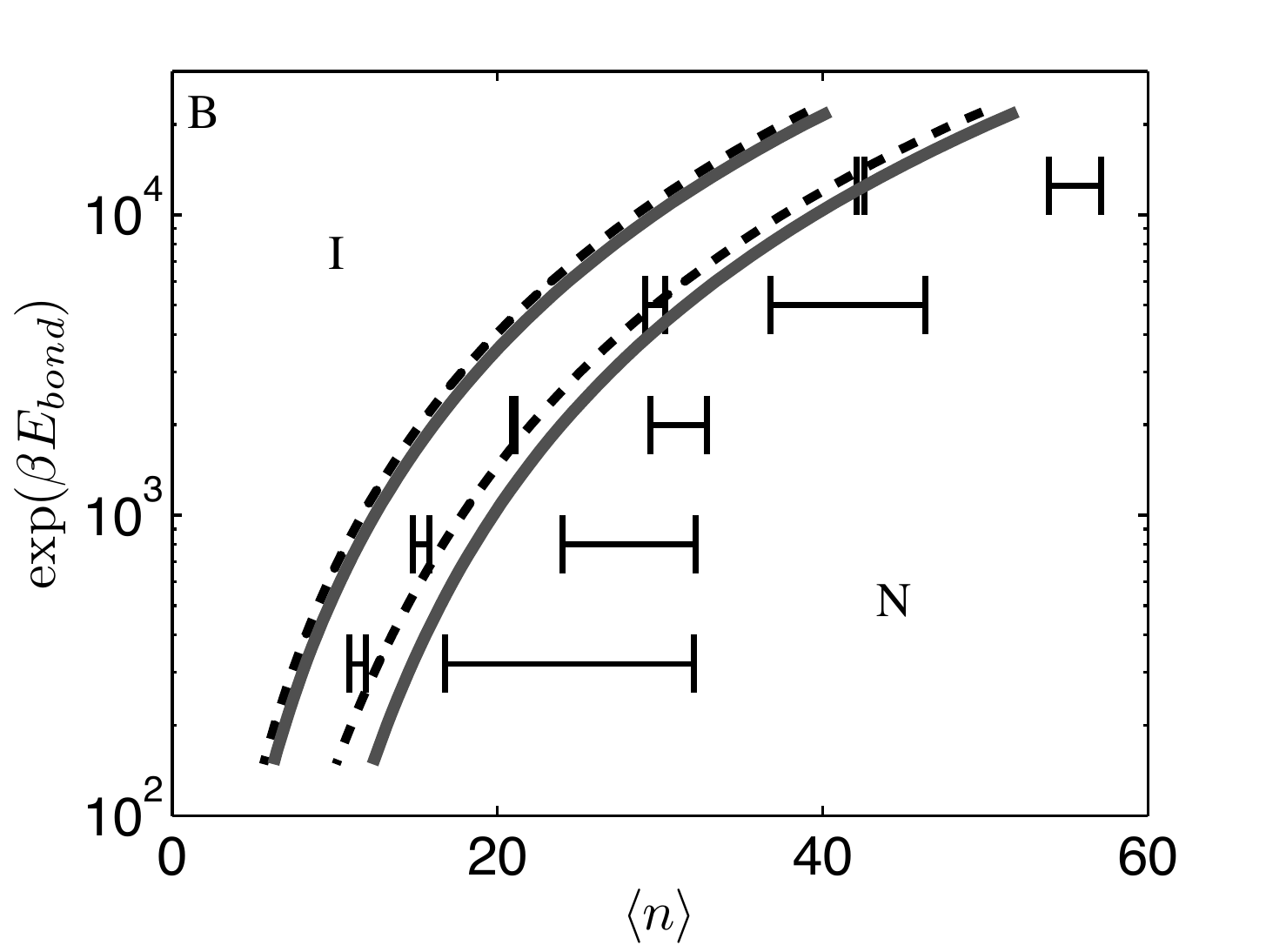}
  \caption{\label{lu_aggregate} Theoretical predictions of mean
    aggregation number in the isotropic and nematic phases at I-N
    coexistence. (A), $\bar{l}_p=1000$. (B), $\bar{l}_p=100$.
    Comparison of theory (solid and dashed curves) to simulation
    results of L\"u and Kindt \cite{lu04} (points).  Curves denote
    borders of regions of I-N coexistence. Solid curves: simple
    second-virial approximation. Dashed curves: Parsons-Lee
    approximation. Points and bars: simulation results of L\"u and
    Kindt \cite{lu04}. Note that for $\bar{l}_p=100$ (B) L\"u and
    Kindt gave ranges of measured values, which we plot using
    horizontal bars, while for $\bar{l}_p=1000$ (A) L\"u and Kindt
    gave only mean values, which we plot using points.}
  \end{figure}

First, we calculated the aggregate length distributions and order
parameter as a function of length to compare with the simulation
results of L\"u and Kindt (Figure \ref{lu_distrib}).  L\"u and Kindt
considered two persistence length values, $\bar{l}_p = 1000$ and
$\bar{l}_p=100$ \cite{lu04}.  Note that $\bar{l}_p = l_p/L$ is the
persistence length divided by the monomer length. (In L\"u and Kindt's
work, the monomers are spheres.) For this figure, we used the the
association constant $K = e^{\beta E_{\rm bond}} = 5000$.

A significant and unexpected result is the calculated biexponential
aggregate length distribution in the nematic phase. This distribution
arises naturally from the free-energy minimization (Figure
\ref{lu_distrib}, top). To our knowledge, our work is the first to
find this biexponential distribution in analytic theory. L\"u and
Kindt's initial analytic work found only a single exponential
distribution \cite{lu04}; in later work they found that assuming a
biexponential distribution improved the theory-simulation agreement
\cite{lu06}.

Overall we find good quantitative agreement between the predictions of
the analytic theory for aggregate length distributions and order
parameter as a function of aggregate length (compare Figure
\ref{lu_distrib} to Figure 2 of reference \cite{lu04}). The aggregate
length distribution is exponential in the isotropic phase and
biexponential in the nematic phase. The nematic phase also shows
significantly longer aggregates.  Stiffer aggregates also tend to be
longer in the nematic phase. The average alignment depends strongly on
aggregate length in the nematic phase, with short aggregates only
weakly aligned and a rapid increase in the length-dependent order
parameter with aggregate length.

\begin{figure}
  \includegraphics[width=0.5
  \textwidth]{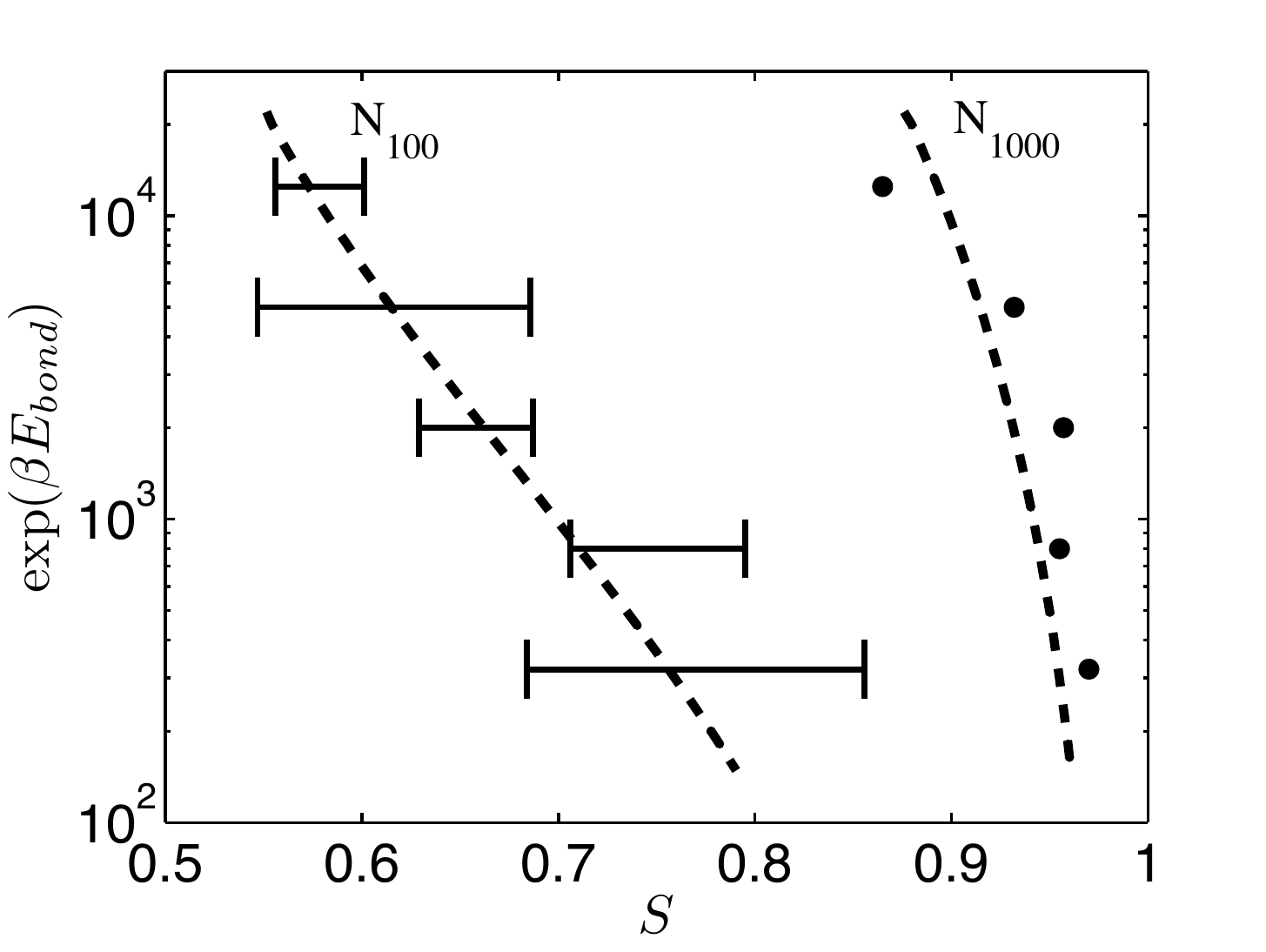}
  \caption{\label{lu_order} Comparison of nematic order parameter at
    I-N coexistence predicted by theory (dashed curves) to simulation
    results of L\"u and Kindt \cite{lu04} (points).  Persistence
    lengths $\bar{l}_p=1000$ and $\bar{l}_p=100$ indicated by
    subscripts.}
\end{figure}

Next we compared our predictions for the aggregate length and nematic
order parameter at I-N coexistence to the simulations of L\"u and
Kindt for a range of monomer binding energies $E_{\rm bond}$.  In
Figure \ref{lu_aggregate} we plot the average aggregation number in
the isotropic and nematic phases at I-N coexistence for a range of
values of $E_{\rm bond}$. The curves are predictions of our analytic
theory and the bars and points are values from the simulations of L\"u
and Kindt \cite{lu04}. Note that for the shorter persistence length
$\bar{l}_p=100$ L\"u and Kindt gave ranges of measured values, which
we plot using horizontal bars, while for the larger persistence length
$\bar{l}_p=1000$ L\"u and Kindt gave only mean values, which we plot
using points. (The same holds for the two figures below in which we
compare to L\"u and Kindt's work). This result illustrates the strong
enhancement of aggregation in the nematic phase. The theory clearly
captures the correct qualitative trends of the simulations: as the
binding energy increases, the aggregates become longer in both the
isotropic and the nematic phases. Quantitatively agreement is
reasonable for the larger persistence length, but for the shorter
persistence length $\bar{l}_p=100$ the predicted values of $\langle n
\rangle$ are off by approximately 50\%. This illustrates the decreased
quantitative accuracy of the theory as the aggregates become less
flexible.

The order parameter in the nematic phase at I-N coexistence is shown
in Figure \ref{lu_order}.  Here our calculations for the order
parameter agree well with the simulation results. As expected, the
stiffer, longer aggregates show a higher average order parameter that
is less sensitive to the binding energy.

\begin{figure}
  \includegraphics[width=0.5
  \textwidth]{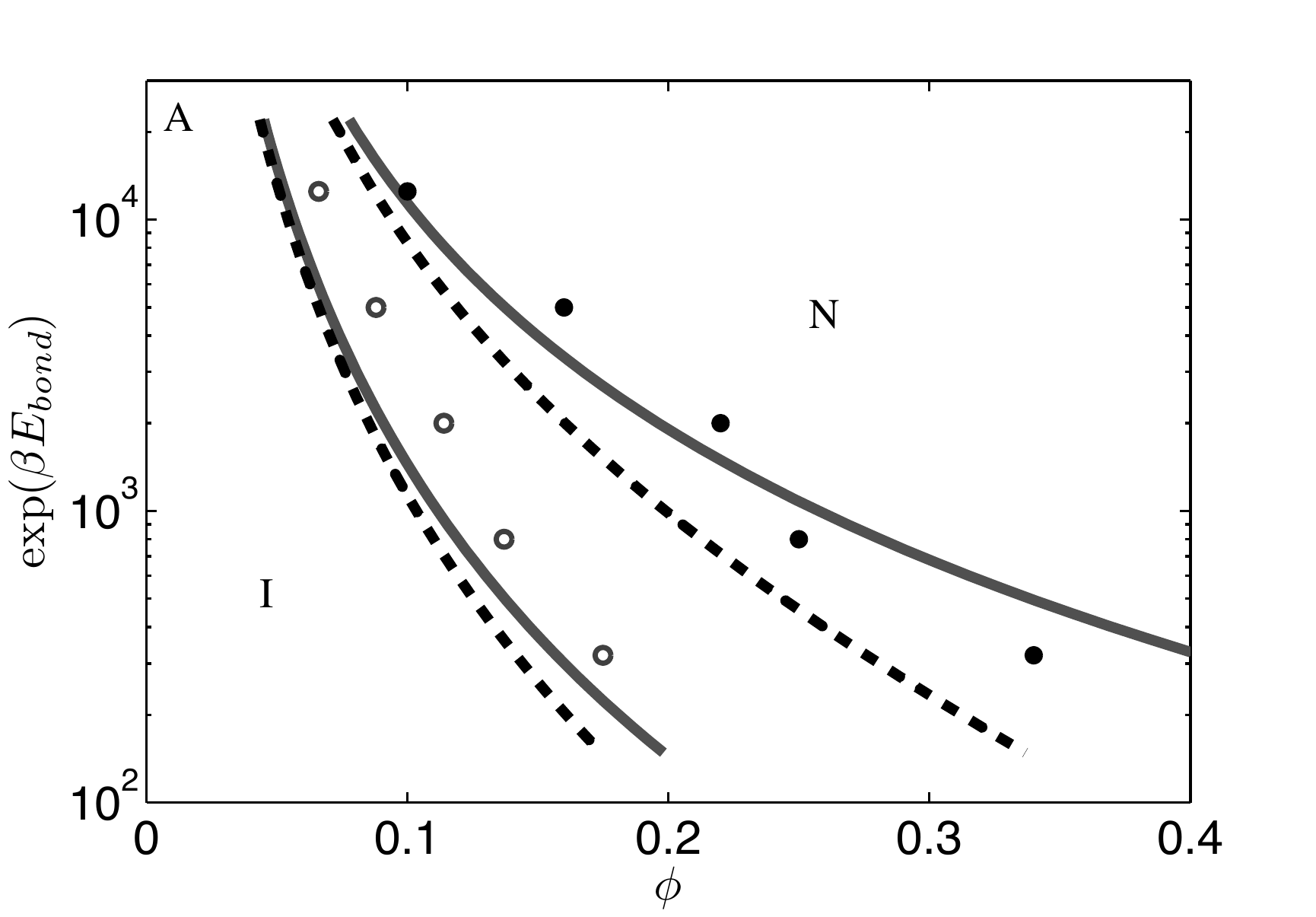}
  \includegraphics[width=0.5
  \textwidth]{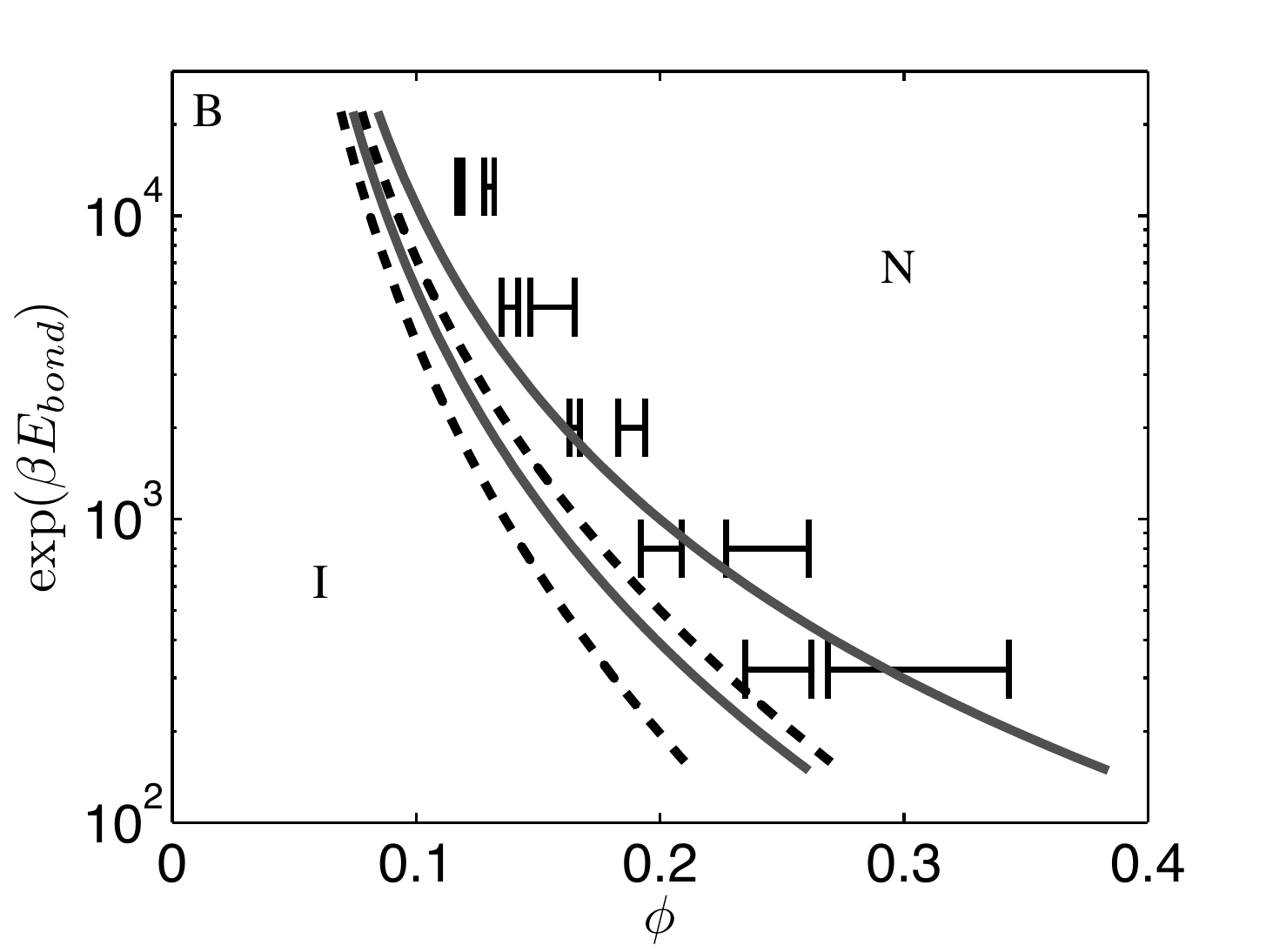}
  \caption{\label{lu_phase} Comparison of theory (solid and dashed
    curves) to simulation results of L\"u and Kindt \cite{lu04}
    (points). (A), $l_p=1000$. (B), $l_p=100$. Curves denote borders
    of regions of I-N coexistence. Solid curves: simple second-virial
    approximation. Dashed curves: Parsons-Lee approximation.}
\end{figure}

\begin{figure}
  \includegraphics[width=0.5
   \textwidth]{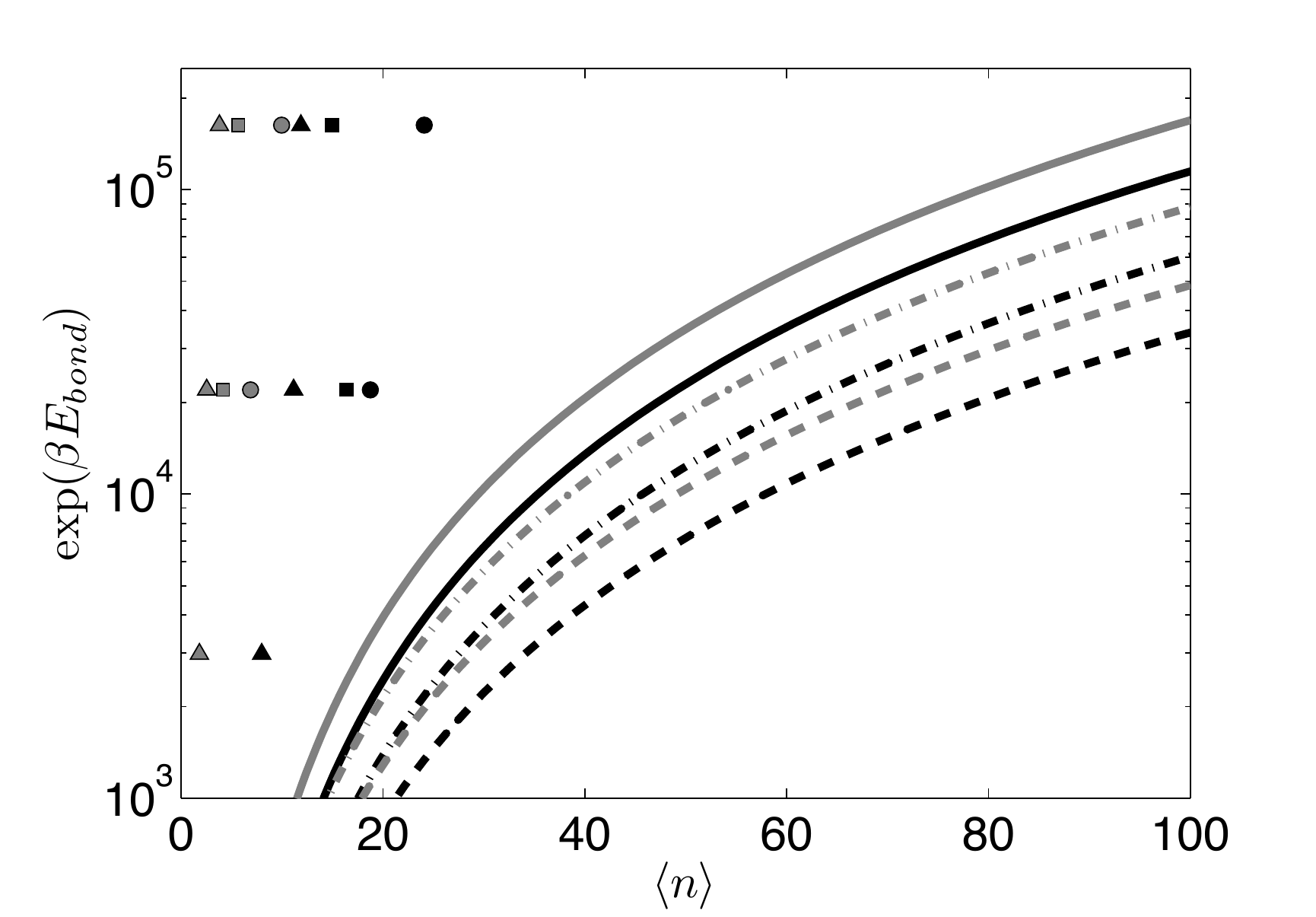}
   \includegraphics[width=0.5
   \textwidth]{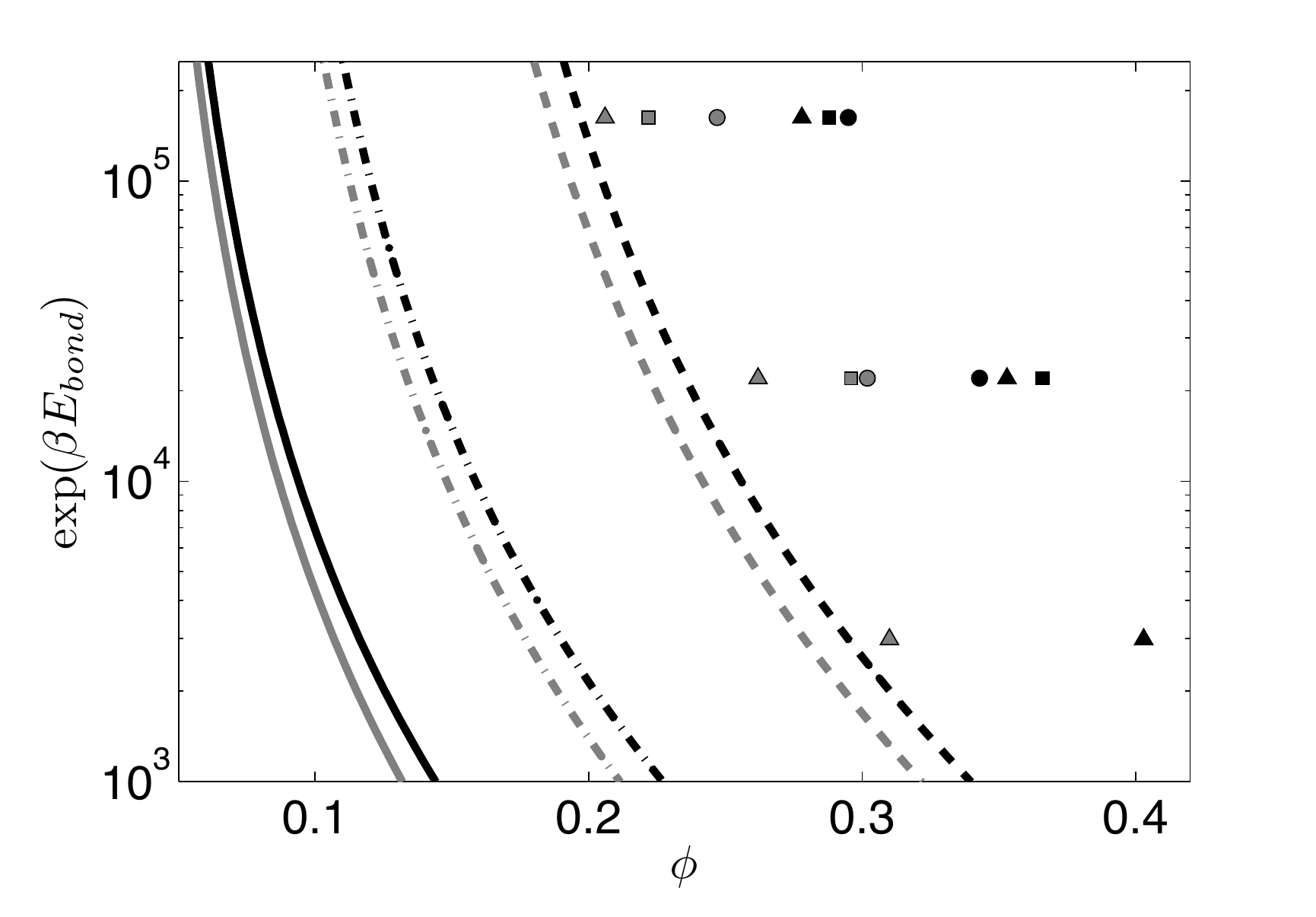}
  \caption{\label{compare_phase} Comparison of I-N coexistence in
    theory (curves) and simulation (points). Top, aggregation number.
    Botton, phase diagarm Curves denote borders of regions of I-N
    coexistence; gray: onset of I-N coexistence; black: upper limit of
    I-N coexistence. Solid curve: $L/D = 2$. Dashed curves: $L/D = 1$.
    Dotted curves: $L/D = 0.5$. In doing the calculations we used the
    empiricial persistence length found from simulations $\bar{l}_p =
    5.07 + 2.14 \beta E_{\rm bond}$. Triangles: $L/D = 2$. Squares:
    $L/D = 1$.  Circles: $L/D = 0.5$.}
\end{figure}

Our final comparison with the simulation results of L\"u and Kindt is
the phase diagrams shown in Figure \ref{lu_phase}. The analytic theory
agrees well with the $\bar{l}_p=1000$ simulation results and less well
with the $\bar{l}_p=100$ simulation results. For $\bar{l}_p=100$ the
analytic theory gives errors of up to about 50\% in the predicted
volume fraction $\phi$ of the coexistence region. Perhaps
surprisingly, we find that the simple second-virial approximation
(solid curves in Figure \ref{lu_phase}) are closer to the simulation
results than the results that include the Parsons-Lee approximation
(dashed curves). 

In our simulations, the ratio of persistence length to monomer length
$\bar{l}_p = L/D$ is significantly less than in the work of L\"u and
Kindt; we find values $\bar{l}_p \approx 20-30$ in the I-N coexistence
range (Table~\ref{averages}). The ratio of mean aggregate length to
persistence length $\ell/l_p \approx 0.1 - 1$ in this range
(Table~\ref{averages}). Because our analytic theory assumes nearly
rigid aggregates ($\ell/l_p \ll 1$) we expect that the theory will
not show good quantitative agreement with the simulations in this
regime. We note that in our simulations the persistence length is not
a control parameter but is determined by the balance of binding energy
and entropy of the aggregates; we found empirically that $\bar{l}_p =
5.07 + 2.14 \beta E_{\rm bond}$ (figure \ref{red_persist_length}). We
used this empirical relationship in the analytic theory.

In Figure \ref{compare_phase} we show the aggregation number and
volume fraction at the borders of I-N coexistence. While the
qualitative trends in the simulations are correctly captured by
theory, there is significant quantitative disagreement. The analytic
theory predicts significantly larger aggregation numbers and
significantly lower volume fractions than found in the simulations.
The biggest discrepancy is with the aggregation numbers, where the
analytic theory predictions are larger than found in simulation by
factors of 2--5. For the volume fraction, the discrepancy is a factor
of 1/3--4. 
However, we do note that although the \textit{position} of
the phase boundaries is not correctly predicted by our theory, their
\textit{slope} is. We note that the slope is affected by the variation
of the persistence length $\bar{l}_p$ with $\beta E_{\rm bond}$:
including this variation in the analytic theory is necessary to
reproduce the slope seen in the simulations.

\begin{figure}
  \includegraphics[width=0.5
   \textwidth]{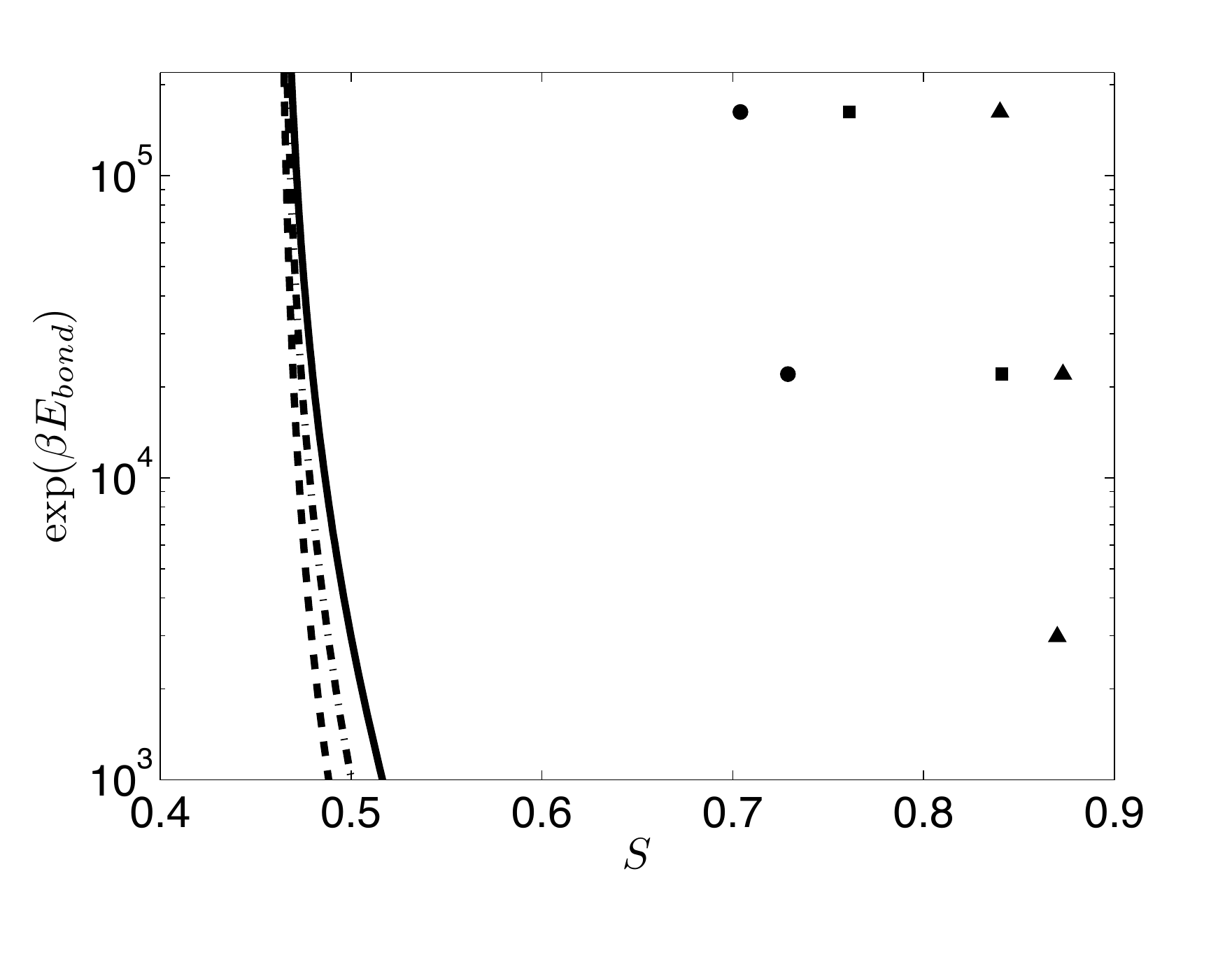}
   \caption{\label{compare_order} Comparison of order parameter in
     theory (curves) and simulation (points) in the nematic phase at
     I-N coexistence. Solid curve: $L/D = 2$. Dashed curves: $L/D =
     1$.  Dotted curves: $L/D = 0.5$. In doing the calculations we
     used the empiricial persistence length found from simulations
     $\bar{l}_p = 5.07 + 2.14 \beta E_{\rm bond}$. Triangles: $L/D =
     2$. Squares: $L/D = 1$.  Circles: $L/D = 0.5$.}
\end{figure}

We also find significant quantitative discrepancy in the order
parameter in the nematic phase at coexistence (Figure
\ref{compare_order}). The theory predicts lower values of $S$ than
found in the simulations by almost a factor of 2.

\section{Conclusion}
\label{conclusion}

Recent experiments on the coupled aggregation and liquid-crystal
ordering of chromonic liquid crystals and short double-stranded
nucleic acids motivated us to consider coarse-grained simulation and
theory of this problem.  In nucleic acid systems in particular, the
monomer aspect ratio can be varied by varying the number of base pairs
per monomer; however, the effects of varying monomer aspect ratio on
the aggregation/LC ordering problem have not been extensively studied
in theoretical work.

In this paper we used both Monte Carlo simulation and analytic theory
to study aggregation and liquid-crystal ordering of a simple model of
sticky cylinders. The model assumes hard cylinders (of length $L$ and
diameter $D$, so the aspect ratio of the monomers can easily be
varied) that experience attractive interactions when they stack end to
end.

We determined approximate phase diagrams by using calculations of the
equation of state in expansion and compression runs; we then studied
the aggregate length distributions, order parameter, and correlation
functions in the different phases.  We find isotropic, nematic,
columnar LC, columnar crystal, and cubatic-like phases are possible
for this system. This family of phases is similar to the results of
Hentschke and colleagues \cite{fodi00,ouyang07} but does show
differences with the work of L\"u and Kindt, who only observed
isotropic and nematic phases \cite{lu04}. Similarly to previous work
we observe the disappearance of the nematic phase (as the monomer
binding energy drops or temperature increases) at an
isotropic-nematic-columnar triple point
\cite{taylor91,fodi00,ouyang07,park08,edwards08}. These phase diagrams
bear a strong qualitative resemblance to those of chromonic liquid
crystals, which show a strong decrease in nematic phase width with
increasing temperature \cite{park08,edwards08}. The location of the
I-N-C triple point appears to be sensitive to the monomer aspect
ratio, occurring at lower $\beta E_{\rm bond}$ for larger $L/D$.

In the simulations we studied three aspect ratios: $L/D = 0.5$, 1, and
2. The same qualitative phase behavior is present for all aspect
ratios. The key differences are that, first, we only observe the
cubatic-like phase for $L/D = 1$. This is consistent with previous
work on hard cylinders \cite{blaak99}. Second, we do not observe the
columnar crystal phase for the lowest aspect ratio of 0.5. Na\"ively
this might be understood because having columns made of a larger
number of shorter monomers makes the increase in entropy allowed by
disorder along the columns more favorable. Third, as noted above, the
nematic phase persists to lower $\beta E_{\rm bond}$ when the aspect
ratio is larger. 
Qualitatively, this makes sense. We find that the mean aggregate
length at I-N coexistence increases with increasing monomer aspect
ratio, which suggests (as in Onsager theory) that the density at
coexistence should also decrease with increasing monomer aspect ratio.
We do observe this decrease, although the trend is not extremely
strong. We also find that the persistence length increases linearly
with increasing monomer aspect ratio; therefore aggregate stiffness
increases with increasing aspect ratio (assuming fixed binding
energy). Both increasing aggregate length and increasing aggregate
stiffness tend to favor nematic order, and therefore tend to broaden
the nematic region.

Our results give evidence that the phase diagram depends on the the
form of the interaction potential, not just the binding energy between
monomers.  The phase behavior is sensitive to the flexibility of the
aggregates, and the aggregate persistence length varies with
temperature in a way that is controlled by the interaction potential.
This suggests that the temperature dependence of the I-N phase
boundary as well as the location of the I-N-C triple point are
influenced by the implicit dependence of $l_p$ on $\beta E_{\rm
  bond}$. This is an interesting direction for future work.

Our simulation results demonstrate a remarkable difference between the
correlation length along aggregates and the length of aggregates in
the nematic and columnar LC phases. Indeed, the aggregate length is
typically 10-100 times larger than the density correlation length
along the columns. This has important implications for attempts to
measure aggregate length distributions: X-ray scattering experiments,
which measure correlation lengths, may not be able to probe aggregate
lengths \cite{park08,joshi09}.

In our computations the aggregate persistence length is determined
from the interaction potential of the monomers; we found a universal
(independent of $L/D$) linear dependence of $\bar{l}_p = l_p/L$ on
$E_{\rm bond}$.

Our extensions of previous theoretical work suggest that quantitative
agreement between analytic theory and simulation for the I-N
transition of aggregates is within reach.  Starting with the
relatively simple second-virial model of aggregation and nematic order
of van der Schoot and Cates \cite{vanderschoot94a}, we added (1) the
Parsons-Lee approximation for higher-density systems, and (2)
expansion of the angular trial function in Legendre polynomials and
numerical solution of the resulting algebraic equations. This
combination significantly improves the quantitative agreement between
the analytic theory and the simulations of L\"u and Kindt \cite{lu04}.

Our analytic theory predicts the biexponential distribution of
aggregate lengths in the nematic phase. Previous analytic work of L\"u
and Kindt assumed a biexponential distribution in the nematic phase,
and based on this assumption found good agreement with their
simulation results \cite{lu06}.  Here the biexponential distribution
arises directly from the free-energy minimization and does not have to
be added to the theory as an assumption.

The comparison between simulation and theory highlights the role of
aggregate flexibility in controling the phase diagram. In our analytic
theory, the aggregates are treated as nearly rigid.  Therefore, our
theoretical results match the simulation results best where the
aggregate persistence length is long ($\bar{l}_p = 1000$).  In this
regime, the quantitative agreement between theory and simulation is
remarkably good, particularly in the prediction of the phase
boundaries, the order parameter, and the biexponential aggregate
length distribution in the nematic phase.  When comparing to the
simulations of L\"u and Kindt with shorter aggregate persistence
length ($\bar{l}_p = 100$) or our simulations ($\bar{l}_p \approx
20-30$), the quantitative agreement between analytic theory and
simulation decreases.  Future improvement to the theory will require
improved treatment of aggregate flexibility.

A major direction for future research is the comparison of our results
to experimental results on chromonic and nucleic-acid liquid crystals,
and refinement of both the simulation model and the analytic theory
based on the comparison. In the future, improvements to the simulation
model and the analytic theory may allow true quantitative agreement
between experiment, simulation, and theory.

\appendix
\section{Free energy minimization in the nematic phase}
\label{calc}

Here we outline the free energy minimization and numerical solution
for the function $M(x)$ introduced in Eq.~(\ref{eq:trial}).  Plugging
the trial function Eq.~(\ref{eq:trial}) into the normalization
condition Eq.~(\ref{eq-rho-norm}) we arrive at the normalization
condition for $\bar{M}(x)\equiv M(x)/M_0$
\begin{equation}
\label{norm-nematic-trial}
\int_{-1}^1 dx  \bar{M}^2(x) =2.
\end{equation}
Minimizing the free energy $f^{(n)}$ (Eq.~(\ref{free-en-M})) with
respect to $\bar{M}(x)$ subject to the normalization constraint gives
\begin{eqnarray}
\frac{\delta (f^{(n)}+\lambda\int_{-1}^1 dx \bar{M}^2(x))}{\delta \bar{M}(x)}=0,
\end{eqnarray}
leading to the equation for $\bar{M}(x)$,
\begin{eqnarray}
\label{M-eq}
&&\bar{M}(x)=\lambda \bar{M}^2(x) -\frac{M_0}{3\bar{l}_p} \bar{M}(x)\partial_x^2 \bar{M}(x)\nonumber\\
&&{}+\frac{4 r\phi M_0}{\pi}\int_{-1}^1 dx^{\prime} \bar{M}^2(x)\bar{M}^2(x^\prime) K(x,x^{\prime}).
\end{eqnarray}
Integrating Eq.~(\ref{M-eq}) over $x$ and using the normalization
condition we obtain the equation for the Lagrange multiplier
$\lambda$:
\begin{equation}
\label{eq-lambda-prime}
\lambda= \frac{1}{2}\left( I_1 + \frac{M_0}{3\bar{l}_p}\, I_2 -
\frac{4\,r\phi M_0}{\pi}\, I_3  \right),
\end{equation}
where the three integrals are
\begin{eqnarray}
I_1 &=& \int_{-1}^1 dx \bar{M}(x),\label{eq-M-I1}\\
I_2 &=& \int_{-1}^1 dx\, \bar{M}(x) \, \partial^2_x\, \bar{M}(x),\label{eq-M-I2}\\
I_3 &=& \int_{-1}^1 dx_1 dx_2\,K(x_1,x_2)\,\bar{M}^2(x_1)\,\bar{M}^2(x_2)\label{eq-M-I3}.
\end{eqnarray} 

To solve equation \eqref{M-eq} and determine the function $\bar{M}(x)$
that minimizes the free energy we seek a series solution: expansion of
$\bar{M}(x)$ in Legendre polynomials turns equation \eqref{M-eq} into
a system of coupled algebraic equations that can straightforwardly be
truncated and solved numerically. We expand $\bar{M}(x)$ in even
Legendre polynomials $P_{2k}(x)$ (note that the nematic phase is
symmetric under $x \to -x$, so only even Legendre polynomials are
allowed):
\begin{equation}
\label{eq-M-expansion}
\bar{M}(x)=\sum_{k=0}^{\infty} m_{2k}\,P_{2k}(x)
\end{equation}
Plugging this series expansion into Eq~(\ref{M-eq}) gives the
equations for the coefficients $m_{2k}$, after using the orthogonality
condition and completeness relation for Legendre polynomials:
\begin{eqnarray}
m_{2k} &=& \frac{4k+1}{2}\left[\lambda\,\alpha_{2k}+\sum_{i,j}\Big
  \{(4 i+1) k_{2j} \Lambda_1\, \alpha_{2i}\,\alpha_{2j}\right. \nonumber\\
&& \left.{}+
(2i)(2i+1) \Lambda_2\, m_{2i}\,m_{2j}\,\Big \} I_{2k,2i,2j} \right].
\label{m-eq}
\end{eqnarray}
The dimensionless parameters $\Lambda_1$ and $\Lambda_2$ are
\begin{eqnarray} 
\Lambda_1 &=&\frac{4\,r\,\phi\, M_0}{\pi},\\
 \Lambda_2 &=& \frac{2M_0}{3\bar{l}_p}.
\end{eqnarray}
The coefficients $k_{2j}$ come from the expansion of the interaction kernel
\begin{equation}
\label{eq:sin-expand}
K(x_1,x_2)=\int_0^{2\pi}\frac{d\phi_1}{2\pi}\frac{d\phi_2}{2\pi}
|\sin\gamma| 
\end{equation}
in Legendre polynomials,
\begin{equation}
K(x_1,x_2) = \sum_{n=0}^{\infty} k_{2n}P_{2n}(x_1)P_{2n}(x_2),
\end{equation}
with $k_0=\pi/4,\ k_2=-5\pi/32$ and
\begin{equation}
\label{coeff-k2n}
k_{2n}= - \frac{\pi(4n+1)(2n-3)!!(2n-1)!!}{2^{2n+2}n!(n+1)!}
\quad\quad (n>1). 
\end{equation}
 The coefficients $\alpha_{2n}$ are
\begin{equation}
\label{alpha-def}
\alpha_{2n}=2\sum_{a,b} m_{2a}\,m_{2b}\, I_{2a,2b,2n},
\end{equation}
and the integral of 3 Legendre polynomials is
\begin{equation}
\label{three-leg}
I_{2k,2i,2j}=\frac{1}{2}\int_{-1}^1 dx P_{2k}(x)P_{2i}(x)P_{2j}(x),
\end{equation}
This integral is known exactly \cite{vanderschoot94a}: it is equal to
0 when $k+i-j<0$ or $k-i+j<0$ or $-k+i+j<0$. In all other cases
\begin{eqnarray}
  &&I_{2k,2i,2j}= \nonumber \\
  &&\frac{(2k+2i-2j)!(2k-2i+2j)!
    (-2k+2i+2j)!}{(2k+2i+2j+1)!} \times \nonumber\\  
  &&\left(\frac{(k+i+j)!}{(k+i-j)!(k-i+j)!(-k+i+j)!} \right)^2.
\end{eqnarray}

The three integrals in Eqs.~(\ref{eq-M-I1})-(\ref{eq-M-I3}) become,
after plugging in the series expansion Eq.~(\ref{eq-M-expansion}),
\begin{eqnarray}
I_1&=& 2\,m_0,\label{I1-coef}\\
I_2&=& -\sum_n \frac{4n(2n+1)}{4n+1}(m_{2n})^2, \label{I2-coef}\\
I_3&=& \sum_{n}k_{2n}(\alpha_{2n})^2.\label{I3-coef}
\end{eqnarray}

We truncated and solved the system of algebraic
equations~(\ref{eq-lambda-prime}), (\ref{m-eq}) and
(\ref{I1-coef})-(\ref{I3-coef}) numerically in Matlab.  We typically
kept 25-30 expansion coefficients for the function $\bar{M}(x)$,
depending on how well the numerical solution converged.

In terms of the expansion coeficients $m_{2k}$ the mean aggregation
number $\langle n\rangle$ in the nematic phase is
\begin{equation}
\langle n \rangle = \frac{2 M_0}{\int_{-1}^1 dx \bar{M}(x)}
=\frac{M_0}{m_0},
\end{equation}
and the order parameter is
\begin{equation}
  S=\frac{\int_{-1}^1 dx P_2(x) \bar{M}^2(x)}{\int_{-1}^1 dx
    \bar{M}^2(x)}= \sum_{i,j}m_{2i}m_{2j} I_{2,2i,2j}. 
\end{equation}

\begin{acknowledgements}
  The authors thank Tommaso Bellini, Noel Clark, Joel Eaves, and Leo
  Radzihovsky for useful discussions. This work was supported by NSF
  MRSEC Grant DMR-0820579, NSF Grant DMR-0847685 to MDB, the ICAM
  Branches Cost Sharing Fund through a fellowship to TK, and by a
  generous grant of computer time from Joel Eaves.
\end{acknowledgements}

\bibliography{aggregates}{}

\begin{thebibliography}{90}%
\makeatletter
\providecommand \@ifxundefined [1]{%
 \@ifx{#1\undefined}
}%
\providecommand \@ifnum [1]{%
 \ifnum #1\expandafter \@firstoftwo
 \else \expandafter \@secondoftwo
 \fi
}%
\providecommand \@ifx [1]{%
 \ifx #1\expandafter \@firstoftwo
 \else \expandafter \@secondoftwo
 \fi
}%
\providecommand \natexlab [1]{#1}%
\providecommand \enquote  [1]{``#1''}%
\providecommand \bibnamefont  [1]{#1}%
\providecommand \bibfnamefont [1]{#1}%
\providecommand \citenamefont [1]{#1}%
\providecommand \href@noop [0]{\@secondoftwo}%
\providecommand \href [0]{\begingroup \@sanitize@url \@href}%
\providecommand \@href[1]{\@@startlink{#1}\@@href}%
\providecommand \@@href[1]{\endgroup#1\@@endlink}%
\providecommand \@sanitize@url [0]{\catcode `\\12\catcode `\$12\catcode
  `\&12\catcode `\#12\catcode `\^12\catcode `\_12\catcode `\%12\relax}%
\providecommand \@@startlink[1]{}%
\providecommand \@@endlink[0]{}%
\providecommand \url  [0]{\begingroup\@sanitize@url \@url }%
\providecommand \@url [1]{\endgroup\@href {#1}{\urlprefix }}%
\providecommand \urlprefix  [0]{URL }%
\providecommand \Eprint [0]{\href }%
\@ifxundefined \urlstyle {%
  \providecommand \doi  [0]{\begingroup \@sanitize@url \@doi}%
  \providecommand \@doi [1]{\endgroup \@@startlink {\doibase
  #1}doi:\discretionary {}{}{}#1\@@endlink }%
}{%
  \providecommand \doi  [0]{doi:\discretionary{}{}{}\begingroup
  \urlstyle{rm}\Url }%
}%
\providecommand \doibase [0]{http://dx.doi.org/}%
\providecommand \Doi [0]{\begingroup \@sanitize@url \@Doi }%
\providecommand \@Doi  [1]{\endgroup\@@startlink{\doibase#1}\@@Doi}%
\providecommand \@@Doi [1]{#1\@@endlink}%
\providecommand \selectlanguage [0]{\@gobble}%
\providecommand \bibinfo  [0]{\@secondoftwo}%
\providecommand \bibfield  [0]{\@secondoftwo}%
\providecommand \translation [1]{[#1]}%
\providecommand \BibitemOpen [0]{}%
\providecommand \bibitemStop [0]{}%
\providecommand \bibitemNoStop [0]{.\EOS\space}%
\providecommand \EOS [0]{\spacefactor3000\relax}%
\providecommand \BibitemShut  [1]{\csname bibitem#1\endcsname}%
\bibitem [{\citenamefont {Hamley}(2007)}]{hamley07}%
  \BibitemOpen
  \bibfield  {author} {\bibinfo {author} {\bibfnamefont {I.~W.}\ \bibnamefont
  {Hamley}},\ }\href@noop {} {\emph {\bibinfo {title} {Introduction to Soft
  Matter}}}\ (\bibinfo  {publisher} {Wiley \& Sons},\ \bibinfo {year}
  {2007})\BibitemShut {NoStop}%
\bibitem [{\citenamefont {Ciferri}(2002)}]{ciferri02}%
  \BibitemOpen
  \bibfield  {author} {\bibinfo {author} {\bibfnamefont {A.}~\bibnamefont
  {Ciferri}},\ }\href@noop {} {\bibfield  {journal} {\bibinfo  {journal}
  {Macromolecular Rapid Communications},\ }\textbf {\bibinfo {volume} {23}},\
  \bibinfo {pages} {511} (\bibinfo {year} {2002})}\BibitemShut {NoStop}%
\bibitem [{\citenamefont {Taylor}\ and\ \citenamefont
  {Herzfeld}(1993)}]{taylor93}%
  \BibitemOpen
  \bibfield  {author} {\bibinfo {author} {\bibfnamefont {M.~P.}\ \bibnamefont
  {Taylor}}\ and\ \bibinfo {author} {\bibfnamefont {J.}~\bibnamefont
  {Herzfeld}},\ }\href@noop {} {\bibfield  {journal} {\bibinfo  {journal}
  {Journal of Physics-Condensed Matter},\ }\textbf {\bibinfo {volume} {5}},\
  \bibinfo {pages} {2651} (\bibinfo {year} {1993})}\BibitemShut {NoStop}%
\bibitem [{\citenamefont {Ciferri}(1999)}]{ciferri99}%
  \BibitemOpen
  \bibfield  {author} {\bibinfo {author} {\bibfnamefont {A.}~\bibnamefont
  {Ciferri}},\ }\href@noop {} {\bibfield  {journal} {\bibinfo  {journal}
  {Liquid Crystals},\ }\textbf {\bibinfo {volume} {26}},\ \bibinfo {pages}
  {489} (\bibinfo {year} {1999})}\BibitemShut {NoStop}%
\bibitem [{\citenamefont {Herzfeld}\ and\ \citenamefont
  {Briehl}(1981)}]{herzfeld81}%
  \BibitemOpen
  \bibfield  {author} {\bibinfo {author} {\bibfnamefont {J.}~\bibnamefont
  {Herzfeld}}\ and\ \bibinfo {author} {\bibfnamefont {R.~W.}\ \bibnamefont
  {Briehl}},\ }\href@noop {} {\bibfield  {journal} {\bibinfo  {journal}
  {Macromolecules},\ }\textbf {\bibinfo {volume} {14}},\ \bibinfo {pages} {397}
  (\bibinfo {year} {1981})}\BibitemShut {NoStop}%
\bibitem [{\citenamefont {Hentschke}\ and\ \citenamefont
  {Herzfeld}(1989)}]{hentschke89}%
  \BibitemOpen
  \bibfield  {author} {\bibinfo {author} {\bibfnamefont {R.}~\bibnamefont
  {Hentschke}}\ and\ \bibinfo {author} {\bibfnamefont {J.}~\bibnamefont
  {Herzfeld}},\ }\href@noop {} {\bibfield  {journal} {\bibinfo  {journal}
  {Journal of Chemical Physics},\ }\textbf {\bibinfo {volume} {90}},\ \bibinfo
  {pages} {5094} (\bibinfo {year} {1989})}\BibitemShut {NoStop}%
\bibitem [{\citenamefont {Aggeli}\ \emph {et~al.}(2003)\citenamefont {Aggeli},
  \citenamefont {Bell}, \citenamefont {Carrick}, \citenamefont {Fishwick},
  \citenamefont {Harding}, \citenamefont {Mawer}, \citenamefont {Radford},
  \citenamefont {Strong},\ and\ \citenamefont {Boden}}]{aggeli03}%
  \BibitemOpen
  \bibfield  {author} {\bibinfo {author} {\bibfnamefont {A.}~\bibnamefont
  {Aggeli}}, \bibinfo {author} {\bibfnamefont {M.}~\bibnamefont {Bell}},
  \bibinfo {author} {\bibfnamefont {L.~M.}\ \bibnamefont {Carrick}}, \bibinfo
  {author} {\bibfnamefont {C.~W.~G.}\ \bibnamefont {Fishwick}}, \bibinfo
  {author} {\bibfnamefont {R.}~\bibnamefont {Harding}}, \bibinfo {author}
  {\bibfnamefont {P.~J.}\ \bibnamefont {Mawer}}, \bibinfo {author}
  {\bibfnamefont {S.~E.}\ \bibnamefont {Radford}}, \bibinfo {author}
  {\bibfnamefont {A.~E.}\ \bibnamefont {Strong}}, \ and\ \bibinfo {author}
  {\bibfnamefont {N.}~\bibnamefont {Boden}},\ }\href@noop {} {\bibfield
  {journal} {\bibinfo  {journal} {Journal of the American Chemical Society},\
  }\textbf {\bibinfo {volume} {125}},\ \bibinfo {pages} {9619} (\bibinfo {year}
  {2003})}\BibitemShut {NoStop}%
\bibitem [{\citenamefont {Ciferri}(2007)}]{ciferri07}%
  \BibitemOpen
  \bibfield  {author} {\bibinfo {author} {\bibfnamefont {A.}~\bibnamefont
  {Ciferri}},\ }\href@noop {} {\bibfield  {journal} {\bibinfo  {journal}
  {Liquid Crystals},\ }\textbf {\bibinfo {volume} {34}},\ \bibinfo {pages}
  {693} (\bibinfo {year} {2007})}\BibitemShut {NoStop}%
\bibitem [{\citenamefont {Lee}(2009){\natexlab{a}}}]{lee09a}%
  \BibitemOpen
  \bibfield  {author} {\bibinfo {author} {\bibfnamefont {C.~F.}\ \bibnamefont
  {Lee}},\ }\href@noop {} {\bibfield  {journal} {\bibinfo  {journal} {Physical
  Review E},\ }\textbf {\bibinfo {volume} {80}} (\bibinfo {year}
  {2009}{\natexlab{a}})}\BibitemShut {NoStop}%
\bibitem [{\citenamefont {Lee}(2009){\natexlab{b}}}]{lee09b}%
  \BibitemOpen
  \bibfield  {author} {\bibinfo {author} {\bibfnamefont {C.~F.}\ \bibnamefont
  {Lee}},\ }\href@noop {} {\bibfield  {journal} {\bibinfo  {journal} {Physical
  Review E},\ }\textbf {\bibinfo {volume} {80}} (\bibinfo {year}
  {2009}{\natexlab{b}})}\BibitemShut {NoStop}%
\bibitem [{\citenamefont {Park}\ \emph {et~al.}(2009)\citenamefont {Park},
  \citenamefont {Han}, \citenamefont {Oh},\ and\ \citenamefont {Kim}}]{park09}%
  \BibitemOpen
  \bibfield  {author} {\bibinfo {author} {\bibfnamefont {J.~S.}\ \bibnamefont
  {Park}}, \bibinfo {author} {\bibfnamefont {T.~H.}\ \bibnamefont {Han}},
  \bibinfo {author} {\bibfnamefont {J.~K.}\ \bibnamefont {Oh}}, \ and\ \bibinfo
  {author} {\bibfnamefont {S.~O.}\ \bibnamefont {Kim}},\ }\href@noop {}
  {\bibfield  {journal} {\bibinfo  {journal} {Macromolecular Chemistry and
  Physics},\ }\textbf {\bibinfo {volume} {210}},\ \bibinfo {pages} {1283}
  (\bibinfo {year} {2009})}\BibitemShut {NoStop}%
\bibitem [{\citenamefont {Jung}\ and\ \citenamefont {Mezzenga}(2010)}]{jung10}%
  \BibitemOpen
  \bibfield  {author} {\bibinfo {author} {\bibfnamefont {J.~M.}\ \bibnamefont
  {Jung}}\ and\ \bibinfo {author} {\bibfnamefont {R.}~\bibnamefont
  {Mezzenga}},\ }\href@noop {} {\bibfield  {journal} {\bibinfo  {journal}
  {Langmuir},\ }\textbf {\bibinfo {volume} {26}},\ \bibinfo {pages} {504}
  (\bibinfo {year} {2010})}\BibitemShut {NoStop}%
\bibitem [{\citenamefont {Khan}(1996)}]{khan96}%
  \BibitemOpen
  \bibfield  {author} {\bibinfo {author} {\bibfnamefont {A.}~\bibnamefont
  {Khan}},\ }\href@noop {} {\bibfield  {journal} {\bibinfo  {journal} {Current
  Opinion in Colloid \& Interface Science},\ }\textbf {\bibinfo {volume} {1}},\
  \bibinfo {pages} {614} (\bibinfo {year} {1996})}\BibitemShut {NoStop}%
\bibitem [{\citenamefont {Charvolin}\ \emph {et~al.}(1979)\citenamefont
  {Charvolin}, \citenamefont {Levelut},\ and\ \citenamefont
  {Samulski}}]{charvolin79}%
  \BibitemOpen
  \bibfield  {author} {\bibinfo {author} {\bibfnamefont {J.}~\bibnamefont
  {Charvolin}}, \bibinfo {author} {\bibfnamefont {A.~M.}\ \bibnamefont
  {Levelut}}, \ and\ \bibinfo {author} {\bibfnamefont {E.~T.}\ \bibnamefont
  {Samulski}},\ }\href@noop {} {\bibfield  {journal} {\bibinfo  {journal}
  {Journal De Physique Lettres},\ }\textbf {\bibinfo {volume} {40}},\ \bibinfo
  {pages} {L587} (\bibinfo {year} {1979})}\BibitemShut {NoStop}%
\bibitem [{\citenamefont {Hendrikx}\ and\ \citenamefont
  {Charvolin}(1981)}]{hendrikx81}%
  \BibitemOpen
  \bibfield  {author} {\bibinfo {author} {\bibfnamefont {Y.}~\bibnamefont
  {Hendrikx}}\ and\ \bibinfo {author} {\bibfnamefont {J.}~\bibnamefont
  {Charvolin}},\ }\href@noop {} {\bibfield  {journal} {\bibinfo  {journal}
  {Journal De Physique},\ }\textbf {\bibinfo {volume} {42}},\ \bibinfo {pages}
  {1427} (\bibinfo {year} {1981})}\BibitemShut {NoStop}%
\bibitem [{\citenamefont {Hendrikx}\ \emph {et~al.}(1983)\citenamefont
  {Hendrikx}, \citenamefont {Charvolin}, \citenamefont {Rawiso}, \citenamefont
  {Liebert},\ and\ \citenamefont {Holmes}}]{hendrikx83}%
  \BibitemOpen
  \bibfield  {author} {\bibinfo {author} {\bibfnamefont {Y.}~\bibnamefont
  {Hendrikx}}, \bibinfo {author} {\bibfnamefont {J.}~\bibnamefont {Charvolin}},
  \bibinfo {author} {\bibfnamefont {M.}~\bibnamefont {Rawiso}}, \bibinfo
  {author} {\bibfnamefont {L.}~\bibnamefont {Liebert}}, \ and\ \bibinfo
  {author} {\bibfnamefont {M.~C.}\ \bibnamefont {Holmes}},\ }\href@noop {}
  {\bibfield  {journal} {\bibinfo  {journal} {Journal of Physical Chemistry},\
  }\textbf {\bibinfo {volume} {87}},\ \bibinfo {pages} {3991} (\bibinfo {year}
  {1983})}\BibitemShut {NoStop}%
\bibitem [{\citenamefont {Boden}\ \emph {et~al.}(1985)\citenamefont {Boden},
  \citenamefont {Bushby},\ and\ \citenamefont {Hardy}}]{boden85}%
  \BibitemOpen
  \bibfield  {author} {\bibinfo {author} {\bibfnamefont {N.}~\bibnamefont
  {Boden}}, \bibinfo {author} {\bibfnamefont {R.~J.}\ \bibnamefont {Bushby}}, \
  and\ \bibinfo {author} {\bibfnamefont {C.}~\bibnamefont {Hardy}},\
  }\href@noop {} {\bibfield  {journal} {\bibinfo  {journal} {Journal De
  Physique Lettres},\ }\textbf {\bibinfo {volume} {46}},\ \bibinfo {pages}
  {L325} (\bibinfo {year} {1985})}\BibitemShut {NoStop}%
\bibitem [{\citenamefont {Boden}\ \emph {et~al.}(1990)\citenamefont {Boden},
  \citenamefont {Clements}, \citenamefont {Jolley}, \citenamefont {Parker},\
  and\ \citenamefont {Smith}}]{boden90}%
  \BibitemOpen
  \bibfield  {author} {\bibinfo {author} {\bibfnamefont {N.}~\bibnamefont
  {Boden}}, \bibinfo {author} {\bibfnamefont {J.}~\bibnamefont {Clements}},
  \bibinfo {author} {\bibfnamefont {K.~W.}\ \bibnamefont {Jolley}}, \bibinfo
  {author} {\bibfnamefont {D.}~\bibnamefont {Parker}}, \ and\ \bibinfo {author}
  {\bibfnamefont {M.~H.}\ \bibnamefont {Smith}},\ }\href@noop {} {\bibfield
  {journal} {\bibinfo  {journal} {Journal of Chemical Physics},\ }\textbf
  {\bibinfo {volume} {93}},\ \bibinfo {pages} {9096} (\bibinfo {year}
  {1990})}\BibitemShut {NoStop}%
\bibitem [{\citenamefont {Quist}\ \emph {et~al.}(1992)\citenamefont {Quist},
  \citenamefont {Halle},\ and\ \citenamefont {Furo}}]{quist92}%
  \BibitemOpen
  \bibfield  {author} {\bibinfo {author} {\bibfnamefont {P.~O.}\ \bibnamefont
  {Quist}}, \bibinfo {author} {\bibfnamefont {B.}~\bibnamefont {Halle}}, \ and\
  \bibinfo {author} {\bibfnamefont {I.}~\bibnamefont {Furo}},\ }\href@noop {}
  {\bibfield  {journal} {\bibinfo  {journal} {Journal of Chemical Physics},\
  }\textbf {\bibinfo {volume} {96}},\ \bibinfo {pages} {3875} (\bibinfo {year}
  {1992})}\BibitemShut {NoStop}%
\bibitem [{\citenamefont {Leaver}\ and\ \citenamefont
  {Holmes}(1993)}]{leaver93}%
  \BibitemOpen
  \bibfield  {author} {\bibinfo {author} {\bibfnamefont {M.~S.}\ \bibnamefont
  {Leaver}}\ and\ \bibinfo {author} {\bibfnamefont {M.~C.}\ \bibnamefont
  {Holmes}},\ }\href@noop {} {\bibfield  {journal} {\bibinfo  {journal}
  {Journal De Physique Ii},\ }\textbf {\bibinfo {volume} {3}},\ \bibinfo
  {pages} {105} (\bibinfo {year} {1993})}\BibitemShut {NoStop}%
\bibitem [{\citenamefont {Furo}\ and\ \citenamefont {Halle}(1995)}]{furo95}%
  \BibitemOpen
  \bibfield  {author} {\bibinfo {author} {\bibfnamefont {I.}~\bibnamefont
  {Furo}}\ and\ \bibinfo {author} {\bibfnamefont {B.}~\bibnamefont {Halle}},\
  }\href@noop {} {\bibfield  {journal} {\bibinfo  {journal} {Physical Review
  E},\ }\textbf {\bibinfo {volume} {51}},\ \bibinfo {pages} {466} (\bibinfo
  {year} {1995})}\BibitemShut {NoStop}%
\bibitem [{\citenamefont {Johannesson}\ \emph {et~al.}(1996)\citenamefont
  {Johannesson}, \citenamefont {Furo},\ and\ \citenamefont
  {Halle}}]{johannesson96}%
  \BibitemOpen
  \bibfield  {author} {\bibinfo {author} {\bibfnamefont {H.}~\bibnamefont
  {Johannesson}}, \bibinfo {author} {\bibfnamefont {I.}~\bibnamefont {Furo}}, \
  and\ \bibinfo {author} {\bibfnamefont {B.}~\bibnamefont {Halle}},\
  }\href@noop {} {\bibfield  {journal} {\bibinfo  {journal} {Physical Review
  E},\ }\textbf {\bibinfo {volume} {53}},\ \bibinfo {pages} {4904} (\bibinfo
  {year} {1996})}\BibitemShut {NoStop}%
\bibitem [{\citenamefont {von Berlepsch}\ \emph {et~al.}(1996)\citenamefont
  {von Berlepsch}, \citenamefont {Dautzenberg}, \citenamefont {Rother},\ and\
  \citenamefont {Jager}}]{vonberlepsch96}%
  \BibitemOpen
  \bibfield  {author} {\bibinfo {author} {\bibfnamefont {H.}~\bibnamefont {von
  Berlepsch}}, \bibinfo {author} {\bibfnamefont {H.}~\bibnamefont
  {Dautzenberg}}, \bibinfo {author} {\bibfnamefont {G.}~\bibnamefont {Rother}},
  \ and\ \bibinfo {author} {\bibfnamefont {J.}~\bibnamefont {Jager}},\
  }\href@noop {} {\bibfield  {journal} {\bibinfo  {journal} {Langmuir},\
  }\textbf {\bibinfo {volume} {12}},\ \bibinfo {pages} {3613} (\bibinfo {year}
  {1996})}\BibitemShut {NoStop}%
\bibitem [{\citenamefont {Angelico}\ \emph {et~al.}(2000)\citenamefont
  {Angelico}, \citenamefont {Ceglie}, \citenamefont {Olsson},\ and\
  \citenamefont {Palazzo}}]{angelico00}%
  \BibitemOpen
  \bibfield  {author} {\bibinfo {author} {\bibfnamefont {R.}~\bibnamefont
  {Angelico}}, \bibinfo {author} {\bibfnamefont {A.}~\bibnamefont {Ceglie}},
  \bibinfo {author} {\bibfnamefont {U.}~\bibnamefont {Olsson}}, \ and\ \bibinfo
  {author} {\bibfnamefont {G.}~\bibnamefont {Palazzo}},\ }\href@noop {}
  {\bibfield  {journal} {\bibinfo  {journal} {Langmuir},\ }\textbf {\bibinfo
  {volume} {16}},\ \bibinfo {pages} {2124} (\bibinfo {year}
  {2000})}\BibitemShut {NoStop}%
\bibitem [{\citenamefont {Pereira}\ \emph {et~al.}(2000)\citenamefont
  {Pereira}, \citenamefont {Palangana}, \citenamefont {Mansanares},
  \citenamefont {da~Silva}, \citenamefont {Bento},\ and\ \citenamefont
  {Baesso}}]{pereira00}%
  \BibitemOpen
  \bibfield  {author} {\bibinfo {author} {\bibfnamefont {J.~R.~D.}\
  \bibnamefont {Pereira}}, \bibinfo {author} {\bibfnamefont {A.~J.}\
  \bibnamefont {Palangana}}, \bibinfo {author} {\bibfnamefont {A.~M.}\
  \bibnamefont {Mansanares}}, \bibinfo {author} {\bibfnamefont {E.~C.}\
  \bibnamefont {da~Silva}}, \bibinfo {author} {\bibfnamefont {A.~C.}\
  \bibnamefont {Bento}}, \ and\ \bibinfo {author} {\bibfnamefont {M.~L.}\
  \bibnamefont {Baesso}},\ }\href@noop {} {\bibfield  {journal} {\bibinfo
  {journal} {Physical Review E},\ }\textbf {\bibinfo {volume} {61}},\ \bibinfo
  {pages} {5410} (\bibinfo {year} {2000})}\BibitemShut {NoStop}%
\bibitem [{\citenamefont {Fischer}\ and\ \citenamefont
  {Callaghan}(2001)}]{fischer01}%
  \BibitemOpen
  \bibfield  {author} {\bibinfo {author} {\bibfnamefont {E.}~\bibnamefont
  {Fischer}}\ and\ \bibinfo {author} {\bibfnamefont {P.~T.}\ \bibnamefont
  {Callaghan}},\ }\href@noop {} {\bibfield  {journal} {\bibinfo  {journal}
  {Physical Review E},\ }\textbf {\bibinfo {volume} {64}} (\bibinfo {year}
  {2001})}\BibitemShut {NoStop}%
\bibitem [{\citenamefont {Goodchild}\ \emph {et~al.}(2007)\citenamefont
  {Goodchild}, \citenamefont {Collier}, \citenamefont {Millar}, \citenamefont
  {Prokes}, \citenamefont {Lord}, \citenamefont {Butts}, \citenamefont
  {Bowers}, \citenamefont {Webster},\ and\ \citenamefont
  {Heenan}}]{goodchild07}%
  \BibitemOpen
  \bibfield  {author} {\bibinfo {author} {\bibfnamefont {I.}~\bibnamefont
  {Goodchild}}, \bibinfo {author} {\bibfnamefont {L.}~\bibnamefont {Collier}},
  \bibinfo {author} {\bibfnamefont {S.~L.}\ \bibnamefont {Millar}}, \bibinfo
  {author} {\bibfnamefont {I.}~\bibnamefont {Prokes}}, \bibinfo {author}
  {\bibfnamefont {J.~C.~D.}\ \bibnamefont {Lord}}, \bibinfo {author}
  {\bibfnamefont {C.~P.}\ \bibnamefont {Butts}}, \bibinfo {author}
  {\bibfnamefont {J.}~\bibnamefont {Bowers}}, \bibinfo {author} {\bibfnamefont
  {J.~R.~P.}\ \bibnamefont {Webster}}, \ and\ \bibinfo {author} {\bibfnamefont
  {R.~K.}\ \bibnamefont {Heenan}},\ }\href@noop {} {\bibfield  {journal}
  {\bibinfo  {journal} {Journal of Colloid and Interface Science},\ }\textbf
  {\bibinfo {volume} {307}},\ \bibinfo {pages} {455} (\bibinfo {year}
  {2007})}\BibitemShut {NoStop}%
\bibitem [{\citenamefont {Kuntz}\ and\ \citenamefont {Walker}(2008)}]{kuntz08}%
  \BibitemOpen
  \bibfield  {author} {\bibinfo {author} {\bibfnamefont {D.~M.}\ \bibnamefont
  {Kuntz}}\ and\ \bibinfo {author} {\bibfnamefont {L.~M.}\ \bibnamefont
  {Walker}},\ }\href@noop {} {\bibfield  {journal} {\bibinfo  {journal} {Soft
  Matter},\ }\textbf {\bibinfo {volume} {4}},\ \bibinfo {pages} {286} (\bibinfo
  {year} {2008})}\BibitemShut {NoStop}%
\bibitem [{\citenamefont {Ciuchi}\ \emph {et~al.}(1994)\citenamefont {Ciuchi},
  \citenamefont {Dinicola}, \citenamefont {Franz}, \citenamefont {Gottarelli},
  \citenamefont {Mariani}, \citenamefont {Bossi},\ and\ \citenamefont
  {Spada}}]{ciuchi94}%
  \BibitemOpen
  \bibfield  {author} {\bibinfo {author} {\bibfnamefont {F.}~\bibnamefont
  {Ciuchi}}, \bibinfo {author} {\bibfnamefont {G.}~\bibnamefont {Dinicola}},
  \bibinfo {author} {\bibfnamefont {H.}~\bibnamefont {Franz}}, \bibinfo
  {author} {\bibfnamefont {G.}~\bibnamefont {Gottarelli}}, \bibinfo {author}
  {\bibfnamefont {P.}~\bibnamefont {Mariani}}, \bibinfo {author} {\bibfnamefont
  {M.~G.~P.}\ \bibnamefont {Bossi}}, \ and\ \bibinfo {author} {\bibfnamefont
  {G.~P.}\ \bibnamefont {Spada}},\ }\href@noop {} {\bibfield  {journal}
  {\bibinfo  {journal} {Journal of the American Chemical Society},\ }\textbf
  {\bibinfo {volume} {116}},\ \bibinfo {pages} {7064} (\bibinfo {year}
  {1994})}\BibitemShut {NoStop}%
\bibitem [{\citenamefont {Franz}\ \emph {et~al.}(1994)\citenamefont {Franz},
  \citenamefont {Ciuchi}, \citenamefont {Dinicola}, \citenamefont {Demorais},\
  and\ \citenamefont {Mariani}}]{franz94}%
  \BibitemOpen
  \bibfield  {author} {\bibinfo {author} {\bibfnamefont {H.}~\bibnamefont
  {Franz}}, \bibinfo {author} {\bibfnamefont {F.}~\bibnamefont {Ciuchi}},
  \bibinfo {author} {\bibfnamefont {G.}~\bibnamefont {Dinicola}}, \bibinfo
  {author} {\bibfnamefont {M.~M.}\ \bibnamefont {Demorais}}, \ and\ \bibinfo
  {author} {\bibfnamefont {P.}~\bibnamefont {Mariani}},\ }\href@noop {}
  {\bibfield  {journal} {\bibinfo  {journal} {Physical Review E},\ }\textbf
  {\bibinfo {volume} {50}},\ \bibinfo {pages} {395} (\bibinfo {year}
  {1994})}\BibitemShut {NoStop}%
\bibitem [{\citenamefont {Spindler}\ \emph {et~al.}(2002)\citenamefont
  {Spindler}, \citenamefont {Olenik}, \citenamefont {Copic}, \citenamefont
  {Romih}, \citenamefont {Cerar}, \citenamefont {Skerjanc},\ and\ \citenamefont
  {Mariani}}]{spindler02}%
  \BibitemOpen
  \bibfield  {author} {\bibinfo {author} {\bibfnamefont {L.}~\bibnamefont
  {Spindler}}, \bibinfo {author} {\bibfnamefont {I.~D.}\ \bibnamefont
  {Olenik}}, \bibinfo {author} {\bibfnamefont {M.}~\bibnamefont {Copic}},
  \bibinfo {author} {\bibfnamefont {R.}~\bibnamefont {Romih}}, \bibinfo
  {author} {\bibfnamefont {J.}~\bibnamefont {Cerar}}, \bibinfo {author}
  {\bibfnamefont {J.}~\bibnamefont {Skerjanc}}, \ and\ \bibinfo {author}
  {\bibfnamefont {P.}~\bibnamefont {Mariani}},\ }\href@noop {} {\bibfield
  {journal} {\bibinfo  {journal} {European Physical Journal E},\ }\textbf
  {\bibinfo {volume} {7}},\ \bibinfo {pages} {95} (\bibinfo {year}
  {2002})}\BibitemShut {NoStop}%
\bibitem [{\citenamefont {Mangenot}\ \emph {et~al.}(2003)\citenamefont
  {Mangenot}, \citenamefont {Leforestier}, \citenamefont {Durand},\ and\
  \citenamefont {Livolant}}]{mangenot03}%
  \BibitemOpen
  \bibfield  {author} {\bibinfo {author} {\bibfnamefont {S.}~\bibnamefont
  {Mangenot}}, \bibinfo {author} {\bibfnamefont {A.}~\bibnamefont
  {Leforestier}}, \bibinfo {author} {\bibfnamefont {D.}~\bibnamefont {Durand}},
  \ and\ \bibinfo {author} {\bibfnamefont {F.}~\bibnamefont {Livolant}},\
  }\href@noop {} {\bibfield  {journal} {\bibinfo  {journal} {Journal of
  Molecular Biology},\ }\textbf {\bibinfo {volume} {333}},\ \bibinfo {pages}
  {907} (\bibinfo {year} {2003})}\BibitemShut {NoStop}%
\bibitem [{\citenamefont {Tiddy}\ \emph {et~al.}(1995)\citenamefont {Tiddy},
  \citenamefont {Mateer}, \citenamefont {Ormerod}, \citenamefont {Harrison},\
  and\ \citenamefont {Edwards}}]{tiddy95}%
  \BibitemOpen
  \bibfield  {author} {\bibinfo {author} {\bibfnamefont {G.~J.~T.}\
  \bibnamefont {Tiddy}}, \bibinfo {author} {\bibfnamefont {D.~L.}\ \bibnamefont
  {Mateer}}, \bibinfo {author} {\bibfnamefont {A.~P.}\ \bibnamefont {Ormerod}},
  \bibinfo {author} {\bibfnamefont {W.~J.}\ \bibnamefont {Harrison}}, \ and\
  \bibinfo {author} {\bibfnamefont {D.~J.}\ \bibnamefont {Edwards}},\
  }\href@noop {} {\bibfield  {journal} {\bibinfo  {journal} {Langmuir},\
  }\textbf {\bibinfo {volume} {11}},\ \bibinfo {pages} {390} (\bibinfo {year}
  {1995})}\BibitemShut {NoStop}%
\bibitem [{\citenamefont {Harrison}\ \emph {et~al.}(1996)\citenamefont
  {Harrison}, \citenamefont {Mateer},\ and\ \citenamefont
  {Tiddy}}]{harrison96}%
  \BibitemOpen
  \bibfield  {author} {\bibinfo {author} {\bibfnamefont {W.~J.}\ \bibnamefont
  {Harrison}}, \bibinfo {author} {\bibfnamefont {D.~L.}\ \bibnamefont
  {Mateer}}, \ and\ \bibinfo {author} {\bibfnamefont {G.~J.~T.}\ \bibnamefont
  {Tiddy}},\ }\href@noop {} {\bibfield  {journal} {\bibinfo  {journal} {Faraday
  Discussions},\ }\textbf {\bibinfo {volume} {104}},\ \bibinfo {pages} {139}
  (\bibinfo {year} {1996})}\BibitemShut {NoStop}%
\bibitem [{\citenamefont {Lydon}(1998)}]{lydon98}%
  \BibitemOpen
  \bibfield  {author} {\bibinfo {author} {\bibfnamefont {J.}~\bibnamefont
  {Lydon}},\ }\href@noop {} {\bibfield  {journal} {\bibinfo  {journal} {Current
  Opinion in Colloid \& Interface Science},\ }\textbf {\bibinfo {volume} {3}},\
  \bibinfo {pages} {458} (\bibinfo {year} {1998})}\BibitemShut {NoStop}%
\bibitem [{\citenamefont {von Berlepsch}\ \emph {et~al.}(2000)\citenamefont
  {von Berlepsch}, \citenamefont {Bottcher},\ and\ \citenamefont
  {Dahne}}]{vonberlepsch00}%
  \BibitemOpen
  \bibfield  {author} {\bibinfo {author} {\bibfnamefont {H.}~\bibnamefont {von
  Berlepsch}}, \bibinfo {author} {\bibfnamefont {C.}~\bibnamefont {Bottcher}},
  \ and\ \bibinfo {author} {\bibfnamefont {L.}~\bibnamefont {Dahne}},\
  }\href@noop {} {\bibfield  {journal} {\bibinfo  {journal} {Journal of
  Physical Chemistry B},\ }\textbf {\bibinfo {volume} {104}},\ \bibinfo {pages}
  {8792} (\bibinfo {year} {2000})}\BibitemShut {NoStop}%
\bibitem [{\citenamefont {Lydon}(2004)}]{lydon04}%
  \BibitemOpen
  \bibfield  {author} {\bibinfo {author} {\bibfnamefont {J.}~\bibnamefont
  {Lydon}},\ }\href@noop {} {\bibfield  {journal} {\bibinfo  {journal} {Current
  Opinion in Colloid \& Interface Science},\ }\textbf {\bibinfo {volume} {8}},\
  \bibinfo {pages} {480} (\bibinfo {year} {2004})}\BibitemShut {NoStop}%
\bibitem [{\citenamefont {Horowitz}\ \emph {et~al.}(2005)\citenamefont
  {Horowitz}, \citenamefont {Janowitz}, \citenamefont {Modic}, \citenamefont
  {Heiney},\ and\ \citenamefont {Collings}}]{horowitz05}%
  \BibitemOpen
  \bibfield  {author} {\bibinfo {author} {\bibfnamefont {V.~R.}\ \bibnamefont
  {Horowitz}}, \bibinfo {author} {\bibfnamefont {L.~A.}\ \bibnamefont
  {Janowitz}}, \bibinfo {author} {\bibfnamefont {A.~L.}\ \bibnamefont {Modic}},
  \bibinfo {author} {\bibfnamefont {P.~A.}\ \bibnamefont {Heiney}}, \ and\
  \bibinfo {author} {\bibfnamefont {P.~J.}\ \bibnamefont {Collings}},\
  }\href@noop {} {\bibfield  {journal} {\bibinfo  {journal} {Physical Review
  E},\ }\textbf {\bibinfo {volume} {72}} (\bibinfo {year} {2005})}\BibitemShut
  {NoStop}%
\bibitem [{\citenamefont {Nastishin}\ \emph {et~al.}(2004)\citenamefont
  {Nastishin}, \citenamefont {Liu}, \citenamefont {Shiyanovskii}, \citenamefont
  {Lavrentovich}, \citenamefont {Kostko},\ and\ \citenamefont
  {Anisimov}}]{nastishin04}%
  \BibitemOpen
  \bibfield  {author} {\bibinfo {author} {\bibfnamefont {Y.~A.}\ \bibnamefont
  {Nastishin}}, \bibinfo {author} {\bibfnamefont {H.}~\bibnamefont {Liu}},
  \bibinfo {author} {\bibfnamefont {S.~V.}\ \bibnamefont {Shiyanovskii}},
  \bibinfo {author} {\bibfnamefont {O.~D.}\ \bibnamefont {Lavrentovich}},
  \bibinfo {author} {\bibfnamefont {A.~F.}\ \bibnamefont {Kostko}}, \ and\
  \bibinfo {author} {\bibfnamefont {M.~A.}\ \bibnamefont {Anisimov}},\
  }\href@noop {} {\bibfield  {journal} {\bibinfo  {journal} {Physical Review
  E},\ }\textbf {\bibinfo {volume} {70}} (\bibinfo {year} {2004})}\BibitemShut
  {NoStop}%
\bibitem [{\citenamefont {Nastishin}\ \emph {et~al.}(2005)\citenamefont
  {Nastishin}, \citenamefont {Liu}, \citenamefont {Schneider}, \citenamefont
  {Nazarenko}, \citenamefont {Vasyuta}, \citenamefont {Shiyanovskii},\ and\
  \citenamefont {Lavrentovich}}]{nastishin05}%
  \BibitemOpen
  \bibfield  {author} {\bibinfo {author} {\bibfnamefont {Y.~A.}\ \bibnamefont
  {Nastishin}}, \bibinfo {author} {\bibfnamefont {H.}~\bibnamefont {Liu}},
  \bibinfo {author} {\bibfnamefont {T.}~\bibnamefont {Schneider}}, \bibinfo
  {author} {\bibfnamefont {V.}~\bibnamefont {Nazarenko}}, \bibinfo {author}
  {\bibfnamefont {R.}~\bibnamefont {Vasyuta}}, \bibinfo {author} {\bibfnamefont
  {S.~V.}\ \bibnamefont {Shiyanovskii}}, \ and\ \bibinfo {author}
  {\bibfnamefont {O.~D.}\ \bibnamefont {Lavrentovich}},\ }\href@noop {}
  {\bibfield  {journal} {\bibinfo  {journal} {Physical Review E},\ }\textbf
  {\bibinfo {volume} {72}} (\bibinfo {year} {2005})}\BibitemShut {NoStop}%
\bibitem [{\citenamefont {Prasad}\ \emph {et~al.}(2007)\citenamefont {Prasad},
  \citenamefont {Nair}, \citenamefont {Hegde},\ and\ \citenamefont
  {Jayalakshmi}}]{prasad07}%
  \BibitemOpen
  \bibfield  {author} {\bibinfo {author} {\bibfnamefont {S.~K.}\ \bibnamefont
  {Prasad}}, \bibinfo {author} {\bibfnamefont {G.~G.}\ \bibnamefont {Nair}},
  \bibinfo {author} {\bibfnamefont {G.}~\bibnamefont {Hegde}}, \ and\ \bibinfo
  {author} {\bibfnamefont {V.}~\bibnamefont {Jayalakshmi}},\ }\href@noop {}
  {\bibfield  {journal} {\bibinfo  {journal} {Journal of Physical Chemistry
  B},\ }\textbf {\bibinfo {volume} {111}},\ \bibinfo {pages} {9741} (\bibinfo
  {year} {2007})}\BibitemShut {NoStop}%
\bibitem [{\citenamefont {Edwards}\ \emph {et~al.}(2008)\citenamefont
  {Edwards}, \citenamefont {Jones}, \citenamefont {Lozman}, \citenamefont
  {Ormerod}, \citenamefont {Sintyureva},\ and\ \citenamefont
  {Tiddy}}]{edwards08}%
  \BibitemOpen
  \bibfield  {author} {\bibinfo {author} {\bibfnamefont {D.~J.}\ \bibnamefont
  {Edwards}}, \bibinfo {author} {\bibfnamefont {J.~W.}\ \bibnamefont {Jones}},
  \bibinfo {author} {\bibfnamefont {O.}~\bibnamefont {Lozman}}, \bibinfo
  {author} {\bibfnamefont {A.~P.}\ \bibnamefont {Ormerod}}, \bibinfo {author}
  {\bibfnamefont {M.}~\bibnamefont {Sintyureva}}, \ and\ \bibinfo {author}
  {\bibfnamefont {G.~J.~T.}\ \bibnamefont {Tiddy}},\ }\href@noop {} {\bibfield
  {journal} {\bibinfo  {journal} {Journal of Physical Chemistry B},\ }\textbf
  {\bibinfo {volume} {112}},\ \bibinfo {pages} {14628} (\bibinfo {year}
  {2008})}\BibitemShut {NoStop}%
\bibitem [{\citenamefont {Park}\ \emph {et~al.}(2008)\citenamefont {Park},
  \citenamefont {Kang}, \citenamefont {Tortora}, \citenamefont {Nastishin},
  \citenamefont {Finotello}, \citenamefont {Kumar},\ and\ \citenamefont
  {Lavrentovich}}]{park08}%
  \BibitemOpen
  \bibfield  {author} {\bibinfo {author} {\bibfnamefont {H.~S.}\ \bibnamefont
  {Park}}, \bibinfo {author} {\bibfnamefont {S.~W.}\ \bibnamefont {Kang}},
  \bibinfo {author} {\bibfnamefont {L.}~\bibnamefont {Tortora}}, \bibinfo
  {author} {\bibfnamefont {Y.}~\bibnamefont {Nastishin}}, \bibinfo {author}
  {\bibfnamefont {D.}~\bibnamefont {Finotello}}, \bibinfo {author}
  {\bibfnamefont {S.}~\bibnamefont {Kumar}}, \ and\ \bibinfo {author}
  {\bibfnamefont {O.~D.}\ \bibnamefont {Lavrentovich}},\ }\href@noop {}
  {\bibfield  {journal} {\bibinfo  {journal} {Journal of Physical Chemistry
  B},\ }\textbf {\bibinfo {volume} {112}},\ \bibinfo {pages} {16307} (\bibinfo
  {year} {2008})}\BibitemShut {NoStop}%
\bibitem [{\citenamefont {Joshi}\ \emph {et~al.}(2009)\citenamefont {Joshi},
  \citenamefont {Kang}, \citenamefont {Agra-Kooijman},\ and\ \citenamefont
  {Kumar}}]{joshi09}%
  \BibitemOpen
  \bibfield  {author} {\bibinfo {author} {\bibfnamefont {L.}~\bibnamefont
  {Joshi}}, \bibinfo {author} {\bibfnamefont {S.~W.}\ \bibnamefont {Kang}},
  \bibinfo {author} {\bibfnamefont {D.~M.}\ \bibnamefont {Agra-Kooijman}}, \
  and\ \bibinfo {author} {\bibfnamefont {S.}~\bibnamefont {Kumar}},\
  }\href@noop {} {\bibfield  {journal} {\bibinfo  {journal} {Physical Review
  E},\ }\textbf {\bibinfo {volume} {80}} (\bibinfo {year} {2009})}\BibitemShut
  {NoStop}%
\bibitem [{\citenamefont {Wu}\ \emph {et~al.}(2009)\citenamefont {Wu},
  \citenamefont {Lal}, \citenamefont {Simon}, \citenamefont {Burton},\ and\
  \citenamefont {Luk}}]{wu09}%
  \BibitemOpen
  \bibfield  {author} {\bibinfo {author} {\bibfnamefont {L.}~\bibnamefont
  {Wu}}, \bibinfo {author} {\bibfnamefont {J.}~\bibnamefont {Lal}}, \bibinfo
  {author} {\bibfnamefont {K.~A.}\ \bibnamefont {Simon}}, \bibinfo {author}
  {\bibfnamefont {E.~A.}\ \bibnamefont {Burton}}, \ and\ \bibinfo {author}
  {\bibfnamefont {Y.~Y.}\ \bibnamefont {Luk}},\ }\href@noop {} {\bibfield
  {journal} {\bibinfo  {journal} {Journal of the American Chemical Society},\
  }\textbf {\bibinfo {volume} {131}},\ \bibinfo {pages} {7430} (\bibinfo {year}
  {2009})}\BibitemShut {NoStop}%
\bibitem [{\citenamefont {Nakata}\ \emph {et~al.}(2007)\citenamefont {Nakata},
  \citenamefont {Zanchetta}, \citenamefont {Chapman}, \citenamefont {Jones},
  \citenamefont {Cross}, \citenamefont {Pindak}, \citenamefont {Bellini},\ and\
  \citenamefont {Clark}}]{nakata07}%
  \BibitemOpen
  \bibfield  {author} {\bibinfo {author} {\bibfnamefont {M.}~\bibnamefont
  {Nakata}}, \bibinfo {author} {\bibfnamefont {G.}~\bibnamefont {Zanchetta}},
  \bibinfo {author} {\bibfnamefont {B.~D.}\ \bibnamefont {Chapman}}, \bibinfo
  {author} {\bibfnamefont {C.~D.}\ \bibnamefont {Jones}}, \bibinfo {author}
  {\bibfnamefont {J.~O.}\ \bibnamefont {Cross}}, \bibinfo {author}
  {\bibfnamefont {R.}~\bibnamefont {Pindak}}, \bibinfo {author} {\bibfnamefont
  {T.}~\bibnamefont {Bellini}}, \ and\ \bibinfo {author} {\bibfnamefont
  {N.~A.}\ \bibnamefont {Clark}},\ }\href@noop {} {\bibfield  {journal}
  {\bibinfo  {journal} {Science},\ }\textbf {\bibinfo {volume} {318}},\
  \bibinfo {pages} {1276} (\bibinfo {year} {2007})}\BibitemShut {NoStop}%
\bibitem [{\citenamefont {Zanchetta}\ \emph
  {et~al.}(2008){\natexlab{a}}\citenamefont {Zanchetta}, \citenamefont
  {Nakata}, \citenamefont {Buscaglia}, \citenamefont {Clark},\ and\
  \citenamefont {Bellini}}]{zanchetta08a}%
  \BibitemOpen
  \bibfield  {author} {\bibinfo {author} {\bibfnamefont {G.}~\bibnamefont
  {Zanchetta}}, \bibinfo {author} {\bibfnamefont {M.}~\bibnamefont {Nakata}},
  \bibinfo {author} {\bibfnamefont {M.}~\bibnamefont {Buscaglia}}, \bibinfo
  {author} {\bibfnamefont {N.~A.}\ \bibnamefont {Clark}}, \ and\ \bibinfo
  {author} {\bibfnamefont {T.}~\bibnamefont {Bellini}},\ }\href@noop {}
  {\bibfield  {journal} {\bibinfo  {journal} {Journal of Physics-Condensed
  Matter},\ }\textbf {\bibinfo {volume} {20}} (\bibinfo {year}
  {2008}{\natexlab{a}})}\BibitemShut {NoStop}%
\bibitem [{\citenamefont {Zanchetta}\ \emph
  {et~al.}(2008){\natexlab{b}}\citenamefont {Zanchetta}, \citenamefont
  {Nakata}, \citenamefont {Buscaglia}, \citenamefont {Bellini},\ and\
  \citenamefont {Clark}}]{zanchetta08b}%
  \BibitemOpen
  \bibfield  {author} {\bibinfo {author} {\bibfnamefont {G.}~\bibnamefont
  {Zanchetta}}, \bibinfo {author} {\bibfnamefont {M.}~\bibnamefont {Nakata}},
  \bibinfo {author} {\bibfnamefont {M.}~\bibnamefont {Buscaglia}}, \bibinfo
  {author} {\bibfnamefont {T.}~\bibnamefont {Bellini}}, \ and\ \bibinfo
  {author} {\bibfnamefont {N.~A.}\ \bibnamefont {Clark}},\ }\href@noop {}
  {\bibfield  {journal} {\bibinfo  {journal} {Proceedings of the National
  Academy of Sciences of the United States of America},\ }\textbf {\bibinfo
  {volume} {105}},\ \bibinfo {pages} {1111} (\bibinfo {year}
  {2008}{\natexlab{b}})}\BibitemShut {NoStop}%
\bibitem [{\citenamefont {Gelbart}\ \emph {et~al.}(1985)\citenamefont
  {Gelbart}, \citenamefont {McMullen},\ and\ \citenamefont
  {Benshaul}}]{gelbart85}%
  \BibitemOpen
  \bibfield  {author} {\bibinfo {author} {\bibfnamefont {W.~M.}\ \bibnamefont
  {Gelbart}}, \bibinfo {author} {\bibfnamefont {W.~E.}\ \bibnamefont
  {McMullen}}, \ and\ \bibinfo {author} {\bibfnamefont {A.}~\bibnamefont
  {Benshaul}},\ }\href@noop {} {\bibfield  {journal} {\bibinfo  {journal}
  {Journal De Physique},\ }\textbf {\bibinfo {volume} {46}},\ \bibinfo {pages}
  {1137} (\bibinfo {year} {1985})}\BibitemShut {NoStop}%
\bibitem [{\citenamefont {McMullen}\ \emph {et~al.}(1985)\citenamefont
  {McMullen}, \citenamefont {Gelbart},\ and\ \citenamefont
  {Benshaul}}]{mcmullen85}%
  \BibitemOpen
  \bibfield  {author} {\bibinfo {author} {\bibfnamefont {W.~E.}\ \bibnamefont
  {McMullen}}, \bibinfo {author} {\bibfnamefont {W.~M.}\ \bibnamefont
  {Gelbart}}, \ and\ \bibinfo {author} {\bibfnamefont {A.}~\bibnamefont
  {Benshaul}},\ }\href@noop {} {\bibfield  {journal} {\bibinfo  {journal}
  {Journal of Chemical Physics},\ }\textbf {\bibinfo {volume} {82}},\ \bibinfo
  {pages} {5616} (\bibinfo {year} {1985})}\BibitemShut {NoStop}%
\bibitem [{\citenamefont {Herzfeld}(1988)}]{herzfeld88a}%
  \BibitemOpen
  \bibfield  {author} {\bibinfo {author} {\bibfnamefont {J.}~\bibnamefont
  {Herzfeld}},\ }\href@noop {} {\bibfield  {journal} {\bibinfo  {journal}
  {Journal of Chemical Physics},\ }\textbf {\bibinfo {volume} {88}},\ \bibinfo
  {pages} {2776} (\bibinfo {year} {1988})}\BibitemShut {NoStop}%
\bibitem [{\citenamefont {Herzfeld}\ and\ \citenamefont
  {Taylor}(1988)}]{herzfeld88b}%
  \BibitemOpen
  \bibfield  {author} {\bibinfo {author} {\bibfnamefont {J.}~\bibnamefont
  {Herzfeld}}\ and\ \bibinfo {author} {\bibfnamefont {M.~P.}\ \bibnamefont
  {Taylor}},\ }\href@noop {} {\bibfield  {journal} {\bibinfo  {journal}
  {Journal of Chemical Physics},\ }\textbf {\bibinfo {volume} {88}},\ \bibinfo
  {pages} {2780} (\bibinfo {year} {1988})}\BibitemShut {NoStop}%
\bibitem [{\citenamefont {Taylor}\ and\ \citenamefont
  {Herzfeld}(1990)}]{taylor90}%
  \BibitemOpen
  \bibfield  {author} {\bibinfo {author} {\bibfnamefont {M.~P.}\ \bibnamefont
  {Taylor}}\ and\ \bibinfo {author} {\bibfnamefont {J.}~\bibnamefont
  {Herzfeld}},\ }\href@noop {} {\bibfield  {journal} {\bibinfo  {journal}
  {Langmuir},\ }\textbf {\bibinfo {volume} {6}},\ \bibinfo {pages} {911}
  (\bibinfo {year} {1990})}\BibitemShut {NoStop}%
\bibitem [{\citenamefont {Taylor}\ and\ \citenamefont
  {Herzfeld}(1991)}]{taylor91}%
  \BibitemOpen
  \bibfield  {author} {\bibinfo {author} {\bibfnamefont {M.~P.}\ \bibnamefont
  {Taylor}}\ and\ \bibinfo {author} {\bibfnamefont {J.}~\bibnamefont
  {Herzfeld}},\ }\href@noop {} {\bibfield  {journal} {\bibinfo  {journal}
  {Physical Review A},\ }\textbf {\bibinfo {volume} {43}},\ \bibinfo {pages}
  {1892} (\bibinfo {year} {1991})}\BibitemShut {NoStop}%
\bibitem [{\citenamefont {Odijk}(1987)}]{odijk87}%
  \BibitemOpen
  \bibfield  {author} {\bibinfo {author} {\bibfnamefont {T.}~\bibnamefont
  {Odijk}},\ }\href@noop {} {\bibfield  {journal} {\bibinfo  {journal} {Journal
  De Physique},\ }\textbf {\bibinfo {volume} {48}},\ \bibinfo {pages} {125}
  (\bibinfo {year} {1987})}\BibitemShut {NoStop}%
\bibitem [{\citenamefont {van~der Schoot}\ and\ \citenamefont
  {Cates}(1994){\natexlab{a}}}]{vanderschoot94a}%
  \BibitemOpen
  \bibfield  {author} {\bibinfo {author} {\bibfnamefont {P.}~\bibnamefont
  {van~der Schoot}}\ and\ \bibinfo {author} {\bibfnamefont {M.~E.}\
  \bibnamefont {Cates}},\ }\href@noop {} {\bibfield  {journal} {\bibinfo
  {journal} {Langmuir},\ }\textbf {\bibinfo {volume} {10}},\ \bibinfo {pages}
  {670} (\bibinfo {year} {1994}{\natexlab{a}})}\BibitemShut {NoStop}%
\bibitem [{\citenamefont {Hentschke}(1991)}]{hentschke91a}%
  \BibitemOpen
  \bibfield  {author} {\bibinfo {author} {\bibfnamefont {R.}~\bibnamefont
  {Hentschke}},\ }\href@noop {} {\bibfield  {journal} {\bibinfo  {journal}
  {Liquid Crystals},\ }\textbf {\bibinfo {volume} {10}},\ \bibinfo {pages}
  {691} (\bibinfo {year} {1991})}\BibitemShut {NoStop}%
\bibitem [{\citenamefont {Hentschke}\ and\ \citenamefont
  {Herzfeld}(1991)}]{hentschke91b}%
  \BibitemOpen
  \bibfield  {author} {\bibinfo {author} {\bibfnamefont {R.}~\bibnamefont
  {Hentschke}}\ and\ \bibinfo {author} {\bibfnamefont {J.}~\bibnamefont
  {Herzfeld}},\ }\href@noop {} {\bibfield  {journal} {\bibinfo  {journal}
  {Physical Review A},\ }\textbf {\bibinfo {volume} {44}},\ \bibinfo {pages}
  {1148} (\bibinfo {year} {1991})}\BibitemShut {NoStop}%
\bibitem [{\citenamefont {van~der Schoot}\ and\ \citenamefont
  {Cates}(1994){\natexlab{b}}}]{vanderschoot94b}%
  \BibitemOpen
  \bibfield  {author} {\bibinfo {author} {\bibfnamefont {P.}~\bibnamefont
  {van~der Schoot}}\ and\ \bibinfo {author} {\bibfnamefont {M.~E.}\
  \bibnamefont {Cates}},\ }\href@noop {} {\bibfield  {journal} {\bibinfo
  {journal} {Europhysics Letters},\ }\textbf {\bibinfo {volume} {25}},\
  \bibinfo {pages} {515} (\bibinfo {year} {1994}{\natexlab{b}})}\BibitemShut
  {NoStop}%
\bibitem [{\citenamefont {van~der Schoot}(1994)}]{vanderschoot94c}%
  \BibitemOpen
  \bibfield  {author} {\bibinfo {author} {\bibfnamefont {P.}~\bibnamefont
  {van~der Schoot}},\ }\href@noop {} {\bibfield  {journal} {\bibinfo  {journal}
  {Macromolecules},\ }\textbf {\bibinfo {volume} {27}},\ \bibinfo {pages}
  {6473} (\bibinfo {year} {1994})}\BibitemShut {NoStop}%
\bibitem [{\citenamefont {van~der Schoot}(1995)}]{vanderschoot95}%
  \BibitemOpen
  \bibfield  {author} {\bibinfo {author} {\bibfnamefont {P.}~\bibnamefont
  {van~der Schoot}},\ }\href@noop {} {\bibfield  {journal} {\bibinfo  {journal}
  {Journal De Physique {II}},\ }\textbf {\bibinfo {volume} {5}},\ \bibinfo
  {pages} {243} (\bibinfo {year} {1995})}\BibitemShut {NoStop}%
\bibitem [{\citenamefont {van~der Schoot}(1996)}]{vanderschoot96}%
  \BibitemOpen
  \bibfield  {author} {\bibinfo {author} {\bibfnamefont {P.}~\bibnamefont
  {van~der Schoot}},\ }\href@noop {} {\bibfield  {journal} {\bibinfo  {journal}
  {Journal of Chemical Physics},\ }\textbf {\bibinfo {volume} {104}},\ \bibinfo
  {pages} {1130} (\bibinfo {year} {1996})}\BibitemShut {NoStop}%
\bibitem [{\citenamefont {Odijk}(1996)}]{odijk96}%
  \BibitemOpen
  \bibfield  {author} {\bibinfo {author} {\bibfnamefont {T.}~\bibnamefont
  {Odijk}},\ }\href@noop {} {\bibfield  {journal} {\bibinfo  {journal} {Current
  Opinion in Colloid \& Interface Science},\ }\textbf {\bibinfo {volume} {1}},\
  \bibinfo {pages} {337} (\bibinfo {year} {1996})}\BibitemShut {NoStop}%
\bibitem [{\citenamefont {Kindt}\ and\ \citenamefont
  {Gelbart}(2001)}]{kindt01}%
  \BibitemOpen
  \bibfield  {author} {\bibinfo {author} {\bibfnamefont {J.~T.}\ \bibnamefont
  {Kindt}}\ and\ \bibinfo {author} {\bibfnamefont {W.~M.}\ \bibnamefont
  {Gelbart}},\ }\href@noop {} {\bibfield  {journal} {\bibinfo  {journal}
  {Journal of Chemical Physics},\ }\textbf {\bibinfo {volume} {114}},\ \bibinfo
  {pages} {1432} (\bibinfo {year} {2001})}\BibitemShut {NoStop}%
\bibitem [{\citenamefont {Lu}\ and\ \citenamefont {Kindt}(2004)}]{lu04}%
  \BibitemOpen
  \bibfield  {author} {\bibinfo {author} {\bibfnamefont {X.~J.}\ \bibnamefont
  {Lu}}\ and\ \bibinfo {author} {\bibfnamefont {J.~T.}\ \bibnamefont {Kindt}},\
  }\href@noop {} {\bibfield  {journal} {\bibinfo  {journal} {Journal of
  Chemical Physics},\ }\textbf {\bibinfo {volume} {120}},\ \bibinfo {pages}
  {10328} (\bibinfo {year} {2004})}\BibitemShut {NoStop}%
\bibitem [{\citenamefont {Lu}\ and\ \citenamefont {Kindt}(2006)}]{lu06}%
  \BibitemOpen
  \bibfield  {author} {\bibinfo {author} {\bibfnamefont {X.~J.}\ \bibnamefont
  {Lu}}\ and\ \bibinfo {author} {\bibfnamefont {J.~T.}\ \bibnamefont {Kindt}},\
  }\href@noop {} {\bibfield  {journal} {\bibinfo  {journal} {Journal of
  Chemical Physics},\ }\textbf {\bibinfo {volume} {125}} (\bibinfo {year}
  {2006})}\BibitemShut {NoStop}%
\bibitem [{\citenamefont {Bast}\ and\ \citenamefont
  {Hentschke}(1996){\natexlab{a}}}]{bast96a}%
  \BibitemOpen
  \bibfield  {author} {\bibinfo {author} {\bibfnamefont {T.}~\bibnamefont
  {Bast}}\ and\ \bibinfo {author} {\bibfnamefont {R.}~\bibnamefont
  {Hentschke}},\ }\href@noop {} {\bibfield  {journal} {\bibinfo  {journal}
  {Journal of Molecular Modeling},\ }\textbf {\bibinfo {volume} {2}},\ \bibinfo
  {pages} {330} (\bibinfo {year} {1996}{\natexlab{a}})}\BibitemShut {NoStop}%
\bibitem [{\citenamefont {Bast}\ and\ \citenamefont
  {Hentschke}(1996){\natexlab{b}}}]{bast96b}%
  \BibitemOpen
  \bibfield  {author} {\bibinfo {author} {\bibfnamefont {T.}~\bibnamefont
  {Bast}}\ and\ \bibinfo {author} {\bibfnamefont {R.}~\bibnamefont
  {Hentschke}},\ }\href@noop {} {\bibfield  {journal} {\bibinfo  {journal}
  {Journal of Physical Chemistry},\ }\textbf {\bibinfo {volume} {100}},\
  \bibinfo {pages} {12162} (\bibinfo {year} {1996}{\natexlab{b}})}\BibitemShut
  {NoStop}%
\bibitem [{\citenamefont {Tieleman}\ \emph {et~al.}(2000)\citenamefont
  {Tieleman}, \citenamefont {van~der Spoel},\ and\ \citenamefont
  {Berendsen}}]{tieleman00}%
  \BibitemOpen
  \bibfield  {author} {\bibinfo {author} {\bibfnamefont {D.~P.}\ \bibnamefont
  {Tieleman}}, \bibinfo {author} {\bibfnamefont {D.}~\bibnamefont {van~der
  Spoel}}, \ and\ \bibinfo {author} {\bibfnamefont {H.~J.~C.}\ \bibnamefont
  {Berendsen}},\ }\href@noop {} {\bibfield  {journal} {\bibinfo  {journal}
  {Journal of Physical Chemistry B},\ }\textbf {\bibinfo {volume} {104}},\
  \bibinfo {pages} {6380} (\bibinfo {year} {2000})}\BibitemShut {NoStop}%
\bibitem [{\citenamefont {Mohanty}\ \emph {et~al.}(2006)\citenamefont
  {Mohanty}, \citenamefont {Chou}, \citenamefont {Brostrom},\ and\
  \citenamefont {Aguilera}}]{mohanty06}%
  \BibitemOpen
  \bibfield  {author} {\bibinfo {author} {\bibfnamefont {S.}~\bibnamefont
  {Mohanty}}, \bibinfo {author} {\bibfnamefont {S.~H.}\ \bibnamefont {Chou}},
  \bibinfo {author} {\bibfnamefont {M.}~\bibnamefont {Brostrom}}, \ and\
  \bibinfo {author} {\bibfnamefont {J.}~\bibnamefont {Aguilera}},\ }\href@noop
  {} {\bibfield  {journal} {\bibinfo  {journal} {Molecular Simulation},\
  }\textbf {\bibinfo {volume} {32}},\ \bibinfo {pages} {1179} (\bibinfo {year}
  {2006})}\BibitemShut {NoStop}%
\bibitem [{\citenamefont {Yakovlev}\ and\ \citenamefont
  {Boek}(2007)}]{yakovlev07}%
  \BibitemOpen
  \bibfield  {author} {\bibinfo {author} {\bibfnamefont {D.~S.}\ \bibnamefont
  {Yakovlev}}\ and\ \bibinfo {author} {\bibfnamefont {E.~S.}\ \bibnamefont
  {Boek}},\ }\href@noop {} {\bibfield  {journal} {\bibinfo  {journal}
  {Langmuir},\ }\textbf {\bibinfo {volume} {23}},\ \bibinfo {pages} {6588}
  (\bibinfo {year} {2007})}\BibitemShut {NoStop}%
\bibitem [{\citenamefont {Turner}\ \emph {et~al.}(2010)\citenamefont {Turner},
  \citenamefont {Yin}, \citenamefont {Kindt},\ and\ \citenamefont
  {Zhang}}]{turner10}%
  \BibitemOpen
  \bibfield  {author} {\bibinfo {author} {\bibfnamefont {D.~C.}\ \bibnamefont
  {Turner}}, \bibinfo {author} {\bibfnamefont {F.~C.}\ \bibnamefont {Yin}},
  \bibinfo {author} {\bibfnamefont {J.~T.}\ \bibnamefont {Kindt}}, \ and\
  \bibinfo {author} {\bibfnamefont {H.~L.}\ \bibnamefont {Zhang}},\ }\href@noop
  {} {\bibfield  {journal} {\bibinfo  {journal} {Langmuir},\ }\textbf {\bibinfo
  {volume} {26}},\ \bibinfo {pages} {4687} (\bibinfo {year}
  {2010})}\BibitemShut {NoStop}%
\bibitem [{\citenamefont {Edwards}\ \emph {et~al.}(1995)\citenamefont
  {Edwards}, \citenamefont {Henderson},\ and\ \citenamefont
  {Pinning}}]{edwards95}%
  \BibitemOpen
  \bibfield  {author} {\bibinfo {author} {\bibfnamefont {R.~G.}\ \bibnamefont
  {Edwards}}, \bibinfo {author} {\bibfnamefont {J.~R.}\ \bibnamefont
  {Henderson}}, \ and\ \bibinfo {author} {\bibfnamefont {R.~L.}\ \bibnamefont
  {Pinning}},\ }\href@noop {} {\bibfield  {journal} {\bibinfo  {journal}
  {Molecular Physics},\ }\textbf {\bibinfo {volume} {86}},\ \bibinfo {pages}
  {567} (\bibinfo {year} {1995})}\BibitemShut {NoStop}%
\bibitem [{\citenamefont {Maiti}\ \emph {et~al.}(2002)\citenamefont {Maiti},
  \citenamefont {Lansac}, \citenamefont {Glaser},\ and\ \citenamefont
  {Clark}}]{maiti02}%
  \BibitemOpen
  \bibfield  {author} {\bibinfo {author} {\bibfnamefont {P.~K.}\ \bibnamefont
  {Maiti}}, \bibinfo {author} {\bibfnamefont {Y.}~\bibnamefont {Lansac}},
  \bibinfo {author} {\bibfnamefont {M.~A.}\ \bibnamefont {Glaser}}, \ and\
  \bibinfo {author} {\bibfnamefont {N.~A.}\ \bibnamefont {Clark}},\ }\href@noop
  {} {\bibfield  {journal} {\bibinfo  {journal} {Liquid Crystals},\ }\textbf
  {\bibinfo {volume} {29}},\ \bibinfo {pages} {619} (\bibinfo {year}
  {2002})}\BibitemShut {NoStop}%
\bibitem [{\citenamefont {Rouault}(1998)}]{rouault98b}%
  \BibitemOpen
  \bibfield  {author} {\bibinfo {author} {\bibfnamefont {Y.}~\bibnamefont
  {Rouault}},\ }\href@noop {} {\bibfield  {journal} {\bibinfo  {journal}
  {European Physical Journal B},\ }\textbf {\bibinfo {volume} {6}},\ \bibinfo
  {pages} {75} (\bibinfo {year} {1998})}\BibitemShut {NoStop}%
\bibitem [{\citenamefont {Fodi}\ and\ \citenamefont
  {Hentschke}(2000)}]{fodi00}%
  \BibitemOpen
  \bibfield  {author} {\bibinfo {author} {\bibfnamefont {B.}~\bibnamefont
  {Fodi}}\ and\ \bibinfo {author} {\bibfnamefont {R.}~\bibnamefont
  {Hentschke}},\ }\href@noop {} {\bibfield  {journal} {\bibinfo  {journal}
  {Journal of Chemical Physics},\ }\textbf {\bibinfo {volume} {112}},\ \bibinfo
  {pages} {6917} (\bibinfo {year} {2000})}\BibitemShut {NoStop}%
\bibitem [{\citenamefont {Ouyang}\ and\ \citenamefont
  {Hentschke}(2007)}]{ouyang07}%
  \BibitemOpen
  \bibfield  {author} {\bibinfo {author} {\bibfnamefont {W.~Z.}\ \bibnamefont
  {Ouyang}}\ and\ \bibinfo {author} {\bibfnamefont {R.}~\bibnamefont
  {Hentschke}},\ }\href@noop {} {\bibfield  {journal} {\bibinfo  {journal}
  {Journal of Chemical Physics},\ }\textbf {\bibinfo {volume} {127}},\ \bibinfo
  {pages} {164501} (\bibinfo {year} {2007})}\BibitemShut {NoStop}%
\bibitem [{\citenamefont {Chatterji}\ and\ \citenamefont
  {Pandit}(2003)}]{chatterji03}%
  \BibitemOpen
  \bibfield  {author} {\bibinfo {author} {\bibfnamefont {A.}~\bibnamefont
  {Chatterji}}\ and\ \bibinfo {author} {\bibfnamefont {R.}~\bibnamefont
  {Pandit}},\ }\href@noop {} {\bibfield  {journal} {\bibinfo  {journal}
  {Journal of Statistical Physics},\ }\textbf {\bibinfo {volume} {110}},\
  \bibinfo {pages} {1219} (\bibinfo {year} {2003})}\BibitemShut {NoStop}%
\bibitem [{\citenamefont {Kroeger}\ \emph {et~al.}(2007)\citenamefont
  {Kroeger}, \citenamefont {Deimede}, \citenamefont {Belack}, \citenamefont
  {Lieberwirth}, \citenamefont {Fytas},\ and\ \citenamefont
  {Wegner}}]{kroeger07}%
  \BibitemOpen
  \bibfield  {author} {\bibinfo {author} {\bibfnamefont {A.}~\bibnamefont
  {Kroeger}}, \bibinfo {author} {\bibfnamefont {V.}~\bibnamefont {Deimede}},
  \bibinfo {author} {\bibfnamefont {J.}~\bibnamefont {Belack}}, \bibinfo
  {author} {\bibfnamefont {I.}~\bibnamefont {Lieberwirth}}, \bibinfo {author}
  {\bibfnamefont {G.}~\bibnamefont {Fytas}}, \ and\ \bibinfo {author}
  {\bibfnamefont {G.}~\bibnamefont {Wegner}},\ }\href@noop {} {\bibfield
  {journal} {\bibinfo  {journal} {Macromolecules},\ }\textbf {\bibinfo {volume}
  {40}},\ \bibinfo {pages} {105} (\bibinfo {year} {2007})}\BibitemShut
  {NoStop}%
\bibitem [{\citenamefont {Blaak}\ \emph {et~al.}(1999)\citenamefont {Blaak},
  \citenamefont {Frenkel},\ and\ \citenamefont {Mulder}}]{blaak99}%
  \BibitemOpen
  \bibfield  {author} {\bibinfo {author} {\bibfnamefont {R.}~\bibnamefont
  {Blaak}}, \bibinfo {author} {\bibfnamefont {D.}~\bibnamefont {Frenkel}}, \
  and\ \bibinfo {author} {\bibfnamefont {B.~M.}\ \bibnamefont {Mulder}},\
  }\href@noop {} {\bibfield  {journal} {\bibinfo  {journal} {Journal of
  Chemical Physics},\ }\textbf {\bibinfo {volume} {110}},\ \bibinfo {pages}
  {11652} (\bibinfo {year} {1999})}\BibitemShut {NoStop}%
\bibitem [{\citenamefont {Ibarra-Avalos}\ \emph {et~al.}(2007)\citenamefont
  {Ibarra-Avalos}, \citenamefont {Gil-Villegas},\ and\ \citenamefont
  {Richa}}]{ibarra-avalos07}%
  \BibitemOpen
  \bibfield  {author} {\bibinfo {author} {\bibfnamefont {N.}~\bibnamefont
  {Ibarra-Avalos}}, \bibinfo {author} {\bibfnamefont {A.}~\bibnamefont
  {Gil-Villegas}}, \ and\ \bibinfo {author} {\bibfnamefont {A.~M.}\
  \bibnamefont {Richa}},\ }\href@noop {} {\bibfield  {journal} {\bibinfo
  {journal} {Molecular Simulation},\ }\textbf {\bibinfo {volume} {33}},\
  \bibinfo {pages} {505} (\bibinfo {year} {2007})}\BibitemShut {NoStop}%
\bibitem [{\citenamefont {Frenkel}\ and\ \citenamefont
  {Smit}(2001)}]{frenkel01}%
  \BibitemOpen
  \bibfield  {author} {\bibinfo {author} {\bibfnamefont {D.}~\bibnamefont
  {Frenkel}}\ and\ \bibinfo {author} {\bibfnamefont {B.}~\bibnamefont {Smit}},\
  }\href@noop {} {\emph {\bibinfo {title} {Understanding Molecular
  Simulation}}},\ \bibinfo {edition} {2nd}\ ed.\ (\bibinfo  {publisher}
  {Academic Press},\ \bibinfo {year} {2001})\BibitemShut {NoStop}%
\bibitem [{\citenamefont {Orkoulas}\ and\ \citenamefont
  {Panagiotopoulos}(1994)}]{orkoulas94}%
  \BibitemOpen
  \bibfield  {author} {\bibinfo {author} {\bibfnamefont {G.}~\bibnamefont
  {Orkoulas}}\ and\ \bibinfo {author} {\bibfnamefont {A.~Z.}\ \bibnamefont
  {Panagiotopoulos}},\ }\href@noop {} {\bibfield  {journal} {\bibinfo
  {journal} {Journal of Chemical Physics},\ }\textbf {\bibinfo {volume}
  {101}},\ \bibinfo {pages} {1452} (\bibinfo {year} {1994})}\BibitemShut
  {NoStop}%
\bibitem [{\citenamefont {Whitelam}\ and\ \citenamefont
  {Geissler}(2007)}]{whitelam07}%
  \BibitemOpen
  \bibfield  {author} {\bibinfo {author} {\bibfnamefont {S.}~\bibnamefont
  {Whitelam}}\ and\ \bibinfo {author} {\bibfnamefont {P.~L.}\ \bibnamefont
  {Geissler}},\ }\href@noop {} {\bibfield  {journal} {\bibinfo  {journal}
  {Journal of Chemical Physics},\ }\textbf {\bibinfo {volume} {127}},\ \bibinfo
  {pages} {154101} (\bibinfo {year} {2007})}\BibitemShut {NoStop}%
\bibitem [{\citenamefont {Whitelam}\ and\ \citenamefont
  {Geissler}(2008)}]{whitelam08}%
  \BibitemOpen
  \bibfield  {author} {\bibinfo {author} {\bibfnamefont {S.}~\bibnamefont
  {Whitelam}}\ and\ \bibinfo {author} {\bibfnamefont {P.~L.}\ \bibnamefont
  {Geissler}},\ }\href@noop {} {\bibfield  {journal} {\bibinfo  {journal}
  {Journal of Chemical Physics},\ }\textbf {\bibinfo {volume} {128}},\ \bibinfo
  {pages} {219901} (\bibinfo {year} {2008})}\BibitemShut {NoStop}%
\bibitem [{\citenamefont {Parsons}(1979)}]{parsons79}%
  \BibitemOpen
  \bibfield  {author} {\bibinfo {author} {\bibfnamefont {J.~D.}\ \bibnamefont
  {Parsons}},\ }\href@noop {} {\bibfield  {journal} {\bibinfo  {journal}
  {Physical Review A},\ }\textbf {\bibinfo {volume} {19}},\ \bibinfo {pages}
  {1225} (\bibinfo {year} {1979})}\BibitemShut {NoStop}%
\bibitem [{\citenamefont {Khokhlov}\ and\ \citenamefont
  {Semenov}(1982)}]{khokhlov82}%
  \BibitemOpen
  \bibfield  {author} {\bibinfo {author} {\bibfnamefont {A.~R.}\ \bibnamefont
  {Khokhlov}}\ and\ \bibinfo {author} {\bibfnamefont {A.~N.}\ \bibnamefont
  {Semenov}},\ }\href@noop {} {\bibfield  {journal} {\bibinfo  {journal}
  {Physica A},\ }\textbf {\bibinfo {volume} {112}},\ \bibinfo {pages} {605}
  (\bibinfo {year} {1982})}\BibitemShut {NoStop}%
\bibitem [{\citenamefont {Cinacchi}\ \emph {et~al.}(2004)\citenamefont
  {Cinacchi}, \citenamefont {Mederos},\ and\ \citenamefont
  {Velasco}}]{cinacchi04}%
  \BibitemOpen
  \bibfield  {author} {\bibinfo {author} {\bibfnamefont {G.}~\bibnamefont
  {Cinacchi}}, \bibinfo {author} {\bibfnamefont {L.}~\bibnamefont {Mederos}}, \
  and\ \bibinfo {author} {\bibfnamefont {E.}~\bibnamefont {Velasco}},\
  }\href@noop {} {\bibfield  {journal} {\bibinfo  {journal} {Journal of
  Chemical Physics},\ }\textbf {\bibinfo {volume} {121}},\ \bibinfo {pages}
  {3854} (\bibinfo {year} {2004})}\BibitemShut {NoStop}%
\bibitem [{\citenamefont {Cuetos}\ \emph {et~al.}(2007)\citenamefont {Cuetos},
  \citenamefont {Martinez-Haya}, \citenamefont {Lago},\ and\ \citenamefont
  {Rull}}]{cuetos07}%
  \BibitemOpen
  \bibfield  {author} {\bibinfo {author} {\bibfnamefont {A.}~\bibnamefont
  {Cuetos}}, \bibinfo {author} {\bibfnamefont {B.}~\bibnamefont
  {Martinez-Haya}}, \bibinfo {author} {\bibfnamefont {S.}~\bibnamefont {Lago}},
  \ and\ \bibinfo {author} {\bibfnamefont {L.~F.}\ \bibnamefont {Rull}},\
  }\href@noop {} {\bibfield  {journal} {\bibinfo  {journal} {Physical Review
  E},\ }\textbf {\bibinfo {volume} {75}},\ \bibinfo {pages} {12} (\bibinfo
  {year} {2007})}\BibitemShut {NoStop}%
\bibitem [{\citenamefont {Carnahan}\ and\ \citenamefont
  {Starling}(1969)}]{carnahan69}%
  \BibitemOpen
  \bibfield  {author} {\bibinfo {author} {\bibfnamefont {N.~F.}\ \bibnamefont
  {Carnahan}}\ and\ \bibinfo {author} {\bibfnamefont {K.~E.}\ \bibnamefont
  {Starling}},\ }\href@noop {} {\bibfield  {journal} {\bibinfo  {journal}
  {Journal of Chemical Physics},\ }\textbf {\bibinfo {volume} {51}},\ \bibinfo
  {pages} {635} (\bibinfo {year} {1969})}\BibitemShut {NoStop}%
\end{thebibliography}%
\bibliographystyle{apsrev4-1}

\end{document}